\documentclass[a4paper,11pt]{article}
\usepackage[utf8]{inputenc}

\usepackage{amsmath,amsfonts,amsthm,amssymb,dsfont}
\usepackage{graphicx,wrapfig,lipsum}
\usepackage[section]{placeins}
\usepackage{indentfirst}
\usepackage{cite}

\usepackage{tikz}
\usetikzlibrary{shapes.geometric,decorations.markings}
\usepackage{subfigure}

\usepackage[hidelinks]{hyperref}
\hypersetup{colorlinks = true, linkcolor = red, linktocpage = true, citecolor = blue}

\usepackage{geometry}
\numberwithin{equation}{section}
\newcommand{\Vol}{\text{Vol}}

\begin{document}

\begin{titlepage}

\begin{center}

%-------------------- Title ---------------------
$\phantom{.}$\\ \vspace{2cm}
\noindent{\Large{\textbf{Confinement in $(1+1)$ dimensions: a holographic perspective from I-branes}}}

\vspace{1cm}

%-------------------- Authors ---------------------
Carlos Nunez$^a$\footnote{c.nunez@swansea.ac.uk}, 
Marcelo Oyarzo$^b$\footnote{moyarzoca1@gmail.com} and Ricardo Stuardo$^a$\footnote{ricardostuardotroncoso@gmail.com}

\vspace{0.5cm}

%-------------------- Afiliation ---------------------
\textit{$^a$Department of Physics, Swansea University, Swansea SA2 8PP, United Kingdom\\
$^b$Departamento de Física, Universidad de Concepción, Casilla 160-C, Concepción, Chile }

\end{center}

\vspace{0.5cm}
%-------------------- Abstract ---------------------
\centerline{\textbf{Abstract}} 

\vspace{0.5cm}

\noindent{In this paper we holographically study the strongly coupled dynamics of the field theory on I-branes (D5 branes intersecting on a line). In this regime, the field theory becomes  $(2 + 1)$ dimensional  with 16 supercharges. The dual background  has an IR singularity. We resolve this singularity by compactifying the theory on a circle, preserving 4 supercharges.  We study various aspects: confinement, symmetry breaking, Entanglement Entropy, etc. We also discuss a black membrane solution and make some comments on the string $\sigma$-model on our backgrounds.}

\vspace{0.5cm}

\flushright{\textit{Dedicated to the memory of Roman Jackiw}}

\vspace*{\fill}

\end{titlepage}

%------------------ Table of Contents ------------------
\newpage

\addtocontents{toc}{\protect\thispagestyle{empty}}

\tableofcontents

\thispagestyle{empty}

%---------------------- Main Body -----------------------
\newpage
\setcounter{page}{1}
\setcounter{footnote}{0}

\section{Introduction}
Shortly after the conjectured duality between super conformal field theories and string/M theory on spaces with an AdS factor  \cite{Maldacena:1997re} -\cite{Witten:1998qj}, these ideas were extended to non-conformal situations \cite{Itzhaki:1998dd}-\cite{Bobev:2018eer}.

In this work, we holographically study one particular two dimensional field theory that at weak coupling is defined on the intersection of two stacks of D5 branes (these are called I-branes). Dynamical features of these theories imply that, as the coupling is increased the field theory turns $(2+1)$-dimensional and doubles the amount of SUSY preserved. This dynamics is very well explained in \cite{Itzhaki:2005tu}.

The string background, holographic dual to this strongly coupled QFT is well understood  at large values of the radial coordinate, where it can be written as an intersection of NS five branes. Good control over the string $\sigma$-model in such background has been developed \cite{Itzhaki:2005tu}, \cite{Giveon:2019twx}. Nevertheless, this background is singular in the IR (at small values of a suitable radial coordinate). In this paper we propose a 'completion' of this dual background, making it trustable at low energies. The new solution is very explicit and simple. It preserves four supercharges and can be thought of as the dual to a $(2+1)$-dimensional field theory that is compactified to $(1+1)$-dimensions. The QFT is a two nodes quiver with Chern-Simons terms, connected by bifundamental matter.

In the bulk of this paper, we study holographically various aspects of the strongly coupled dynamics of this QFT. We define a suitable gauge coupling, that suggest a low energy confining behaviour. Theta angles and the breaking of $U(1)$-R symmetries are  discussed, together with an estimate of a density of degrees of freedom as a function of energy.  Maldacena-Wilson loops are calculated (again indicating confinement). Also, 't Hooft loops and Entanglement Entropy on a strip are calculated, discussing how the non-local UV dynamics of the system impacts of these observables.

We also briefly touch upon two aspects that will be further developed in future publications: we present a black membrane solution, obtained as analytic continuation of our new background and calculate some characteristic quantities.
Also, we shortly discuss some aspects of the string $\sigma$-model on our backgrounds.

The paper is organised as follows. In Section \ref{geometry} we present the supergravity backgrounds studied in the rest of the paper. In Section \ref{QFTobs} we propose  the QFT dual to these backgrounds, with various characteristic observables calculated. This proposal is sharpened in Section \ref{otherobs}, where the IR confining behaviour is determined and the influence of the high energy LST dynamics on observables like 't Hooft loops and Entanglement Entropy is discussed. Some aspects of the string $\sigma$-model are discussed in Section \ref{sigmamodelsection}. The conclusions and future lines of research suggested by this paper are written in Section \ref{concl}. Various appendices are included, these discuss in great detail the very many interesting technical aspects needed in the main body of this work.

\section{The supergravity backgrounds}\label{geometry}

In this section we write the supergravity backgrounds studied. The first background already appears in the bibliography \cite{Khuri:1993ii,Cowdall:1998bu},  the second background is new. We refer the reader to Appendices \ref{appendix1} and \ref{appendix2} for detailed derivations.
The associated charges are  studied. A black membrane solution is obtained as a bonus, by performing analytic continuations. Some characteristic observables of the black membrane are calculated.
\subsection{Background I}
To describe the backgrounds we
use the coordinates  $\left( t,x,  \varphi, r ,{\theta_A},{\phi_A},{\psi_A}%
,{\theta_B},{\phi_B},{\psi_B}\right) $. We  set  $\alpha'=g_s=1$ and define two sets of left-invariant forms of $SU(2)$,
\begin{eqnarray}
& & \hat{\omega}_1= \cos {\psi_A} d{\theta_A} +\sin{\psi_A}\sin\theta_A d{\phi_A},\;\;\;\;\; \tilde{\omega}_1= \cos {\psi_B} d {\theta_B} +\sin {\psi_B}\sin\theta_B d{\phi_B},\nonumber\\
& & \hat{\omega}_2= -\sin{\psi_A} d{\theta_A} +\cos {\psi_A}\sin\theta_A d{\phi_A},\;\;\tilde{\omega}_2= -\sin {\psi_B} d{\theta_B} +\cos {\psi_B} \sin\theta_B d {\phi_B},\nonumber\\
& & \hat{\omega}_3= d{\psi_A} +\cos{\theta_A}d{\phi_A},\;\;\;\;\;\;\;\; ~~~~~~~~~\;\;\;\;\; \tilde{\omega}_3= d{\psi_B} +\cos {\theta_B}d {\phi}_B.\nonumber
\end{eqnarray}
In terms of these we present the first background. The string frame metric, the Ramond three form $F_3$, the potential $C_2$ and the  dilaton $\Phi$  read,
\begin{eqnarray}
ds^{2}_{st} &=&r\left\{ -dt^{2}+dx^{2}+ \frac{(e_A^2+e_B^2) d{\varphi}^2}{2}+ \frac{8~ dr^2}{r^{2} (e_A^2+e_B^2) }%
+\frac{2}{e_{A}^{2}}\left[ \hat{%
\omega}_{1}^{2}+\hat{\omega}_{2}^{2}+ \hat{\omega}_{3}^{2}\right] 
 +\frac{2}{e_{B}^{2}}\left[ \tilde{\omega}_{1}^{2}+\tilde{\omega}%
_{2}^{2}+ \tilde{\omega}_{3}^{2}\right]\right\} \ ,\nonumber \\
F_{3} &=& dC_2= -\frac{2}{e_{A}^{2}}\hat{\omega}_{1}\wedge \hat{\omega}%
_{2}\wedge \hat{\omega}_{3} -\frac{2}{e_{B}^{2}}\tilde{\omega}_{1}\wedge \tilde{\omega}_{2}\wedge \tilde{\omega}_3\ ,   \label{background1}\\
C_2&=&
 -\frac{2}{e_{A}^{2}} \psi_A \sin {\theta_A}d{\theta_A}%
\wedge d{\phi_A} - \frac{2}{e_{B}^{2}} \psi_B \sin {\theta_B}d{\theta_B}%
\wedge d{\phi_B} .\nonumber\\
\Phi  &=&\log r\ .  \notag
%H_{3} &=&0.
\end{eqnarray}%
Here $\left(e_A,e_B\right)$ are parameters which are fixed when imposing charge quantisation.  The coordinate $\varphi$ is non-compact and could be rescaled to absorb the prefactor  $\frac{(e_A^2+e_B^2) }{2}$.

There are two three-spheres labelled by $\hat{S}^{3}$ and $\tilde{S}^3$ respectively parametrised by the  Euler angles $[{\theta_A}, {\phi_A}, {\psi_A}]$ and $[{\theta_B},{\phi_B},{\psi_B}]$. The range for these angles are  $\theta
_{A,B}\in \lbrack 0,\pi \lbrack $  $\psi _{A,B} \in \left[0,4\pi\right]$, and $\phi _{A,B}\in \left[
0,2\pi \right] $.

The Ricci scalar for the metric in eq.(\ref{background1}) is
{\begin{equation}
R= -\frac{3\left( e_A^2 +e_B^2\right)}{2r},\label{riccisingular}
\end{equation}}
indicating a singularity at $r=0$. This is the singular behaviour found close to a stack of D5 branes--see for example equation (3.38) in the paper \cite{Lozano:2019emq}. In fact, the background can be understood as the backreaction of two stacks of D5 branes that intersect along the non-compact coordinates $(t, x)$ and extend respectively over $(y_1,y_2,y_3,y_4)$ and $(w_1,w_2,w_3,w_4)$, which can be written as radial coordinates and three spheres. After backreaction (at strong coupling), the two stacks share the directions $[t,x,\varphi]$ wrap the spheres $\hat{S}^3[\theta_A,\phi_A,\psi_A]$ and $\tilde{S}^3[\theta_B,\phi_B,\psi_B]$, as  we find  in the background (\ref{background1}). See details in Appendix \ref{appendix28}.

The singular behaviour at $r\sim 0$ indicates the need of a description in terms of other variables. On the other hand for large values of the radial coordinate $r$, the growth of the dilaton and the string coupling $(g_s\sim e^{\Phi})$ requires an S-duality and the description of the system is in terms of an intersection of NS five branes. The system is then dual to two Little String Theories (LST) that intersect along $(t,x,\varphi)$ each one wrapping the spheres $\hat{S}^3$ and $\tilde{S}^3$. 
We further elaborate on this background in Section \ref{sigmamodelsection} and Appendix \ref{appendix28}.

We are interested in resolving the singular behaviour at $r=0$ of the background  in eq.(\ref{background1}), making the solution in terms of D5 branes trustable in the ''IR-regime''. The goal is to write a trustable dual description for a strongly coupled QFT (that is UV-completed by a LST).  We are also interested in preserving some amount of SUSY for stability purposes.
\subsection{Background II}
 We write below a solution to the equations of motion of Type IIB supergravity that resolves the singular behaviour  by compactifying the coordinate $\varphi$, with a precise period. A fibration between the spheres and the coordinate $\varphi$ is also needed. This solution reads,
\begin{eqnarray}
ds^{2}_{st} &=&r\left\{ -dt^{2}+dx^{2}+f_{s}\left( r\right) d\varphi ^{2}+\frac{4}{r^{2}f_{s}\left( r\right) }%
dr^{2}+\frac{2}{e_{A}^{2}}\left[ \hat{%
\omega}_{1}^{2}+\hat{\omega}_{2}^{2}+\left( \hat{\omega}_{3}-e_{A}Q_{A}\zeta
(r)d\varphi \right) ^{2}\right] \right.   \notag \\
&&\left. +\frac{2}{e_{B}^{2}}\left[ \tilde{\omega}_{1}^{2}+\tilde{\omega}%
_{2}^{2}+\left( \tilde{\omega}_{3}-e_{B}Q_{B}\zeta \left( r\right)
d\varphi \right) ^{2}\right] \right\} \ ,  \label{background 211} \\
F_{3} &=& dC_2= 2 \zeta'(r)dr\wedge d\varphi \wedge \left( 
\frac{Q_{A}}{e_{A}}\hat{\omega}_{3}+\frac{Q_{B}}{e_{B}}\tilde{\omega}%
_{3}\right) +\frac{2}{e_{A}^{2}}\hat{\omega}_{1}\wedge \hat{\omega}%
_{2}\wedge \left( e_{A}Q_{A}\zeta (r)d\varphi -\hat{\omega}_{3}\right)  \notag\\
&&+\frac{2}{e_{B}^{2}}\tilde{\omega}_{1}\wedge \tilde{\omega}_{2}\wedge
\left( e_{B}Q_{B}\zeta (r)d\varphi -\tilde{\omega}_{3}\right) \ , \notag \\
C_2&=&
{\psi_A}\left( \frac{2Q_{A}}{e_{A}}\zeta ^{\prime }\left(
r\right) dr\wedge d\varphi -\frac{2}{e_{A}^{2}}\sin {\theta_A}d{\theta_A}%
\wedge d{\phi_A}\right) +\frac{2}{e_{A}}\cos {\theta_A}Q_{A}\zeta
\left( r\right) d\varphi \wedge d{\phi_A}\nonumber \\
&&+{\psi_B}\left( \frac{2Q_{B}}{e_{B}}\zeta ^{\prime }\left( r\right)
dr\wedge d\varphi -\frac{2}{e_{B}^{2}}\sin {\theta_B}d{\theta_B}%
\wedge d{\phi_B}\right) +\frac{2}{e_{B}}\cos {\theta_B}Q_{B}\zeta
\left( r\right) d\varphi \wedge d{\phi_B}\ .\nonumber\\
\Phi  &=&\log r\ .  \notag
%H_{3} &=&0.
\end{eqnarray}%
Here $\left(e_A, Q_A, e_B, Q_B\right)$ are parameters. The functions $f_s(r), \zeta(r)$ are given by
\begin{eqnarray}
f_{s}(r) &=&\frac{e_{A}^{2}+e_{B}^{2}}{2}-\frac{m}{r^{2}}-\frac{2\left(
Q_{A}^{2}+Q_{B}^{2}\right) }{r^{4}}\equiv \frac{e_{A}^{2}+e_{B}^{2}}{2r^{4}}%
(r^{2}-r_{+}^{2})(r^{2}-r_{-}^{2})\,, \label{eq2.2} \\
\zeta \left( r\right)  &=&\frac{1}{r^{2}}-\frac{1}{r_{+}^{2}}\ , \;\;\;
 r_{\pm }^{2}=\frac{m\pm \sqrt{%
m^{2}+4(Q_{A}^{2}+Q_{B}^{2})(e_{A}^{2}+e_{B}^{2})}}{e_{A}^{2}+e_{B}^{2}}.\label{functionsbackground}
\end{eqnarray}%
If the parameter $m=0$ and $e_A Q_B=\pm e_B Q_A$, the background preserves four supercharges. For the SUSY study and the details of the construction of the background in eqs.(\ref{background 211})-(\ref{functionsbackground}), see Appendices \ref{appendix1} and \ref{appendix2}. 

Note that the circle parametrised by the angle $\varphi$ shrinks smoothly at $r=r_+$ if we choose its periodicity to be 
\begin{eqnarray}\label{EuclideanPeriod}
& & \varphi\sim \varphi +L_\varphi,
%\nonumber\\
~~    L_\varphi\! =\! \frac{8 \pi}{r_+f^{\prime}_s(r_+)}\! =\! \frac{4 \pi}{e_A^2+e_B^2}\left( 1+\frac{m}{\sqrt{m^2+4 (e_A^2+e_B^2)(Q_A^2+Q_B^2) }}   \right) .        \label{periodvarphi}
\end{eqnarray}
In the BPS limit, the Ricci scalar associated with the geometry in eq.(\ref{background 211}) is 
\begin{equation}
R=-\frac{(e_A^2+e_B^2)}{2 e_A^2 r^5}\left(4 Q_A^2 + 3 e_A^2 r^4 \right).\label{Ricci}
\end{equation}
That is bounded for all the range of the radial coordinate $[r_+,\infty)$.

Notice that in the case $Q_A=Q_B=m=0$, the background in eq.(\ref{background 211}) becomes that in eq.(\ref{background1}). On the other hand, in the limit $r\to\infty$ since $\zeta(r)$ is non-vanishing, the background \eqref{background 211} is the same as \eqref{background1} up to a large gauge transformation that cancels the fibration. In some of the observables we discuss below, we can perform a regularisation that takes away the effects on the observable that come from the background (\ref{background1}) from the same observable computed in the background (\ref{background 211}). Also, in what follows, for any object, $\xi _{A,B}$, (like a D-brane) extended along the spheres we use the notation $\xi_{A}\equiv \hat{\xi}$ and $\xi _{B}\equiv \tilde{\xi}$. 

\subsection{Conserved charges }\label{sectioncharges}

To calculate the D5 brane charges associated with the Ramond Field $F_3$ in eq.(\ref{background 211}), we define the three-cycles,
   \begin{equation}
        \mathcal{M}_{A} = (\psi_{A},\theta_{A},\phi_{A}) ,~~~~\mathcal{M}_{B} = (\psi_{B},\theta_{B},\phi_{B}).\label{3-cycles}
    \end{equation}
    Let us call the A-stack of branes to be  the one extended along the coordinates $[t,x,\varphi, \theta_A,\phi_A,\psi_A]$ and analogously for the B-stack.
   To calculate the number of branes in the A-stack, we need to integrate $F_3$ over the three cycle $\mathcal{M}_B$--as this is orthogonal to the A-brane stack. Analogously, the number of branes in the B-stack will be obtained by integrating $F_3$ over $\mathcal{M}_A$.
 
 Setting  $\alpha'=g_s=1$,  the quantisation condition for Dp-branes is 
    \begin{eqnarray}
    & &  {(2\pi)^{7-p} g_s \alpha'^{\frac{7-p}{2}}} N_{Dp}= \int_{\Sigma_{8-p}} F_{8-p},~~\text{leads to}~~ N^{i}_{D5} = \frac{1}{(2\pi)^{2}}\int_{\mathcal{M}_{i}} F_{3}.\nonumber
    \end{eqnarray}
After choosing a convenient orientation for the three-cycles (equivalently, changing the sign of $F_3$) we find for the D5 charges,
    \begin{equation}
         N_{A}= \frac{8}{e^{2}_{B}} ,~~~~
         N_{B} = \frac{8}{e^{2}_{A}} .\label{charges}
         %N^{1}_{D5} &= -\frac{8}{e^{2}_{A}}-\frac{8}{e^{2}_{B}} 
         %= N^{1}_{D5}+N^{2}_{D5} ,\\
         %N^{1}_{D5} &= -\frac{8}{e^{2}_{A}}+\frac{8}{e^{2}_{B}}  = N^{1}_{D5}-N^{2}_{D5}
    \end{equation}
   This implies a quantisation condition for the parameters $(e_A, e_B)$. We could have chosen different three cycles, leading to the same  conditions. The result is the same for the either of the backgrounds in eqs.(\ref{background1}), (\ref{background 211}).
    
%To calculate fluxes, we could have used the cycles
%\begin{equation}
 %      \mathcal{M}_{3} = (\psi_{A},\theta_{A},\phi_{A})\bigg|_{\psi_{A}=\psi_{B},\theta_{B}=\theta_{A},\phi_{B}=\phi_{A}},~~~~    \mathcal{M}_{4} = (\psi_{A},\theta_{A},\phi_{A})\bigg|_{\psi_{A}=\psi_{B},\theta_{B}=\theta_{A},\phi_{B}=2\pi-\phi_{A}},\nonumber
    %   \end{equation}
%obtaining in this case,
%\begin{equation}
%N^{3}_{D5} = \frac{8}{e^{2}_{A}}-\frac{8}{e^{2}_{B}} 
   %      = N^{A}_{D5}+N^{B}_{D5} ,~~~
      %   N^{4}_{D5} = \frac{8}{e^{2}_{A}}+\frac{8}{e^{2}_{B}}  = N^{A}_{D5}-N^{B}_{D5}.\nonumber
        % \end{equation}

\subsection{Bonus: a black membrane solution}

Let us consider our new  background in eq.(\ref{background 211})\footnote{The material in this section arose in discussion with Juan Maldacena, whom we gratefully acknowledge.}. Performing a  double Wick rotation%
\begin{eqnarray}
\varphi \rightarrow  it\ ,\qquad Q_{A,B}\rightarrow -iQ_{A,B}\ , ~~ t \rightarrow iy,
\end{eqnarray}%
we find a black membrane configuration, which in Einstein frame reads
\begin{eqnarray}
ds_{E}^{2} &=&\sqrt{r}\left\{ dy^{2}+dx^{2}-f_{bh}\left( r\right) dt^{2}+\frac{4}{%
r^{2}f_{bh}\left( r\right) }dr^{2}+\frac{2}{e_{A}^{2}}\left[ \hat{\omega}\notag
_{1}^{2}+\hat{\omega}_{2}^{2}+\left( \hat{\omega}_{3}-e_{A}Q_{A}\zeta \left(
r\right) dt\right) ^{2}\right] \right.  \\
&&+\left. \frac{2}{e_{B}^{2}}\left[ \tilde{\omega}_{1}^{2}+\tilde{\omega}%
_{2}^{2}+\left( \tilde{\omega}_{3}-e_{B}Q_{B}\zeta \left( r\right) dt\right)\notag
^{2}\right] \right\} \ , \\
F_{3} &=&dC_{2}=2\zeta ^{\prime }\left( r\right) dr\wedge dt\wedge \left( 
\frac{Q_{A}}{e_{A}}\hat{\omega}_{3}+\frac{Q_{B}}{e_{B}}\tilde{\omega}%
_{3}\right) +\frac{2}{e_{A}^{2}}\hat{\omega}_{1}\wedge \hat{\omega}%
_{2}\wedge \left( e_{A}Q_{A}\zeta \left( r\right) dt-\hat{\omega}_{3}\right) \notag
\\
&&+\frac{2}{e_{B}^{2}}\tilde{\omega}_{1}\wedge \tilde{\omega}_{2}\wedge
\left( e_{B}Q_{B}\zeta \left( r\right) dt-\tilde{\omega}_{3}\right) \ ,\notag \\
\Phi  &=&\log \left( r\right) \ .\label{blackhole1}
\end{eqnarray}%
where%
\begin{eqnarray}
f_{bh}\left( r\right)  &=&\frac{e_{A}^{2}+e_{B}^{2}}{2}-\frac{m}{r^2}+\frac{%
2\left( Q_{A}^{2}+Q_{B}^{2}\right) }{r^{4}}\equiv \frac{e_{A}^{2}+e_{B}^{2}}{%
2r^{4}}\left( r^{2}-r_{+}^{2}\right) \left( r^{2}-r_{-}^{2}\right) ,\nonumber\\
\zeta \left( r\right)  &=&\frac{1}{r^{2}}-\frac{1}{r_{+}^{2}}\ ,\quad r_{\pm
}^{2}=\frac{m\pm \sqrt{m^{2}-4\left( Q_{A}^{2}+Q_{B}^{2}\right) \left(
e_{A}^{2}+e_{B}^{2}\right) }}{e_{A}^{2}+e_{B}^{2}} \ . \label{funciones blackhole1}
\end{eqnarray}%
In general $f_{bh}\left( r\right) $ has two real roots $r_{\pm }$.
The extremal black membrane is obtained when $r_{+}=r_{-}$ that is,%
\begin{eqnarray}
& & m^{2}=4\left( Q_{A}^{2}+Q_{B}^{2}\right) \left( e_{A}^{2}+e_{B}^{2}\right) \, ~~r_+^2=r_-^2=2\sqrt{\frac{Q_A^2+Q_B^2}{e_A^2+e_B^2}},\nonumber\\
& & 
f_{bh}\left( r\right)  =\frac{e_{A}^{2}+e_{B}^{2}}{2r^{4}}\left(
r^{2}-r_{+}^{2}\right) ^{2}=
%&=&\frac{e_{A}^{2}+e_{B}^{2}}{2r^{4}}\left( r^{2}-\frac{m}{\left(
%e_{A}^{2}+e_{B}^{2}\right) }\right) ^{2}\ , \\
\frac{e_{A}^{2}+e_{B}^{2}}{2r^{4}}\left( r^{2}-2\sqrt{\frac{\left(
Q_{A}^{2}+Q_{B}^{2}\right) }{\left( e_{A}^{2}+e_{B}^{2}\right) }}\right)
^{2}\ ,
\end{eqnarray}%
The preservation of SUSY imposes
$
e_{A}Q_{A}\pm e_{B}Q_{B}=0\ .
$
 In the Einstein frame, the BPS 
 extremal background with $Q_{B}=\frac{e_{A}}{e_{B}}Q_{A}$ reads,
\begin{eqnarray}
& & ds_{E}^{2} =\sqrt{r}\left\{ dy^{2}+dx^{2}+\frac{4}{r^{2}\frac{e_{A}^{2}+e_{B}^{2}%
}{2}r_{+}^{4}\zeta \left( r\right) ^{2}}dr^{2}+\frac{2}{e_{A}^{2}}\left[ 
\hat{\omega}_{1}^{2}+\hat{\omega}_{2}^{2}+\hat{\omega}_{3}^{2}-2\hat{\omega}%
_{3}e_{A}Q_{A}\zeta dt\right] \right.  \nonumber \\
&&\left. +\frac{2}{e_{B}^{2}}\left[ \tilde{\omega}_{1}^{2}+\tilde{\omega}%
_{2}^{2}+\tilde{\omega}_{3}^{2}-2\tilde{\omega}_{3}e_{A}Q_{A}\zeta dt\right]
\right\} \ .\nonumber \\
& & 
F_{3} =2d\left[ \zeta \left( r\right) dt\wedge \left( \frac{Q_{A}}{e_{A}}%
\hat{\omega}_{3}+\frac{Q_{B}}{e_{B}}\tilde{\omega}_{3}\right) \right]  -\frac{2}{e_{A}^{2}}\hat{\omega}_{1}\wedge \hat{\omega}_{2}\wedge \hat{\omega}_{3}-\frac{2}{e_{B}^{2}}\tilde{\omega}_{1}\wedge \tilde{\omega}%
_{2}\wedge \tilde{\omega}_{3}, \nonumber\\
& & \Phi= \log(r).
\end{eqnarray}%
Note that $g_{tt}=0$, thus the vector $\partial _{t}$ is null.
Also, in this configuration $F_{3}$ has both "electric" and magnetic
parts. Nevertheless, the integrating of the magnetic part of $F_7=\star F_3$ does not lead to charge of D1 brane.

In what follows we consider the non-BPS black membrane background \eqref{blackhole1}. This configuration is rotating along the directions $\partial_{\psi_A}$ and $\partial_{\psi_A}$. The coordinates used in \eqref{blackhole1} correspond to a rotating frame, this is a consequence of the fact that the fibrations do not decay at infinity. We can move to a non-rotating frame by doing a large gauge transformation which cancels the constant term of $\zeta(r)$ in \eqref{funciones blackhole1}. 

We will work in the non-rotating frame at infinity  and also for simplicity in the computation of the charges we shift the dilaton by a constant and the $F_3$ by a factor
\begin{equation}
    \zeta(r)=\frac{1}{r^2} ,\qquad \Phi \to \Phi-2\log\left(\frac{e_A^2+e_B^2}{2}\right) ,\qquad F_3 \to \frac{e_A^2+e_B^2}{2}F_3
\end{equation}
This spacetime is asymptotically locally flat and has the same causal structure as the Reissner-Nordström spacetime. Nevertheless, 
 the spacetime is asymptotically conformal to Minkowski four times $S^3 \times S^3$. We compute the conserved charges associated to the spacetime by using the Noether-Wald method \cite{Wald:1993nt}. The expression for the charges are given in Appendix \ref{appendixCharges}. The energy, angular momentum associated to $\partial_{\psi_A}$ and $\partial_{\psi_B}$, the temperature and the entropy for this configuration are given by
\begin{eqnarray}
E &=&\frac{2m}{e_{A}^{3}e_{B}^{3}r_{0}^{2}\kappa ^{2}}\left( 16\pi
^{2}\right) ^{2}L_{x}L_{y}\ , \\
J_{A} &=&-\frac{8Q_{A}}{e_{A}^{4}e_{A}^{3}\kappa ^{2}}\left( 16\pi
^{2}\right) ^{2}L_{x}L_{y}\ , \quad
J_{B} =-\frac{8Q_{B}}{e_{A}^{4}e_{A}^{3}\kappa ^{2}}\left( 16\pi
^{2}\right) ^{2}L_{x}L_{y}\ ,  \\
T &=&\frac{e_{A}^{2}+e_{B}^{2}}{16\pi }-\frac{4\left(
Q_{A}^{2}+Q_{B}^{2}\right) }{16\pi r_{+}^{4}}\ , \quad
S =\frac{2r_{0}^{2}}{e_{A}^{3}e_{B}^{3}G_{10}}\left( 16\pi \right)
^{2}L_{x}L_{y}\,\,\ .
\end{eqnarray}
According to our normalisation $\kappa^2=8\pi G_{10}$ where $G_{10}$ is the Newton constant in ten dimensions. These quantities satisfy the first law of thermodynamics as expected
\begin{equation}
dE=TdS+\Omega _{A}dJ_{A}+\Omega _{B}dJ_{B}\ ,
\end{equation}
where the angular velocities are
\begin{equation}
\Omega _{A}= \frac{e_A Q_A}{r_+^2}\,\, , \qquad \Omega _{B}= \frac{e_B Q_B}{r_+^2} \,\, .
\end{equation}
This background can be understood as a four dimensional planar black hole with electric charges, that was firstly found in \cite{Klemm:1998in}. Thus, we have shown that the presence of the electric charges in four dimensions corresponds to rotations of the branes in ten dimensions. The asymptotic form of the metric leaves an ambiguity in the normalisation of the time-like Killing vector at infinity that appears in the computation of the energy and the temperature. Therefore, the temperature in ten dimensions is the same as in four dimensions up to a numerical factor.
%\textcolor{red}{aca ponemos los resultados de temperatura, entropia, masa, angular momentum y dejamos para un apendice la derivacion de esos resultados, indicando las formulas que se usan, etc.}

We leave this black membrane background here, as it is not the focus of the rest of this work. We move into computing conserved charges for the backgrounds I and II in eqs.(\ref{background1}) and (\ref{background 211})

\section{A proposal for the dual field theory and its observables} \label{QFTobs}
Here we present a proposal for the field theory dual to our new background in eq.(\ref{background 211}). It is convenient to first discuss the field theory dual to
the D5-D5 intersection and the background in eq.(\ref{background1}).
\subsection{The holographic dual to the Background I}
We start discussing the field theory on I-branes.
The result in eq.(\ref{charges}) indicates the presence of two stacks of D5 branes, with $N_A$ and $N_B$ being the number of branes on each stack. 
When taken at weak coupling these stacks intersect over two dimensions. In \cite{Green:1996dd} it was shown that when two stacks of branes intersect along $(4k+2)$, being the transverse dimensions a multiple of four (in our case $k=0$, D5 stacks intersect in two dimensions  and have eight transverse directions) the massless spectrum contains chiral fermions, arising from the open strings connecting the branes. These fermions give rise to gauge (and gravitational) anomalies on the intersection. The anomalies are cancelled by anomaly inflow from the 'bulk of the brane'. This implies that the D-branes world-volume action must contain a Chern-Simons term.

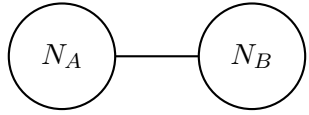
\begin{figure}[h!]
    \centering
    \begin{tikzpicture}
    
        %--------NODES--------------------
        \node (NA) at (-1.25,0) [circle,draw,thick,minimum size=1.4cm] {$N_{A}$};
        \node (NB) at (1.25,0) [circle,draw,thick,minimum size=1.4cm] {$N_{B}$};

        %-------Hypers ------------------
        \draw [thick] (-0.55,0) -- (0.55,0);
    \end{tikzpicture}
    \caption{Dual Theory in ($1+1$) dimensions with chiral fermions running in the links. This encodes the field theory at weak coupling. The inflow from the bulk of the branes is understood.}
    \label{figurequiver}
\end{figure}

When studied at weak coupling the D5-D5 system preserve chiral supercharges. We have two gauge groups $SU(N_A)\times SU(N_B)$ with chiral fermions transforming in the $({\bf N_A, \bar{N}_B})$ representation, the system has $SO(1,1)$ Poincare symmetry. The anomaly is cured by inflow from the bulk of the D5 branes.  In other words, the dynamics of the intersection is not decoupled from the brane dynamics.  The system preserves eight SUSYs \cite{Green:1996dd}. The weakly coupled field theory is summarised by the quiver in Figure \ref{figurequiver}.
%\textcolor{red}{hacer la figura y poner mas material sobre la QFT}. 
As the couplings grow large, the above description breaks down and is replaced by a description in terms of the background in eq.(\ref{background1}). This is carefully described in \cite{Itzhaki:2005tu}. In the strong coupling regime, the system preserves $SO(2,1)$ with $SO(4)\times SO(4)$ R-symmetry and sixteen SUSYs. The three dimensional field  theory has gauge groups with Chern Simons terms $SU(N_A)_{N_B} \times SU(N_B)_{N_A}$ \cite{Itzhaki:2005tu}, \cite{Lin:2005nh}.
At strong coupling, these stacks intersect in the coordinates $(t,x,\varphi)$. One of the stacks extends along $(\theta_A,\phi_A,\psi_A)$ whilst the other does it over $(\theta_B,\phi_B,\psi_B)$. 
%This suggest that at low energies (compared with the inverse size of $\hat{S}^3_A$ and $\tilde{S}^3_B$) a three dimensional field theory describes the dynamics of the system. 
%
The papers  \cite{Itzhaki:2005tu}, \cite{Lin:2005nh}, argue that the field theory is gapped.

Notice that the background in eq. (\ref{background1}) is not trustable for all the range of the radial coordinate. In fact, for large values of $r$, the dilaton becomes large the type IIB system is better described by performing an S-duality and describing the dynamics in terms of the Little String Theory on the two stacks of NS five branes.
That is, the field theory above described has a non-field theoretical UV completion.

 On the other hand, for $r\to 0$ the Ricci scalar in eq.(\ref{riccisingular}) diverges and the background is not trustable. We then need to replace the description by the one given by the configuration in eq.(\ref{background 211}). 

%Note that as the energy grows, the field theory is UV completed by a Little String Theory.
%
%
%
\subsection{The field theory dual to the Background II.}
Here, we analyse the background in eq.(\ref{background 211}). The supergravity solution is smooth, hence the dual QFT is strongly coupled.

In fact, the Ricci scalar  in eq.(\ref{Ricci}) is finite for $r>r_+$ and the string coupling (proportional to $e^{\Phi}$) is bounded below some value $r< r_*$.  The value of $r_*$ is determined by observing that the dilaton in eq.(\ref{background 211}) can be changed by $\Phi=\Phi_0+ \log r$ at the cost of rescaling the Ramond form $F_3\to e^{-\Phi_0} F_3$. These scalings make the string coupling $g_s=e^{\Phi}= r e^{\Phi_0}$. The value $r_*\sim e^{-\Phi_0}$  (for which $g_s\sim 1$), can be made arbitrarily large by suitably choosing the parameter $\Phi_0$. Notice that a chosen large  and negative $\Phi_0$, makes the charges of the D5 branes larger.

We then conclude that the background is trustable in a large region of the radial coordinate $[r_+,r_*)$. Consequently, in a large regime of energies the dual field theory is strongly coupled. At  very high energies, when the string coupling becomes large we should S-dualise arriving to a configuration of intersecting and wrapped NS-five branes. The high energy behaviour of the field theory is UV-completed  in terms of a Little String Theory. 

We now discuss the flowing to lower energies. In this case, the coordinate $\varphi$ is compactified, shrinks to  zero size and one ends with a QFT with less SUSY, smaller R-symmetry and effectively in ($1+1$) dimensions.
%
%
%Note also that the coordinate $\varphi$ shrinks, leading to an effective two-dimensional field theory  description of the dynamics. 
%The low energy and weakly coupled dynamics consists of two gauge groups $SU(N_A)\times SU(N_B)$. Open strings stretched between the two stacks describe the dynamics of bifundamental fields (chiral fermions among them). The low energy field theory is then %\textcolor{red}{El anomaly inflow aun lo tenemos cuando ponemos el cigarro? De ser asi, es la teoria realmente 1+1?}
%
Whilst for large values of the radial coordinate, the backgrounds I and  II  coincide at leading order, a relevant operator is deforming the field theory dual to background II. This deformation is associated with the subleading terms, proportional to the parameters $Q_A, Q_B$. As we lower the energy (still at strong coupling) 
a Kaluza-Klein spectrum of massive modes  arises due to the compactification of the branes on $\varphi$.

At energies around the scale set by $r_+$, the QFT dual to the background in eq.(\ref{background 211}) should be a $(1+1)$ dimensional QFT. This should be the reduction of the Yang-Mills-Chern-Simons $SU(N_A)_{N_B}\times SU(N_B)_{N_A}$ to ($1+1$), preserving four supercharges. The QFT is expected to be gapped, confine and break part of the R-symmetry.

In what follows, we start the study of this interesting field theory. We do so by calculating observables of the two-dimensional QFT using probes of the background in eqs. (\ref{background 211})-(\ref{functionsbackground}).
These probes inform us about gauge couplings, theta-angles, symmetry breaking, confinement, etc. We present a quantity that indicates the number of degrees of freedom (density of states in terms of the energy).

\subsection{Gauge coupling }\label{couplingstheta}
To study the background using D-branes probes, it is first useful to set our conventions for the Dirac-Born-Infeld-Wess-Zumino (DBIWZ) action, describing the dynamics of branes in our background.

Consider a ten dimensional  manifold $M_{10}$ equipped with a metric tensor $G_{\mu\nu}$, Neveu-Schwarz two form $B_{\mu\nu}$, dilaton $\Phi$ and Ramond potentials encoded in the poly-form ${\cal C}$.  In this space, there is an embedded manifold $\Sigma$ of dimension  $(p+1)$ with $(9-p)$
space-like normal vectors. This embedded manifold hosts a Dp-brane. We denote the coordinates on the
Dp-brane as $X^{M}$ with $M=0,1,\dots ,p$. and the induced metric $g_{MN}$.
The action of a single  Dp-brane is the Dirac-Born-Infeld-Wess-Zumino action 
given by
\begin{eqnarray}
& & S_{Dp, DBI}\left[ g_{MN},{\cal F}_{MN}\right] =T_{p}\int d^{p+1}xe^{-\Phi }\sqrt{-\det
\left( g_{MN}+ {\cal F}_{MN}\right) }\ .\label{DBI}\\
& & S_{Dp, WZ}\left[ {\cal C},{\cal F}_{MN}\right]= - T_{p}\int_\Sigma {\cal C}\wedge e^{- {\cal F}_{MN}}\label{WZ}
\end{eqnarray}
Here ${\cal F}_{MN}= B_{MN} +2\pi \alpha' F_{MN} $, where $B_{MN}$ is the pull-back of the background Neveu-Schwarz two-form on $\Sigma$, and $F_{MN}$ is an Abelian gauge field strength defined on the brane.  The tension of the Dp brane is $T_p=\frac{1}{(2\pi)^{7-p}}$, in our chosen units. For the case of our backgrounds, we have $B_{MN}=0$ and   the poly-form ${\cal C}= C_2$  given in eq.(\ref{background 211}), and/or its electric dual $C_6$, given in Appendix \ref{appendix2}. 

With our choice of units  ($g_s=\alpha'=1$)  we can perform a small field $F_{MN}$ expansion of the action in eq.(\ref{DBI}), equivalent to a small-$\alpha'$ expansion. We  obtain an effective action for
the Dp-brane, %
\begin{equation}
S_{Dp, DBI}=T_{p}\int d^{p+1}xe^{-\Phi }\sqrt{-\det g_{MN}}\left[ 1-\frac{1}{4}%
\left( 2\pi \right) ^{2}F^{BC}F_{BC}+\mathcal{O}\left(F^3\right) \right] \   \label{effective action Dp brane}
\end{equation}%
where $F^{BC}=g^{BM}g^{CN}F_{MN}$. The Wess-Zumino part of the action in eq.(\ref{WZ}) contains a finite number of terms. 
Specialising for our backgrounds with $B_{MN}=0$ we have for any D$_p$ brane probe,
%that contains the magnetic Ramond potential $C_2$ (and its electric dual $C_5$) we can only have,
\begin{eqnarray}
S_{Dp,WZ}\!=\!\! 
-\! T_{p}\!\!\int_{\Sigma_{p+1}}\!\! C_{p+1}\!  -\! 2\pi C_{p-1}\!\wedge\! F_2 +\!\frac{(2\pi )^2}{2}\! C_{p-3}\!\wedge\! F_2^2 -\! \frac{(2\pi)^3}{6}\! C_{p-5}\wedge\! F_2^3+\!\frac{(2\pi)^4}{24} \!C_{p-7}\wedge\! F_2^4.\label{WZ2} 
\end{eqnarray}

In what follows we study the backgrounds in eqs. (\ref{background1}), (\ref{background 211})-(\ref{functionsbackground}) with various probe  branes  in Type IIB. 
\\
\\
The first probe is a D5 brane that extends on the directions $[t,x,\varphi, \theta_A,\phi_A,\psi_A]$. This is like a probe that extends where the A-stack originally was. We will switch on an electric field on its worldvolume. 

\underline{\bf Calculation in the Background I}

Let us start by performing the probe calculation in the background of eq.(\ref{background1}). The dual QFT is $(2+1)$-dimensional, as the coordinate $\varphi$ is not compact. By expanding the Born-Infeld action, we find a Maxwell term, with coupling (the details of this calculation are spelled out below)
\begin{equation}\label{coupling2+1}
\frac{1}{g_{YM,A}^2}= (2\pi)^{4} T_{\hat{D}_{5}} \sqrt{\frac{1}{N_{A}} + \frac{1}{N_{B}} }  N^{\frac{3}{2}}_{B}
\end{equation}
Similarly, the Wess-Zumino term gives,
\begin{equation}
S_{D5,WZ}= -2\pi^2 T_{\hat{D}_{5}} \int C_2\wedge F_2\wedge F_2=- \int_{S^3_A} F_3 \int_{t,x,\varphi} A_1\wedge F_2= -N_B  \int_{t,x,\varphi} A_1\wedge F_2.\label{CS2+1}
\end{equation}
We have performed an integration by parts, used the quantisation condition in eq.(\ref{charges}) and set $T_5=\frac{1}{(2\pi)^2}$ (in our units).
There is a similarly symmetric calculation for a D5 probe along $[t,x,\varphi, \theta_B,\phi_B,\psi_B]$. In agreement with the field theory picture discussed above, we find  two gauge groups with Yang-Mills Chern-Simons dynamics, $SU(N_A)_{N_B} \times SU(N_B)_{N_A}$, with fixed gauge couplings. This is exactly in agreement with the field theory expectations  \cite{Itzhaki:2005tu}, that  we summarised in the previous section.

\underline{\bf Calculation in the Background II }

Let us now study the case for which the $(2+1)$ QFT has been compactified along the $\varphi$-direction and we are dealing with a $(1+1)$ dimensional QFT. 
We work with the background in eqs.(\ref{background 211})-(\ref{functionsbackground}), and follow the calculation above, by first writing the induced metric on the D5 brane,
\begin{equation}
ds_{\hat{D}_{5}}^{2}=r\left\{ -dt^{2}+dx^{2}+\left[ f_{s}\left( r\right)
+2Q_{B}^{2}\zeta \left( r\right) ^{2}\right] d\varphi ^{2}+\frac{2}{e_{A}^{2}%
}\left[ \hat{\omega}_{1}^{2}+\hat{\omega}_{2}^{2}+\left( \hat{\omega}%
_{3}-e_{A}Q_{A}\zeta (r)d\varphi \right) ^{2}\right] \right\} \ .  \label{inducedD5}
\end{equation}%
From here we calculate
\begin{equation}
e^{-\Phi}\sqrt{-\det g_{MN}^{\left( \hat{D}_{5}\right) }}=r^{2}\left( \frac{2}{%
e_{A}^{2}}\right) ^{3/2}\sin {\theta_A}\sqrt{f\left( r\right)
+2Q_{B}^{2}\zeta \left( r\right) ^{2}}.\nonumber
\end{equation}
The effective action
for the brane  in eq. (\ref{effective action Dp brane}) reads,
\begin{eqnarray}
S_{\hat{D}_{5}, BI} &=&-T_{\hat{D}_{5}} \int dt  dx  d\varphi d{\theta_A}d{\phi_A}d{\psi_A}\left( \frac{2}{e_{A}^{2}}\right) ^{3/2}r^{2}\sin {\theta_A}\sqrt{f\left( r\right) +2Q_{B}^{2}\zeta\left( r\right) ^{2}}
\left(1-\frac{(2\pi)^2}{4} F^{BC}F_{BC} \right)\ ,  \notag \\
&=&
T_{\hat{D}_{5}}L_{\varphi }\left(
4\pi \right) ^{2}\left( \frac{2}{e_{A}^{2}}\right) ^{3/2}r^{2}\sqrt{f\left(
r\right) +2Q_{B}^{2}\zeta \left( r\right) ^{2}}\int dtdx\left(1 -\frac{(2\pi)^2}{4}%
F^{BC}F_{BC}\right) \ .
\end{eqnarray}%
We turn on $F_{tx}$ so that
\begin{equation}
F^{BC}F_{BC}=2F_{tx}F_{tx}g^{tt}g^{xx}=2\frac{1}{r^{2}}F_{tx}F_{tx}\eta
^{tt}\eta ^{xx}=\frac{1}{r^{2}}F_{\mu \nu }^{2}\ .
\end{equation}%
From here we identify the Yang-Mills coupling for this probe D5,
\begin{equation}
\frac{1}{g_{YM,A}^{2}}=  8\pi^{4} T_{\hat{D}_{5}} N^{\frac{3}{2}}_{B}L_{\varphi }\sqrt{f\left( r\right) +2Q_{B}^{2}\zeta \left( r\right) ^{2}}.
\end{equation}%
If the D5 brane probes the SUSY preserving background, we impose $Q_{B}=\frac{e_{B}}{e_{A}}Q_{A}$ and $m =0$ on the parameters appearing in  eqs.(\ref{functionsbackground})-(\ref{periodvarphi}). This implies $r_{\pm}^2=\pm \frac{2Q_{A}}{e_{A}} $ and $L_{\varphi }=4\pi
/(e_{A}^{2}+e_{B}^{2})$. Together with eq.(\ref{eq2.2}) and the quantisation condition (\ref{charges}), the gauge coupling reduces to\footnote{For large values of the radial coordinate $r$ and decompactifying $\varphi$, this result almost reduces to that in eq.(\ref{coupling2+1}). The difference in the factor of 2 inside the square root is due to the fact that the fibration does not vanish at infinity.}
\begin{equation}
\frac{1}{\hat{g}_{YM,A}^{2} }=  8\pi^{4} T_{\hat{D}_{5}} N^{\frac{3}{2}}_{B}L_{\varphi } \sqrt{ 
     \frac{8}{N_{A}} + \frac{4}{N_{B}}  - \frac{4\sqrt{2N_{B}}Q_{A}}{N_{A} r^{2}}  -\frac{2Q^{2}_{A}}{r^{4}}  }\ ,\label{couplinga}
\end{equation}%
with limiting values 
\begin{equation}
\frac{1}{\hat{g}_{YM,A}^{2}}=\left\{ 
\begin{array}{cc}
(2\pi)^{4} T_{\hat{D}_{5}} N^{\frac{3}{2}}_{B}L_{\varphi } \sqrt{ 
     \frac{2}{N_{A}} + \frac{1}{N_{B}}} & ,\quad r\rightarrow
\infty  \\ 
0 & ,\quad r\rightarrow r_{+}
\end{array}
\right. \ .\label{couplingaa}
\end{equation}
In other words, the gauge coupling grows very large at low energies and asymptotes to a constant value for high energies. As  discussed, at very high energies the field theory is best described in terms of a Little String Theory (LST). This  calculation above refers to the gauge coupling $g_{YM,A}$. Other interactions in the QFT may become large at high energies, in such a way that the field theory is strongly coupled in the UV. This is in agreement with the background in eq.(\ref{background 211}) being weakly curved for all values of the radial coordinate $[r_+,\infty)$. 
%When the dilaton becomes large we should S-dualise, describing the system in terms of 
%stacks of NS five branes, making the microscopic description not field-theoretical, but in terms of a Little String Theory.

Had we studied a D5 probe extended along $[t,x,\varphi, \theta_B,\phi_B,\psi_B]$, with an electric field $F_{tx}$ switched on the brane, the result would be,
\begin{equation}
\frac{1}{g_{YM,B}^{2}}=8\pi^{4} T_{\hat{D}_{5}} N^{\frac{3}{2}}_{A}\sqrt{f\left( r\right) +2Q_{A}^{2}\zeta \left( r\right) ^{2}}.
\end{equation}%
Obviously we have an expression similar to eq.(\ref{couplinga}), for the gauge coupling of the second gauge group $\frac{1}{g_{YM,B}^{2} }$. We should use in this case that $Q_A=\frac{e_A Q_B}{e_B}$
\subsection{Theta angle}
From the viewpoint of the  $(2+1)$ dimensional QFT,  represented by Background I in eq.(\ref{background1}), we can consider dimensionally reducing the Chern-Simons term obtained in eq.(\ref{CS2+1}). We then obtain a theta-term proportional to $N_B \oint_\varphi A_\varphi$ for the QFT on the A-stack. 

For the $(1+1)$ viewpoint, additional probes calculate  the $\Theta$-angle  of each  gauge group. Let us use a D3 probe, extended along
$[t,x,\theta_A,\Phi_A]$ with an electric field  $F_{tx}$ switched on. We study the Wess-Zumino term following eq.(\ref{WZ2}) and using the  two-form potential pulled-back on this D3 probe 
\begin{equation}
C_{2}|_{D_{3}}=-\frac{2}{e_{A}^{2}} {\psi_A}\sin {\theta_A}d{\theta_A}
\wedge d{\phi_A}\ .  \label{C2 conf}
\end{equation}%
Replacing this  $C_{2}$ in eq.(\ref{WZ2}) and using that $F_2= F_{tx} dt\wedge dx$ we find the Wess-Zumino term  for this probe is (note that $C_4=0$ in the background),
\begin{equation}
S_{WZ,D3}=-T_{{3}} 2\pi \int C_{2}\wedge F_{2} = \frac{T_{{3}} 16 \pi^2 }{e_{A}^{2}}{\psi_A}\int F_{tx}dt \wedge  dx\ \ .\label{WZD3}
\end{equation}%
The  $\Theta$-angle associated with the gauge group should be identified according to ,
\begin{eqnarray}
& &S_{WZ,D3}= \frac{\Theta_{A}}{4\pi^2}\int dt dx F_{tx},~~\longrightarrow~~ \Theta_A =\frac{T_{D_{3}}64\pi^4 }{e_{A}^{2}}{\psi_A}= \frac{\psi_A N_B}{2}\ .
\end{eqnarray}
We have used $16 \pi^4 T_{D3}=1 $  (in our units) and the quantisation condition in eq.(\ref{charges}). Notice that the periodic identification $\Theta_A\sim \Theta_A + 2 k \pi$ implies that the angle $\psi_A$ gets quantised to the values 
\begin{equation}
\Delta\psi_A= \frac{4 k \pi}{N_B}, ~~\text{with }~~k=0,1,2...., N_B-1.\label{deltapsia}
\end{equation}
 For $k=N_B$ we have $\Delta \psi_A=4\pi$, covering the full circle.

Had we considered the D3 probe extended along $[t,x, \theta_B,\phi_B]$ with all other coordinates fixed, we would have found (the calculation is exactly symmetrical),
$\Theta_B = \frac{\psi_B N_A}{2}.$

Another way of understanding the R-symmetry breaking would be to consider euclidean D1 branes wrapping $[\theta_A,\phi_A]$.
The Wess-Zumino action contributes to the partition function via
\begin{equation}
Z_{D1}\sim e^{\frac{i}{2\pi}\int C_2}.
\end{equation}
This contribution should not depend on a (large) gauge transformation parameter $\epsilon_A$, that appears as  we change $\psi_A\to \psi_A+\epsilon_A$. We enforce
\begin{equation}
\left(\frac{i}{2\pi}\right) \left(\frac{- 2\epsilon_A}{e_A^2}\right)\int \sin\theta_A d\theta_A d\phi_A= 2 i k \pi ,~~\text{leads to}~~\epsilon_A= \frac{4 k \pi}{N_B}.\label{rbreakfinal}
\end{equation}
  
These results indicate that the background's continuous isometries transforming $\psi_{A,B}\to \psi_{A,B} + \epsilon_{A,B}$, with ($\epsilon_{A,B}$ being constants), are actually broken. In fact, the allowed changes are $\Delta\psi_{A,B}=  \frac{4 k \pi}{N_{B,A}}$, with $k=0,1,2...., N_{B,A}-1$, which should be interpreted as the  breaking of the two field theory global symmetries $U(1)_{A,B}$ into discrete subgroups.

The argument used to derive eqs.(\ref{deltapsia})-(\ref{rbreakfinal}) is not airtight. It uses a two manifold that  has a boundary and at the same time a gauge choice is made for the potential $C_2$ in eq.(\ref{C2 conf}). It would be nice to have an argument for R-symmetry breaking that is explicitly gauge invariant. In the coming section we present a different holographic perspective on the $U(1)_{A,B}$ breaking with this property.

\subsection{\texorpdfstring{$U(1)_{A,B}$}{U(1)-A,B} symmetry breaking pattern}\label{r-symmetry}

Some supersymmetric field theories exhibit a classical $U(1)$ R-symmetry that is quantum mechanically broken to a discrete subgroup. The symmetry breaking can be understood diagrammatically or in terms of instantons. The supergravity dual to the given field theory should encode this, but the mechanism should not involve  instantons (as that are very suppressed  in supergravity). The fact that the Ramond potentials are not gauge invariant under the $U(1)$ R-symmetry is key. The breaking of the global R-symmetry in the field theory manifest as spontaneous breaking in supergravity. The vector field in the bulk, dual of the R-symmetry current acquires a mass. We find this below for our background of eq.(\ref{background 211}). The argument that follows is gauge invariant at all steps. All along this section, we set $m=0$ and focus only in the BPS case.

The $U(1)$ symmetry of the metric is perturbed, the Lagrangian for this fluctuation is described by the usual $F_{\mu\nu}^2$-term. For the perturbation to be consistent, the Ramond fields must be also perturbed, this contributes to the mass term for the fluctuation. The massive gauge field is understood as  symmetry breaking in supergravity.

In the previous section, we hinted at a breaking of the isometries represented by  the Killing vectors $\partial_{\psi_{A,B}}$. To better understand the breaking of the field theory global $U(1)$ symmetries associated with the translations in $\psi_{A,B}$ we proceed as explained in \cite{Klebanov:2002gr}, \cite{Gursoy:2003hf}. We give full details in Appendix \ref{RSymmetryBreaking}. In the holographic background, we gauge the isometry by replacing, both in the metric and in the Ramond potential of eq.(\ref{background 211})
    \begin{equation}
        d\psi_{A,B} \rightarrow d\psi_{A,B} + A_{A,B},\quad
        \psi_{A,B} \rightarrow \psi_{A,B} + \epsilon_{A,B}.
    \end{equation}
Such that a change $ \psi_j \rightarrow \psi_j + \epsilon_j$ is compensated by  $A_j\rightarrow A_j + d \epsilon_j$. 
We then study the Lagrangian for the gauge fields $A_{A,B}$, by replacing these changes in the string frame Lagrangian.
The metric perturbation changes the Ricci scalar to
    \begin{equation}
      e^{-2\Phi}  R \rightarrow e^{-2\Phi }\left(R  -\frac{1}{4}\frac{2r}{e^{2}_{A}}F^{2}_{A}
                           -\frac{1}{4}\frac{2r}{e^{2}_{B}}F^{2}_{B} \right).
    \end{equation}
The kinetic term for $F_{3}$ changes as,
    \begin{equation}
    \begin{aligned}
        \frac{1}{12}F^{\mu\nu\lambda}F_{\mu\nu\lambda} \rightarrow
        &\frac{1}{12}F^{\mu\nu\lambda}F_{\mu\nu\lambda} 
        + \frac{1}{2}\frac{4Q^{2}_{A} +e^{2}_{A} r^{4}}{e^{2}_{A}r^{6}}\left(A_{A}-d\epsilon_{A}\right)^{2}
        + \frac{1}{2}\frac{4Q^{2}_{B} +e^{2}_{B} r^{4}}{e^{2}_{B}r^{6}}\left(A_{B}-d\epsilon_{B}\right)^{2} \\
        &+\frac{4Q_{A}Q_{B}}{e_{A}e_{B}r^{6}} \left(A_{A}-d\epsilon_{A}\right)\cdot\left(A_{B}-d\epsilon_{B}\right).
    \end{aligned}
    \end{equation}
These perturbations around an isometry direction are usually known to be consistent. It is interesting to note  that the kinetic term for the gauge field $A_{(A,B)}$ comes only from the Ricci scalar. The information of $\epsilon_{A,B}$ on the other hand, comes only from  $F^{2}_{3}$. Defining
    \begin{equation}
        W_{(A,B)} = A_{A,B} - d\epsilon_{A,B},
    \end{equation}
and imposing the BPS condition $e_{A}Q_{B}= e_{B}Q_{A}$, the Lagrangian for the perturbation reads,
 %   \begin{equation}
    %    \mathcal{L} = -\frac{1}{4}\frac{2r}{e^{2}_{A}}F^{2}_{A}
       %                    -\frac{1}{4}\frac{2r}{e^{2}_{B}}F^{2}_{B}
          %                 +\frac{1}{2}\frac{4Q^{2}_{A} +e^{2}_{A} r^{4}}{e^{2}_{A}r^{6}}(W^{(A)}_{\mu}W^{(A)\mu}+W^{(B)}_{\mu}W^{(B)\mu})
             %              \pm \frac{4Q^{2}_{A}}{e^{2}_{A}r^{6}} W^{(A)}_{\mu} W^{(B)\mu},
    %\end{equation}
%
%which can be rewritten as
    \begin{equation}
        \mathcal{L} = -\frac{1}{4}\frac{2}{r e^{2}_{A}}F^{2}_{A}
                           -\frac{1}{4}\frac{2}{r e^{2}_{B}}F^{2}_{B}
                           +\frac{1}{2r^{2}} (W^{(A)}_{\mu}W^{(A)\mu}+W^{(B)}_{\mu}W^{(B)\mu})
                           +\frac{1}{2}\frac{4Q^{2}_{A}}{e^{2}_{A}r^{6}}(W^{(A)}_{\mu}+ W^{(B)}_{\mu})^{2}.
    \end{equation}
%In the particular case when $e_{A}=e_{B}$, we change variables
  %  \begin{equation}
    %\begin{aligned}
     %   W_{A} \rightarrow \frac{\sqrt{2}}{2}W_{A} - \frac{\sqrt{2}}{2}W_{B}, ~~~
       % W_{B} \rightarrow \frac{\sqrt{2}}{2}W_{A} + \frac{\sqrt{2}}{2}W_{B}
        %    \end{aligned}
   % \end{equation}
%The Lagrangian for the gauge field reads
   % \begin{equation}
      %  \mathcal{L} = -\frac{1}{4}\frac{2}{r e^{2}_{A}}(F^{2}_{A}+F^{2}_{B})
         %                  +\frac{1}{2r^{2}} (W^{(A)}_{\mu}W^{(A)\mu}+W^{(B)}_{\mu}W^{(B)\mu})
            %               +\frac{4Q^{2}_{A}}{e^{2}_{A}r^{6}}W^{(A)}_{\mu}W^{(A)\mu}.
    %\end{equation}
We interpret this result  as follows. There are two $U(1)$ global symmetries in the QFT, holographically they are represented by the invariance of the metric under changes in $\psi_A$ and $\psi_B$. These global symmetries are broken to  discrete groups $\mathbb{Z}_{N_B}$ and $\mathbb{Z}_{N_A}$, as indicated by eq.(\ref{deltapsia}). This breaking is an effect of the lack of invariance of the gauge potential $C_2$. The breaking of the global symmetries is addressed in this section without appeal to the Ramond potentials, by observing that gauging the metric isometries $\partial_{\psi_{A,B}}$ leads to a breaking of the gauge symmetry, by a mass term. These mass terms are dependent on the radial coordinate.

 Contrary to what happens for the duals to ${\cal N}=1$ SYM, the metric in eq.(\ref{background 211}) does not break these discrete isometries to $\mathbb{Z}_2$. In other words, there is not a radial-regime in the metric that explicitly breaks the isometry  $\partial_{\psi_{A,B}}$. We interpret this result as the (anomalous) breaking of the two $U(1)_{A,B}$ in the QFT not being followed by a further spontaneous breaking. One might argue that VEVs are not allowed in a two dimensional QFT, hence no further breaking can take place by the formation of a condensate. This argument is not completely rigorous, as our QFT is two dimensional in the far IR, but get UV completed around the confining scale to a higher dimensional QFT.
 
 We study now a different observable that gives an approximate idea of the number of degrees of freedom as a function of the energy (a density of states).

\subsection{Holographic central charge}\label{hcc}
Consider a generic holographic background dual to a QFT in $(d + 1)$ spacetime dimensions, with metric and dilaton given by
    \begin{equation}\label{cani}
        ds^{2} = a(r,y^{i}) \left[ dx^{2}_{1,d}+b(r) dr^{2}\right]+ g_{ij}(r,y^i)dy^{i} dy^{j},\quad \Phi(r,y^i).
    \end{equation}
Following \cite{Macpherson:2014eza}, we define quantities $V_{int}, {H}$ according to
    \begin{equation}\label{vint}
        V_{int}= \int dy^i \sqrt{e^{-4 \Phi} a(r, y^i)^d \det[g_{ij}]},\quad {H}= V_{int}^2.\nonumber
    \end{equation}
From these we define the holographic central charge (or free energy),
    \begin{equation}\label{chol}
        c_{hol} = d^{d} \frac{b(r)^{\frac{d}{2}} H^{\frac{2d+1}{2}}}{G^{(10)}_{N} (H')^{d}},
    \end{equation}
where $G^{(10)}_{N} = 8\pi^6$ is (in our conventions), the ten-dimensional Newton constant. 

The holographic central charge of eq.(\ref{chol}) makes perfect sense for backgrounds with an AdS-factor. In those cases the quantity in eq.(\ref{chol}) is a number depending on the parameters of the background and it was successfully matched with the free energy of the CFT. In contrast for our case, without an AdS-factor we use eq.(\ref{chol}) to give an indication of the number of degrees of freedom of the QFT.
% It is interesting to note that this quantity is invariant under redefinitions of the radial coordinate.

Let us first compute the quantities in eqs.(\ref{cani})-(\ref{chol})
 for the solution in eq.(\ref{background1}), dual to a $(2+1)$ dimensional QFT. We find, \footnote{We thank the referee for pointing out a typo in the original version of this expression.}
 \begin{eqnarray}
 & & d=2,~~a(r, y^i)=r, ~~ b(r)= \frac{8}{(e_A^2+e_B^2) r^2},~~V_{int}={\cal N} r^2,~~ {\cal N}= \frac{8 (4\pi)^4}{e_A^3e_B^3}, \nonumber\\
 & & H= {\cal N}^2 r^4,~~~~c_{hol}= \frac{2{\cal N} }{ G_N (e_A^2+e_B^2)} r^2.\label{chol2+1}
 \end{eqnarray}
 This quantity diverges at large energies, hinting at a UV completion in terms of a system in higher dimensions (a LST). Also, it vanishes for $r=0$, indicating a gapped system. 
 Notice nevertheless that the calculation should not be trusted close to $r=0$, as the background (\ref{background1}) is singular there.
 
 Let us now calculate for the background in eq.(\ref{background 211}), as a dual to a ($1+1$)-dim QFT. We find
\begin{eqnarray}
& & d=1,~~a(r,y^i)= r,~~~ b(r)= \frac{4}{r^2 f_s(r)},\nonumber\\
& &V_{int}= {\cal N} r^2\sqrt{f_s(r)},\;\;\;\;
\hat{H}= {\cal N}^2 r^4 f_s(r),~~~ {\cal N}=(4\pi)^4\frac{8}{e_A^3 e_B^3} L_\varphi,\nonumber\\
& & c_{hol}= \frac{{\cal N}}{2 G_N}\frac{f_s(r) r^{2}}{\left( f_s(r) +\frac{r}{4} f'_s(r) \right)}.\label{chol211}
\end{eqnarray}
%We interpret this result as follows. 
At high energies
the number of degrees of freedom grows unbounded (as $r^{2}$), signalling the UV completion in terms of a decompatified QFT in higher dimensions.  At very low energies the number of degrees of freedom vanish, as $f_s(r_+)=0$.  In this case, the calculation is trustable, hence the gapped character of the system is clear.

A related calculation can be done that encodes the fact that we can think our field theory as a three dimensional QFT with anisotropies (or a QFT with a flow across dimensions). We follow the treatment described in section 8.2 of the paper \cite{Bea:2015fja}, see also \cite{Merrikin:2022yho}. Using the notation of \cite{Bea:2015fja}  we have,
\begin{eqnarray}
& &d=2,~ \alpha_0=\alpha_1=r,~~\alpha_2=r f_s(r),~~\beta= \frac{4}{r^2 f_s^{3/2}},~~H= {\cal N}^2 r^4 f_s(r),~{\cal N}=(\frac{2}{e_A e_B})^3 (4\pi)^4,\nonumber\\
& & c_{flow}= \frac{{\cal N}}{G_N}\frac{r^2 f_s}{(f_s(r)+ \frac{r}{4} f_s(r)')^2}
.\label{cflow}
\end{eqnarray}
The result in eq.(\ref{cflow}) has a similar interpretation. At low energies we have no degrees of freedom, at high energies an unbounded growth in the degrees of freedom indicates the UV completion. Note that  the growth at high energies for the flow-central-charge--eq.(\ref{cflow}), is slower than it is for the case in which we consider the QFT to be two dimensional, see eq. (\ref{chol211}). This same feature occurs when considering flows between conformal points in different dimensions.

These results suggest that our QFT generates a mass gap at low energies. The behaviour of the gauge couplings at low energies, see eq.(\ref{couplingaa}) suggest that the QFT is also confining. To ascertain the  confining behaviour we calculate Maldacena-Wilson and 't Hooft loops, that provide order parameters for confinement. We also investigate the Entanglement Entropy (EE), which gives information about the interplay between a confining IR and the non-local UV dynamics of the QFT.

\section{Maldacena-Wilson, 't Hooft loops and Entanglement Entropy}\label{otherobs}
In this section we  calculate different observables to learn more about the proposed field theory.
We start calculating the Maldacena-Wilson loops \cite{Maldacena:1998im,Rey:1998ik}, in order to test the above proposal that at low energies the QFT presents a mass gap and confines. We start with a summary of the formalism to compute Maldacena-Wilson loops. This same formalism is then adapted for the study of 't Hooft loops and Entanglement Entropy.
 \subsection{General comments on Maldacena-Wilson loops and similar probes}\label{wilsongeneralsection}
We start summarising general results pertaining  holographic Wilson loops. We follow the treatment of \cite{Sonnenschein:1999if}, \cite{Nunez:2009da}. This generic treatment is also useful for the study of other probes that reduce to an 'effective string' in the background. Hence it will apply to 't Hooft loops, Entanglement Entropy, as we discuss below.

Consider  a generic holographic  background of the form 
    \begin{equation}
        ds^{2}=-g_{tt}dt^{2}+g_{xx}d\vec{x}^{2}+g_{rr }
            dr^{2}+g_{ij}d\theta ^{i}d\theta ^{j}\ .\label{wilson1}
    \end{equation}
We assume that $g_{tt}$, $g_{xx}$, $g_{rr }$ depend only on the radial coordinate $r$. As usual, we propose a string embedding (parametrised in terms of $(\tau,\sigma)$ the worldsheet coordinates), which leads to a Nambu-Goto action for the F1-string of the form,
\begin{eqnarray}
& & t=\tau,~~~x=x(\sigma),~~~r=r(\sigma).\nonumber\\
& & S_{NG}= T_{F1} \int d\tau d\sigma \sqrt{g_{tt}(r) g_{xx}(r)x'^2 + g_{tt}(r)g_{rr}(r) r'^2}.\label{NGwilson}
\end{eqnarray}
From this action, the equations for the string moving in the generic background reduce to (see \cite{Nunez:2009da} for a detailed derivation)
    \begin{equation}
        \frac{dr }{d\sigma }=\pm \frac{dx}{d\sigma }V_{eff}\left( r \right) \, .\label{eqmov}
    \end{equation}
  We defined the effective potential 
    \begin{equation}
        V_{eff}\left( r \right) =\frac{F\left( r \right) }{CG\left( r
            \right) }\sqrt{F^{2}\left( r \right) -C^{2}}\ ,~~ F^2\left( r \right) =g_{tt}g_{xx},~~G^2\left( r \right)=g_{tt}g_{rr}.\label{potdef}
    \end{equation}
     The constant $C=\frac{F^2x'}{\sqrt{F^2 x'^2+ G^2 r'^2}}$ is obtained from one of the equations of motion. In the simpler case in which we take $x(\sigma)=\sigma$ we find eq.(\ref{eqmov}) from the conserved Hamiltonian. In that case $C= F(r_0)$, being $r_0$ the turning point of the string satisfying $r'(\sigma)=0$ (these are called {\it U-shaped} embeddings). We set $C=F(r_0)$ in what follows.  We enumerate below a set of properties of the U-shaped embeddings.

\begin{enumerate}
  \item{This formalism applies to  an open string whose end points are at $r \to \infty$, where  we add a D-brane.  Dirichlet boundary conditions  for the string at $r \rightarrow \infty $ require that $V_{eff}|_{r \rightarrow \infty}\sim \infty $. }
  \item{We compute the separation between the two ends of the string on the D-brane, which can be thought as the separation between a quark-antiquark pair. The energy of the pair of quarks calculated from the Nambu-Goto action needs regularisation, implemented by   subtracting the mass of two non-dynamical strings extended along the whole range of the radial coordinate $[r_+, \infty)$. 

 The separation and energy are given as functions of $r_0$ (the distance from the origin of the radial coordinate, $r_+$, to the position of the turning point of the string). The expressions for these quantities are,
    \begin{align}
        L_{QQ}\left( r_{0}\right) &=2\int_{r_{0}}^{+\infty }
        \frac{dz}{V_{eff}(z) }\, , \label{QQ separation}  \\
    E_{QQ}\left( r_{0}\right) &=F\left( r_{0}\right) L_{QQ}\left( r_{0}\right)
            +2\int_{r_{0}}^{+\infty }dz\frac{G\left( z\right) }{F\left( z\right) }
            \sqrt{F\left( z\right) ^{2}-F\left( r_{0}\right) ^{2}}-2\int_{r_{+}}^{+\infty }dz\
            G\left( z\right) \label{QQ energy} \, .
    \end{align}
    }
  \item{  
To obtain a finite contribution coming from the upper limit of the $Q\bar{Q}$ pair separation in \eqref{QQ separation}, a further restriction on the behaviour of the effective potential at infinity is needed. See \cite{Nunez:2009da} for a derivation,
    \begin{equation}
        V_{eff}|_{r \rightarrow +\infty}\sim r^{\beta}\, ,\text{\quad with } \beta >1\, .\label{restriction}
    \end{equation}
    }
\item{
When expanded close to the end of the space, which is at $r=r_+$ in the case of the background in eq.(\ref{background 211}), we find $V_{eff}\sim (r-r_+)^\gamma$. If $1 \leq \gamma$, the separation between the pair becomes infinite--see  \cite{Nunez:2009da}. Otherwise (if $\gamma<1$) we have screening behaviour.}
\item{
There is an analytic relation between $E_{QQ}$ and $L_{QQ}$ which is
\begin{equation}
    \frac{dE_{QQ}}{dr_{0}}=F(r_{0}) \frac{dL_{QQ}}{d r_{0}}~~\longrightarrow~~   \frac{dE_{QQ}}{dL_{QQ}}=F(r_{0}).\label{wilsonfinal}
\end{equation}
We have  inverted the relation \eqref{QQ separation} as $r_{0}=r_{0}(L_{QQ})$.
}
\item{
%Let us write an {\it approximate} expression for the separation of the quark-antiquark pair.
 For generic backgrounds, the evaluation of the integral in eq.(\ref{QQ separation}) need not be simple nor have an expression in terms of elementary functions. Nevertheless, the quantity
\begin{equation}
\hat{L}_{QQ}(r_0)= \pi \frac{G}{F'}\Big|_{r_0},\label{approxLQQ}
\end{equation}
provides a reasonable approximation to eq.(\ref{QQ separation}).
We check this approximate expression for different observables studied below.
}
\item{
Following \cite{Faedo:2014naa,Faedo:2013ota} we define
\begin{equation}
Z(r_0)= \frac{d}{dr_0} \hat{L}_{QQ}(r_0)= \pi \frac{d}{dr_0}\left( \frac{G(r_0)}{F'(r_0)}\right).\label{ZZr}
\end{equation}
The stability of the U-shaped string embedding in eq.(\ref{NGwilson}) is guaranteed if $Z(r_0)<0$ \cite{Faedo:2014naa,Faedo:2013ota}. This is valid for any observable that can be reduced to an effective string action of the form (\ref{NGwilson})\footnote{
It was shown in \cite{Faedo:2014naa}-\cite{Avramis:2006nv} that  embeddings  for which $Z(r_0)\geq 0$ do not satisfy two physically well motivated criteria \cite{Bachas:1985xs}:
The force between the quark and the antiquark  is always attractive and positive $\frac{dE_{QQ}}{dL_{QQ}}>0$. It  is also a non-increasing function of the separation $\frac{d^2 E_{QQ}}{d L_{QQ}^2}\leq 0$.
The proposal in \cite{Faedo:2014naa,Faedo:2013ota} is that the two criteria above are equivalent to the stability of the U-shaped embedding or conversely  $Z(r_0)<0$.}}
\item{
In calculations like those in eqs.(\ref{QQ separation})-(\ref{QQ energy}) we can introduce a cutoff  $r_{UV}$ to regulate divergences coming from the upper limit in the integrals. We  
define a quantity analog to (\ref{QQ separation}),
\begin{equation}
L_{QQ}(r_0,r_{UV})= 2\int_{r_0}^{r_{UV}} \frac{dz}{V_{eff}(z)}.
\end{equation}
Following \cite{Faedo:2014naa,Faedo:2013ota} we calculate
\begin{equation}
L_{a}= \lim_{r_0\to \infty} \lim_{r_{UV}\to\infty} L_{QQ}(r_0, r_{UV})~~\text{and}~~L_{b}= \lim_{r_{UV}\to \infty} \lim_{r_{0}\to r_{UV}} L_{QQ}(r_0, r_{UV}).\label{LaLb}
\end{equation}
Whilst $L_b=0$ by definition,  it is sometimes the case that $L_a$ is nonzero. 
If in this case the U-shaped configuration is unstable ($Z(r_0)>0$), there exist 'short configurations' (that appear very close to the cutoff). These short configurations are energetically favoured and induce a phase transition in the observable calculated  \cite{Faedo:2014naa}, \cite{Faedo:2013ota}, \cite{Kol:2014nqa}.
}
\end{enumerate}

The existence of these short configurations that appear when introducing the UV-cutoff is not an artefact  of the cutoff. They indicate that the string embedding we proposed in eq.(\ref{NGwilson}) is not capturing the dynamically favoured configuration. 
In the study of the 't Hooft loop and the Entanglement Entropy we encounter these short configurations, that cure the problem of the instability of the embedding and introduce a phase transition in the observable. We take the instability of the configuration (cured by the introduction of a cutoff) as an indication that the dynamics of the LST is driving the observable. 
%
%About regularisation of these observables: rather than introducing a cutoff we could subtract the result for the observable calculated in the background of eq.(\ref{background 211}) minus the same observable computed in the background of eq.(\ref{background1}). %We do this in Appendix \ref{appendix3}.
\\
 Let us now apply this general treatment to our background in eq.(\ref{background 211}).
\subsection{Maldacena-Wilson loops in our background}
Let us  apply the expressions in eqs.(\ref{wilson1})-(\ref{approxLQQ}) to our background in eq.(\ref{background 211}).
The relevant functions are 
\begin{equation}
F(r)=r,~~~~ G(r)=\frac{2}{\sqrt{f_{s}(r)}},~~~~~V_{eff}(r)= \sqrt{\frac{e_A^2+e_B^2}{8} } \left( \frac{1}{r_0 ~ r} \right) \sqrt{(r^2-r_0^2)(r^2-r_+^2)(r^2-r_-^2)}.\label{functionswilson}
\end{equation}
The condition in eq.(\ref{restriction}) to have  Dirichlet boundary conditions for the string at $r\to\infty$, is satisfied. When expanded close to $r_0=r_+$, $V_{eff}\sim (r-r_+)$, indicating that $L_{QQ}$ diverges when $r_0$ approaches the end of the space $r_+$. This points to  confining behaviour. 
%In fact, the quark-anti-quark pair separation grows unbounded, accompanied by the QCD-string carrying linearly growing energy.

Another quick way of determining a confining behaviour is to check the function $F(r_+)$, that intuitively represents the tension of a QCD-string at low energies in the QFT (close to the end of the space of the geometry). In this case we find $F(r_+)=r_+$,  pointing to confining behaviour. In fact, a finite QCD-string tension leads to an energy growing with the separation of the quark pair.
{Note also that the approximate formula for the separation of the quark pair eq.(\ref{approxLQQ}), gives for the functions in eq.(\ref{functionswilson})
\begin{equation}
\hat{L}_{QQ}(r_0)=\frac{2 \pi}{\sqrt{f_s(r_0)}}.\label{lqqapp}
\end{equation}
For $r_0\to\infty$ this gives $\hat{L}_{QQ}= \pi\sqrt{\frac{8}{e_A^2+e_B^2}}$, which refers to a characteristic length of the UV completion (the Little String Theory scale). On the other hand, for $r_0\sim r_+$, the approximate length diverges and $\frac{1}{\sqrt{r_0-r_+}}$. This indicates that the quark-anti-quark pair can be infinitely separated. According to eq.(\ref{QQ energy}), this gives  an energy that scales linearly with the separation,  another signal of a confining behaviour.}

The quantity $Z(r_0)$ in eq.(\ref{ZZr}) reads,
\begin{equation}
Z(r_0)= \frac{2\sqrt{2} \pi r_0}{\sqrt{e_A^2+e_B^2}} \frac{\left(  -r_0^2(r_+^2+ r_-^2) +2 r_+^2r_-^2\right)}{\left[ (r_0^2-r_+^2)(r_0^2-r_-^2)\right]^{3/2}}.\label{ZZrwilson}
\end{equation}
By inspection, one find that $Z(r_0)<0$ in all the range. Hence the U-shaped  embeddings are stable and no phase transition is expected for the Maldacena-Wilson loops.

\begin{figure}
\centering
    \begin{minipage}{.5\textwidth}
    \centering
    \includegraphics[width=1.0\linewidth]{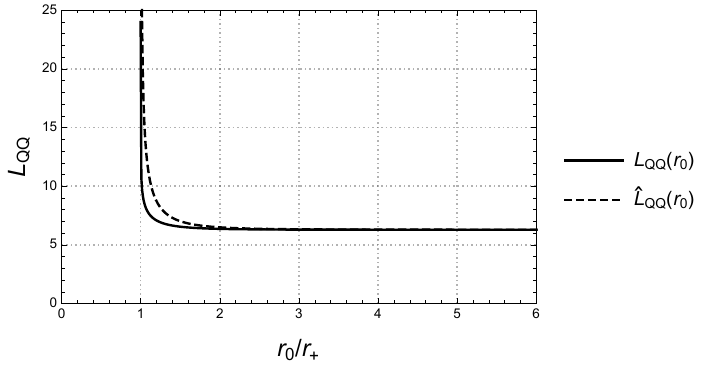}
    \end{minipage}
    \begin{minipage}{.45\textwidth}
    \centering
    \includegraphics[width=.9\linewidth]{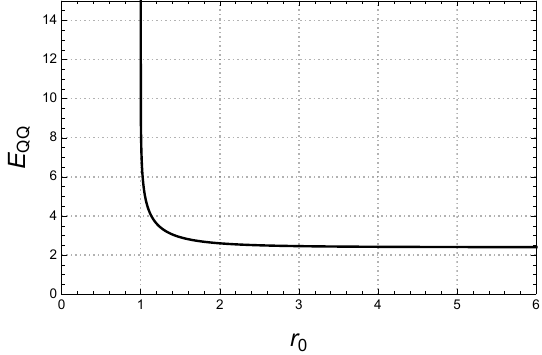}
    \end{minipage}
    \begin{minipage}{.5\textwidth}
    \centering
    \includegraphics[width=.9\linewidth]{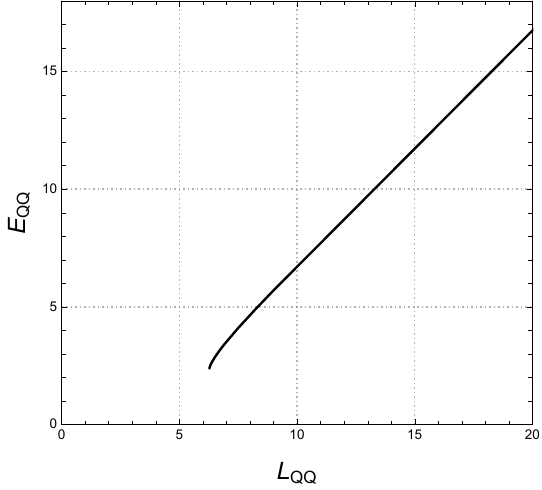}
    \end{minipage}
    \begin{minipage}{.4\textwidth}
    \centering
    \includegraphics[width=.9\linewidth]{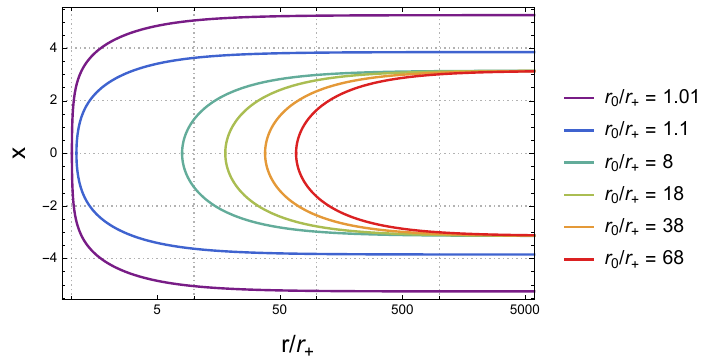}
    \end{minipage}
\caption{Upper left: Plot comparing the exact expression for the quark separation $L_{QQ}$ in eq.(\ref{lqq2}) with the approximate one $\hat{L}_{QQ}$ in eq.(\ref{lqqapp}). The minimal separation as $r_0$ grows large hints at a LST behaviour.  Upper right: Plot of $E_{QQ}(r_0)$  with $r_0$ in units of $r_+$. The plot is made in the BPS limit with $m=0$. Bottom left: Parametric plot of $E_{QQ}(L_{QQ})$ in the BPS bound $m=0\,,\, r_+=e_B=e_A=1$. Bottom right: the profiles of different strings as they explore the bulk. The longer the separation $L_{QQ}$, the more the string approaches $\frac{r_0}{r_+}\sim 1$. This is usual of the backgrounds dual to a confining  QFT behaviour.}\label{figura1a}
\end{figure}

The rigorous way of determining the low energy behaviour of the QFT (either confining or screening) is to analyse the expressions for the distance and the energy of the quarks pair. 
The distance between the quark-antiquark pair  and its energy are written from eqs.(\ref{QQ separation})-(\ref{QQ energy})
\begin{eqnarray}
& & L_{QQ}\left( r_{0}\right) = \sqrt{\frac{32 }{e_{A}^2+ e_B^2 }} r_0 \int_{r_{0}}^{\infty }z
\sqrt{\frac{1}{\left( z^{2}-r_{+}^{2}\right) \left( z^{2}-r_{-}^{2}\right)
\left( z^{2}-r_{0}^{2}\right) }}dz\ ,\label{Lqq} \\
& & E_{QQ} \left( r_{0}\right) =r_{0} L_{QQ} \left( r_{0}\right) +\sqrt{\frac{32}{e_{A}^2+ e_B^2 } } 
\Big[
 \int_{r_{0} }^{\infty } dz \sqrt{\frac{z^{2}\left( z^{2}-r_{0}^{2}\right) }{ \left( z^{2}-r_{+}^{2}\right) \left( z^{2}-r_{-}^{2}\right) } }  -\nonumber\\
 & & 
 \int_{r_{+} }^{\infty } dz \frac{z^{2}}{\sqrt{\left( z^{2}-r_{+}^{2}\right) \left( z^{2}-r_{-}^{2}\right) }}
 \Big]\ .\label{Eqq}
\end{eqnarray}%
The integrals can be performed analytically and expressed in terms of in terms of Elliptic integrals of the first kind. It is interesting to study this in the BPS limit, when $m=0 $ and $r_+^2=- r_-^2$. See Appendix \ref{appendix3} for a detailed study. Let us quote the explicit results for the separation and energy of the quark pair in eqs.(\ref{Lqq})-(\ref{Eqq}). Defining
the elliptic integrals,
\begin{eqnarray}
& & {\bf K}(x)=\int_0^{\frac{\pi}{2}} \frac{d\theta}{\sqrt{1-x \sin^2\theta}},\;\;\;\; {\bf E}(x)= \int_0^{\frac{\pi}{2}} \sqrt{1 -x \sin^2\theta} d\theta,\label{definitionsEK}
\end{eqnarray}
we can write the explicit expressions for the separation and energy of the quark-antiquark pair, which read
\begin{equation}
L_{QQ}\left( r_{0}\right) =2\sqrt{\frac{8}{e_{A}^{2}+e_{B}^{2}}}\frac{%
r_{0}/r_{+}}{\sqrt{r_{0}^{2}/r_{+}^{2}-1}}\mathbf{K}\left( \frac{%
1-r_{-}^{2}/r_{+}^{2}}{1-r_{0}^{2}/r_{+}^{2}}\right) \ ,\label{lqq2}
\end{equation}%
\begin{eqnarray}
E_{QQ}\left( r_{0}\right)  &=&2r_{+}\sqrt{\frac{8}{e_{A}^{2}+e_{B}^{2}}}%
\left[ \frac{r_{0}^{2}/r_{+}^{2}}{\sqrt{r_{0}^{2}/r_{+}^{2}-1}}\mathbf{K}%
\left( \frac{1-r_{-}^{2}/r_{+}^{2}}{1-r_{0}^{2}/r_{+}^{2}}\right) \right. \nonumber\\
&&\qquad \qquad \left. -\mathbf{E}\left( \frac{1-r_{-}^{2}/r_{+}^{2}}{%
1-r_{0}^{2}/r_{+}^{2}}\right) \sqrt{r_{0}^{2}/r_{+}^{2}-1}+C\left(
r_{-}/r_{+}\right) \right] ,  \label{eqq2}
\end{eqnarray}%
where%
\begin{eqnarray}
& & C\left( r_{-}/r_{+}\right)  =\mathbf{E}\left( r_{-}^{2}/r_{+}^{2}\right)
+\lambda _{-}\mathbf{K}\left( 1-r_{-}^{2}/r_{+}^{2}\right) +i\lambda _{-}%
\mathbf{K}\left( r_{-}^{2}/r_{+}^{2}\right)  -\left( 1-r_{-}^{2}/r_{+}^{2}\right) \mathbf{K}\left(
r_{-}^{2}/r_{+}^{2}\right) \ .  \notag\\
& & \lambda_-^2=-\frac{r_-^2}{r_+^2}.\notag
\end{eqnarray}
See Appendix \ref{appendix3} for a careful derivation of these expressions. We plot these results in  Figure \ref{figura1a}.  The various panels of Figure \ref{figura1a} show some conventional and other less conventional behaviours. First, note that the expression for  $\hat{L}_{QQ}$ in eq.(\ref{lqqapp}) very well approximates the exact expression in eq.(\ref{lqq2}). As is usual, the concavity of the curve $E_{QQ}(L_{QQ})$ is 'downwards', indicating that the Nambu-Goto string  configuration of eq.(\ref{functionswilson}) is stable, as confirmed by the $Z(r_0)$ in eq.(\ref{ZZrwilson}). Note also that for large separations between the quark pair $L_{QQ}$, the energy grows linearly (signalling confinement). What is less conventional is that there is a minimal separation, given by the Little String Theory scale. This indicates that the far UV of the QFT  dynamics is not field theoretical (but has the dynamics of the LST). The plot of the strings profiles confirms this. Indeed,  strings that barely explore the bulk (with $\frac{r_0}{r_+}$ large) show a minimal fixed separation between the quark pair. On the other hand, the strings that explore deeper into the bulk display a bigger quark separation and carry higher energy.

Let us now focus on a second interesting observable, the 't Hooft loop. In the next section we propose a string-like object (for the gauge theory observer) with magnetic charge. This characteristically is represented by a Dp brane that wraps a $(p-1)$ cycle in the internal space. Once the Born-Infeld action for this Dp brane is written and integrals over the internal space are performed, we arrive at an action for the 'effective string'. This action is studied with the same formalism as that used for Wilson loops, described in Section \ref{wilsongeneralsection}.

%###################

\subsection{'t Hooft loops}
The 't Hooft loop can be calculated by proposing an object with magnetic charge that effectively appears like a string for the (1+1) dimensional QFT. The ends of this magnetic string appear as a pair of monopoles of oposite charge.  To study this object, we adapt the formulas
summarised in eqs.(\ref{wilson1})-(\ref{ZZr}) for the effective magnetic string.

We propose to calculate the 't Hooft loop by studying the effective magnetic string obtained when extending a D5 brane along the directions $[t,x,\varphi,\theta_A,\phi_A,\psi_A]$, with $r(x)$. There is an analog magnetic string for the second gauge group, for the D5 in the configuration  $[t,x,\varphi,\theta_B,\phi_B,\psi_B]$, with $r(x)$. We do not discuss the latter object, as its result is analog to the one obtained below. Note that this object becomes string-like when we consider the size of the $S^1_\varphi$ to be small enough to not admit excitations. 

The induced metric for the above D5 is,
\begin{eqnarray}
ds^{2} &=&r\left\{ -dt^{2}+\left( 1+\frac{4}{r^{2}f_{s}\left( r\right) }
r^{\prime 2}\right) dx^{2}+\left( f_{s}\left( r\right) +2Q_{B}^{2}\zeta
\left( r\right) ^{2}\right) d\varphi ^{2}\right. \nonumber \\
&&\left. +\frac{2}{e_{A}^{2}}\left[ \hat{\omega}_{1}^{2}+\hat{\omega}%
_{2}^{2}+\left( \hat{\omega}_{3}-e_{A}Q_{A}\zeta (r)d\varphi \right) ^{2}%
\right] \right\} \ .
\end{eqnarray}%
Then we calculate,
\begin{equation}
e^{-\Phi}\sqrt{-\det g_{MN}}=\left( \frac{2}{e_{A}^{2}}\right) ^{3/2}r^{2}\sin {\theta_A}\sqrt{\left( 1+\frac{4r^{\prime 2}}{r^{2}f\left( r\right) }\right)
\left( f\left( r\right) +2Q_{B}^{2}\zeta \left( r\right) ^{2}\right) }\ .
\end{equation}%
The action for this D5 is,
\begin{equation}
S_{\hat{D}_{5}}\left[ r\right] =T_{{D}_{5}}L_{\varphi }\left( 4\pi
\right) ^{2}\left( \frac{2}{e_{A}^{2}}\right) ^{3/2}\int dtdx\sqrt{\left(
f_{s}\left( r\right) +2Q_{B}^{2}\zeta \left( r\right) ^{2}\right)
r^{4}+4r^{2}\left( 1+\frac{2Q_{B}^{2}\zeta \left( r\right) ^{2}}{f_{s}\left(
r\right) }\right) r^{\prime 2}}\ .  \notag
\end{equation}%
Comparing with (\ref{NGwilson})-(\ref{potdef}) we identify%
\begin{eqnarray}
{F}^{2} =\left( f_s(r) +2 Q_B^2\zeta(r)^2\right) r^4 \,\ ,~~~ {G}^{2}=\frac{4}{%
r^{2}f_{s}\left( r\right) }{F}^{2}\ , ~~T_{eff}  =T_{{D}_{5}}L_{\varphi }L_{t}\left( 4\pi \right) ^{2}\left( 
\frac{2}{e_{A}^{2}}\right) ^{3/2}\ .\label{thooftloopqunatities}
\end{eqnarray}%
Where $T_{eff}$ is the effective tension of the magnetic string.
Following eq.(\ref{potdef}),  the effective potential 
\begin{equation}
V_{eff}\left( r\right) =\frac{r^{3}}{2C}\sqrt{f_{s}\left( r\right) \left(
f_{s}\left( r\right) +2Q_{B}^{2}\zeta \left( r\right) ^{2}-\frac{C^{2}}{r^{4}%
}\right) }.\label{veffmm}
\end{equation}%
The constant $C= F(r_0)= \sqrt{\left( f_s(r_0) +2 Q_B^2\zeta^2(r_0)\right) }r_0^2  $. In the asymptotic region the effective potential in eq.(\ref{veffmm}) scales as $
V_{eff}\left( r\rightarrow \infty \right) \sim r^{3}$, satisfying the condition in eq.(\ref{restriction}). 

The intuitive criteria discussed in the previous section applied for this case indicate a screening behaviour. In fact $F(r_+)=0$ and $V_{eff}\sim(r-r_+)^{1/2}$, signalling a vanishing tension of the magnetic QCD-string in the IR, and a finite maximal separation, after which the monopoles are screened (more about this is discussed below).

{We  perform a similar analysis using the {\it approximate} expression for the separation between the monopole-anti-monopole pair $\hat{L}_{MM}$. Replacing in eq.(\ref{approxLQQ}) the functions in eq.(\ref{thooftloopqunatities}), we find an involved expression that asymptotes as,
\begin{equation}
\hat{L}_{MM}(r_0\to\infty)\sim \pi\sqrt{\frac{2}{e_A^2+ e_B^2}},\;\;\;\;\; \hat{L}_{MM}(r_0\to r_+ )\sim 8\pi \sqrt{\frac{r_+(r_0-r_+)}{(e_A^2+e_B^2)(r_+^2-r_-^2)}}.\label{aproxLmm}
\end{equation}
These asymptotic behaviours indicate that at high energies in the field theory, the pair is separated by a maximum distance characteristic of the UV completion (Little String Theory). In this sense, the magnetic string behaves oppositely to the electric one used to compute the Wilson loop (which shows a minimum separation). On the other hand, at low energies the separation decreases to zero (again, oppositely to the electric string case). We calculate $Z(r_0)$ in eq.(\ref{ZZr}) for this configuration and find that is positive in all the range. This indicates the instability of the U-shaped embeddings.  A more general embedding that the one proposed here should drive the dynamics. Instead of finding this more complicated embedding,  below we introduce a UV-cutoff. New short configurations appear close to a cutoff, that dominate the dynamics and produce a phase transition (to a deconfining behaviour).
}

\begin{figure}
\centering
    \begin{minipage}{.5\textwidth}
    \centering
    \includegraphics[width=1.0\linewidth]{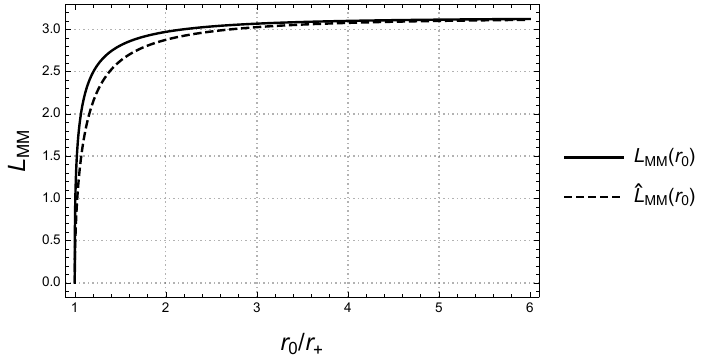}
    \end{minipage}
    \begin{minipage}{.45\textwidth}
    \centering
    \includegraphics[width=.9\linewidth]{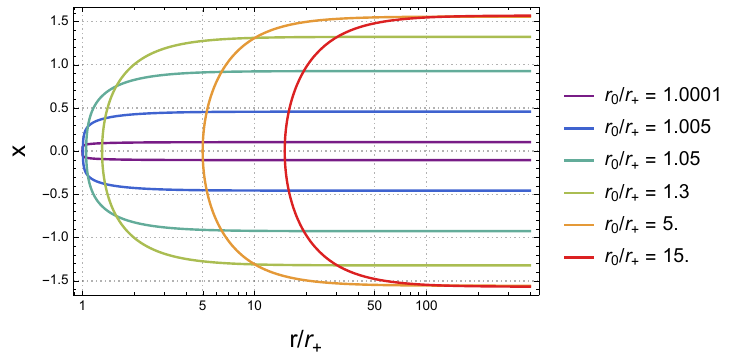}
    \end{minipage}
%
 %   \begin{minipage}{.5\textwidth}
   % \centering
%    5\includegraphics[width=.9\linewidth]{Wilson_EQQvsLQQ.pdf}
    %\end{minipage}
%
%
    %\begin{minipage}{.4\textwidth}
    %\centering
    %\includegraphics[width=.9\linewidth]{Wilson_String_Profile.pdf}
    %\end{minipage}
%
\caption{Left: Plot comparing the exact expression for the quark separation $L_{MM}$ in eq.(\ref{Lmmexact}) with the approximate one $\hat{L}_{MM}$ in eq.(\ref{aproxLmm}), both in the BPS limit. Right: the profiles of different strings as they enter the bulk. Strings with small separation of the monopole pair penetrate deeper into the bulk. There is a maximum separation for the pair of monopoles, associated with the Little String Theory scale. %The energy of the configuration grows as the separation of the monopole pair grows. This points to a phase transition to a disconnected configuration.  This behaviour is usual of the backgrounds dual to a screening behaviour for the magnetically charged objects
}\label{figura2}
\end{figure}

More formally, we write expressions for the separation between the monopole-anti-monopole pair as those in eqs.(\ref{QQ separation})-(\ref{QQ energy}). A careful analysis of these integrals is performed in Appendix \ref{appendix3}. Let us quote the exact expression for the separation for the monopole-anti-monopole pair. We leave the study of  the expression for the energy between the pair of monopoles for Appendix \ref{appendix3}). In terms of the elliptic integral of the first kind
\begin{equation}
{\bf F}\left(y| x\right)= \int_0^y \frac{d\theta}{\sqrt{1-x \sin^2\theta}},\label{definitionelliptic2}
\end{equation}
and the definition for $\bf{K}(x)$ in eq.(\ref{definitionsEK}), working in the BPS limit and defining $\eta =e_{A}/e_{B}$, we find
\begin{eqnarray}
L_{MM}^{BPS}\left( r_{0}\right)  &=&\frac{2}{e_{B}}\sqrt{\frac{2\left(
r_{0}^{2}/r_{+}^{2}-1\right) \left( \left( \eta ^{2}+2\right)
r_{0}^{2}/r_{+}^{2}+\eta ^{2}\right) }{\left( \eta ^{2}+1\right) \left(
r_{0}^{2}/r_{+}^{2}\left( \eta ^{2}+2\right) -1\right) }}\nonumber \\
&&\times \left[ \mathbf{F}\left( \left. \arcsin \sqrt{\frac{2\left(
-1+\left( 2+\eta ^{2}\right) r_{0}^{2}/r_{+}^{2}\right) }{\left(
2r_{0}^{2}/r_{+}^{2}+\eta ^{2}\left( 1+r_{0}^{2}/r_{+}^{2}\right) \right) }}%
\right\vert \frac{\left( 1+r_{0}^{2}/r_{+}^{2}\right) \left(
2r_{0}^{2}/r_{+}^{2}+\eta ^{2}\left( 1+r_{0}^{2}/r_{+}^{2}\right) \right) }{%
-4+4\left( 2+\eta ^{2}\right) r_{0}^{2}/r_{+}^{2}}\right) \right.   \notag \\
&&\qquad \left. +i\mathbf{K}\left( 1-\frac{\left(
1+r_{0}^{2}/r_{+}^{2}\right) \left( 2r_{0}^{2}/r_{+}^{2}+\eta ^{2}\left(
1+r_{0}^{2}/r_{+}^{2}\right) \right) }{-4+4\left( 2+\eta ^{2}\right)
r_{0}^{2}/r_{+}^{2}}\right) \right] \ . \label{Lmmexact}
\end{eqnarray}%
In Figure \ref{figura2} we compare this exact expression with the approximating function whose asymptotics we write in eq.(\ref{aproxLmm}).  As stated above, opposite to the Wilson loop case, we encounter a maximal separation for the monopole pair. This maximal separation is associated with the Little String Theory scale. We  also plot the string profiles as they enter the bulk. We observe a different behaviour to that found in the case of the Wilson loop--compare with  the lower right panel of Figure \ref{figura1a}.

For magnetic strings that barely explore the bulk, the separation between the monopole pair is large (equal to the LST scale). As we decrease the separation between the monopole pair, the magnetic string dives into $r_0\to r_+$. These unconventional behaviours, together with the instability of the string embedding--note that $E_{MM}(L_{MM})$  has upwards concavity (and $Z(r_0)$ is positive), indicate the presence of a second 'disconnected' configuration for which the pair of monopoles separate without energy  expense, which is indicative of screening. The transition to a disconnected configuration is dynamically favoured.

%\textcolor{red}{explicar mejor lo que ocurre con thooft loops, poner un cutoff, mencionar las sutilezas de las integrales para la energia--apendice, etc}

To avoid the 't Hooft loop to be driven by the UV (LST) dynamics and to  realise explicitly these short configurations, we introduce a hard cutoff in the radial direction and recalculate things. Doing so, the behaviour changes qualitatively.

\begin{figure}
\centering
    \begin{minipage}{.5\textwidth}
    \centering
    \includegraphics[width=1.0\linewidth]{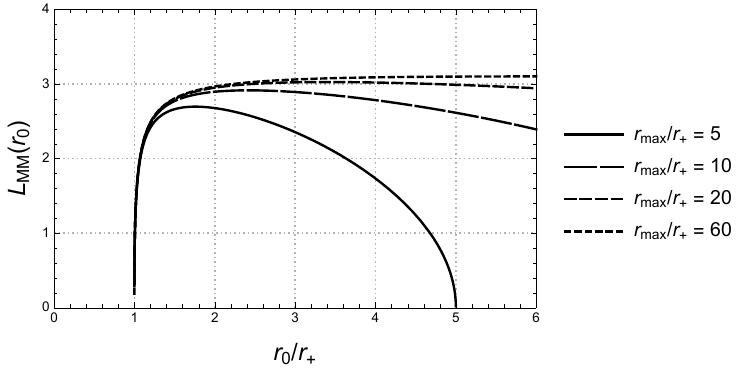}
    \end{minipage}
    \begin{minipage}{.45\textwidth}
    \centering
    \includegraphics[width=.9\linewidth]{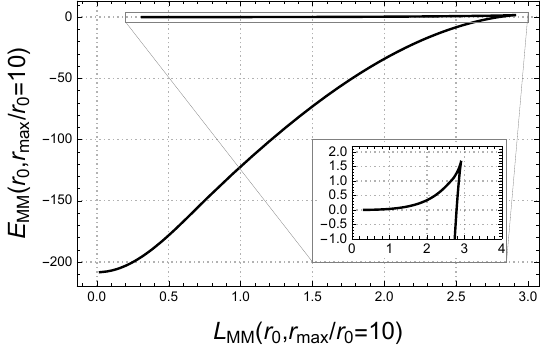}
    \end{minipage}
%
 %   \begin{minipage}{.5\textwidth}
   % \centering
%    5\includegraphics[width=.9\linewidth]{Wilson_EQQvsLQQ.pdf}
    %\end{minipage}
%
%
    %\begin{minipage}{.4\textwidth}
    %\centering
    %\includegraphics[width=.9\linewidth]{Wilson_String_Profile.pdf}
    %\end{minipage}
%

\caption{Left: after introducing a cutoff at $r_{MAX}$, the figure shows the double-valued character of $L_{MM}$. Removing the cutoff recovers the LST behaviour (on the left panel of Figure \ref{figura2}). Right: with the UV cutoff at $r_{MAX}=20$, we see that $E_{MM}$ as a function of $L_{MM}$  has the 'upwards' concavity (indicating stability). The presence of the phase transition to a disconnected configuration is observed.}\label{figura6}
\end{figure}

In fact, the separation between the pair of monopoles does not show a maximum value, instead we find a 'double valued' behaviour as displayed in  the left panel of Figure \ref{figura6}.
This  leads to a phase transition in the curve $E_{MM}(L_{MM})$. Note that the curve now has the correct 'downwards' concavity, indicating that the configuration is stable. This phase transition is the physical manifestation of the magnetic string suddenly changing into two disconnected magnetic strings that move without energy expense. This is deconfinement for the pair monopole anti-monopole. We are finding confinement for the quark-anti-quark pair and screening for a pair of monopoles. These behaviours are consistent with the (electric) confining behaviour of the dual QFT.

The introduction of the UV cutoff might seem unsatisfactory. Here we use it as a device to show that the correct five brane embedding must be more elaborated than the one we proposed above. It is also used to avoid the LST overtaking the dynamics.  The effect of the cutoff is clear considering the integral needed to calculate $L(r_0)$. This integral vanishes for $r_0\to r_{MAX}$. This produces a double-valued $L(r_0)$ and a consequent phase transition.

For the Entanglement Entropy a very similar behaviour occurs. We study this next.

\subsection{Entanglement Entropy}

The Entanglement Entropy (EE) between two regions for field theories with a string dual can be calculated as shown in  \cite{Ryu:2006bv,Ryu:2006ef}. The method is to find  a minimal area eight-surface ($\Sigma_8$, a codimension-two surface to which we refer below as RT surface) such that the boundary of the surface coincides with the two entangled regions. We focus on the case in which one of the regions is a strip of size $L_{EE}$ and the other region is the complement.
The EE between these regions is given in \cite{Ryu:2006bv}-\cite{Klebanov:2007ws}, minimising the quantity
\begin{equation}
S_{EE}=\frac{1}{4 G_N}\int_{\Sigma_8} d^8\sigma \sqrt{e^{-4\Phi} \det[g_{\Sigma_8}]}.\label{SEEdef}
\end{equation}
%String backgrounds dual to confining theories characteristically have cycles that shrink smoothly at the IR-end of the spacetime. In the case of our background in eq.(\ref{background 211}), the cycle parametrised by the coordinate $\varphi$ shrinks at $r=r_+$. The presence of shrinking cycles imply a behaviour for the EE different from that of Wilson loops. 

There are various eight-surfaces that minimise $S_{EE}$ in eq.(\ref{SEEdef}). Due to this, in some cases there is a phase transition between different extremal surfaces. 
It was suggested in \cite{Klebanov:2007ws} that a  criterion for confinement is the presence of a phase transition in the EE. This proposal was critically analysed in  \cite{Kol:2014nqa,Jokela:2020wgs}. It was found in \cite{Kol:2014nqa}, that for the case of field theories that confine, but have a non-local high energy behaviour, the phase transition in the EE is absent. The point is subtle, as introducing a UV cutoff or UV-completing the QFT to avoid the non-locality, recovers the phase transition. %This is analogous to the behaviour of the 't Hooft loop studied in the previous section.

It is in this way that the EE can serve as an order parameter for confinement, but also as a tool to diagnose non-locality in the UV-behaviour of the QFT (when used  together with a confining Wilson loop).
In  \cite{Kol:2014nqa} it was found that introducing  a UV-cutoff implies the existence of new configurations realising the phase transition (and resolving a stability issue with the original eight-surface). Below, we perform an analysis of these features in our background of eq.(\ref{background 211}).

We follow the approach of \cite{Ryu:2006bv,Ryu:2006ef}, in particular the treatment for non-AdS backgrounds developed in  \cite{Kol:2014nqa,Klebanov:2007ws}.
We calculate the Entanglement Entropy on a strip by computing the area of an eight-surface $[x,\varphi, \theta_A,\phi_A,\psi_A,\theta_B,\phi_B,\psi_B]$ with $r=r(x)$ in the background of eq.(\ref{background 211}). 
%Separating the space in an interval of size $L$ and the complement, we calculate the entanglement between these two regions. 
The induced metric on the RT eight-surface, its determinant and the Entanglement Entropy are,
\begin{eqnarray}
ds^{2}_{st} &=&r\left\{ dx^{2}(1+\frac{4 r'^2}{r^2 f_s(r)}) +f_{s}\left( r\right) d\varphi ^{2}+\frac{2}{e_{A}^{2}}\left[ \hat{%
\omega}_{1}^{2}+\hat{\omega}_{2}^{2}+\left( \hat{\omega}_{3}-e_{A}Q_{A}\zeta
(r)d\varphi \right) ^{2}\right] \right.   \notag \\
& & \left. +\frac{2}{e_{B}^{2}}\left[ \tilde{\omega}_{1}^{2}+\tilde{\omega}%
_{2}^{2}+\left( \tilde{\omega}_{3}^{2}-e_{B}Q_{B}\zeta \left( r\right)
d\varphi \right) ^{2}\right] \right\} \ ,  \label{8surface}\\
 \sqrt{e^{-4\Phi}\det[g_8]}&=& \left(  \frac{8}{e_A^3e_B^3}\right)\sqrt{r^4 f_s(r) + 4 r^2 r'^2}\sin\theta_A\sin\theta_B.\nonumber\\
 S_{EE}&=& \frac{1}{4 G_N}\int d^8x \sqrt{e^{-4\Phi}\det[g_8]}=\left(  \frac{2 (4\pi)^4 L_\varphi}{e_A^3e_B^3 G_N}\right)\int_{-L/2}^{L/2} dx \sqrt{r^4 f_s(r) + 4 r^2 r'^2}.\label{EE211}
\end{eqnarray}
From eqs.(\ref{NGwilson})-(\ref{potdef}), this implies
\begin{equation}
F(r)= r^2\sqrt{f_s(r)},~~~~G(r)= 2 r.\label{FGEE}
\end{equation}
To minimise the $S_{EE}$ above, we follow the usual conserved Hamiltonian treatment. The Entanglement Entropy needs to be regularised by the area of two eight-surfaces that hang straight from infinity. Then, computing the regulated area for a surface that turns around at $r_0$, we find for the length of the interval and the Entanglement Entropy,
\begin{eqnarray}
& & L= 4 r_0^2 \sqrt{f_s(r_0)} \int_{r_0}^\infty \frac{dr}{\sqrt{r^2 f_s(r) \left( r^4 f_s(r) - r_0^4 f_s(r_0) \right)}}, \label{LEE}\\
& & S_{EE}= \frac{{\cal N}}{G_N} \left[\int_{r_0}^\infty \sqrt{\frac{r^6 f_s(r)}{r^4 f_s(r) -r_0^4 f_s(r_0) }} dr -\int_{r_+}^\infty  r dr \right].\label{SEE}
\end{eqnarray}
%The reader can compare with the general expressions in equations (2.5)-(2.10) of the paper \cite{Kol:2014nqa}, using the map 
%\begin{equation}
%\alpha=r,~\beta= \frac{4}{r^2 f_s(r)},~~~H= {\cal N}^2 r^4 f_s(r), ~~{\cal N}= \frac{8 (4\pi)^4 L_\varphi}{e_A^3 e_B^3}. \label{map}
%\end{equation}
As in eq.(\ref{approxLQQ}), we can write a simple expression that approximates $L_{EE}$ in eq.(\ref{LEE})---see \cite{Kol:2014nqa},
\begin{equation}
\hat{L}_{EE}= \frac{\pi G(r_0)}{F'(r_0)}=2\pi\frac{H(r)\sqrt{\beta(r)}}{H'(r)}\Big|_{r_0},~~\text{with}~~H(r)= {\cal N}^2 r^4 f_s(r),~~\beta(r)=\frac{4}{r^2 f_s(r)}.\label{LEEapprox}
\end{equation}
Using eq.(\ref{LEEapprox}) we find
\begin{equation}
\hat{L}_{EE}= \left(\pi\sqrt{\frac{8}{e_A^2+e_B^2} }\right)
\frac{\sqrt{(r_0^2-r_+^2)(r_0^2-r_-^2)}}{\left(2r_0^2-r_+^2-r_-^2\right)} .\label{LEEE}
\end{equation}
This function is monotonous, going from a vanishing value at $r_0=r_+$ to a constant value at $r_0\to\infty$. This behaviour prevents the possibility of phase transitions, which require that for a given $L_{EE}$ there are two possible values of $r_0$.
In fact, the conditions for the presence of a phase transition (see section 2.4 of the work  \cite{Kol:2014nqa}) are not satisfied, in particular equations (2.26)-(2.29) of \cite{Kol:2014nqa} imply $j=2$ preventing a phase transition. The absence of a phase transition in a confining model was interpreted in  \cite{Kol:2014nqa} as an effect of the non-locality of the completion of the QFT, in this case, by a LST.

Since $Z(r_0)$  defined in eq.(\ref{ZZr}) gives
\begin{equation}
Z(r_0)= \frac{\sqrt{8} \pi (r_+^2-r_-^2)^2}{r_0 f_s(r_0) (r_+^2+r_-^2-2r_0^2)^2}>0,\label{ZZrEE}
\end{equation}
the proposed embedding is unstable. Upon the introduction of a cutoff, new surfaces appear as found in \cite{Kol:2014nqa}, \cite{Barbon:2008ut},\cite{Barbon:2008sr}. These cure the instability problem of the embedding and give place to the phase transition, in agreement with confinement.
% is recovered when introducing a UV-cutoff. In that case, new configurations appear that minimise  $S_{EE}$. The way these arise is as explained in 

The treatment in the papers \cite{Kol:2014nqa}, \cite{Barbon:2008ut}, \cite{Barbon:2008sr}  applies to our background, even when the IR dynamics is different, the UV dynamics is similarly driven by a Little String Theory. 

\begin{figure}
\centering
    \begin{minipage}{.5\textwidth}
    \centering
    \includegraphics[width=1.0\linewidth]{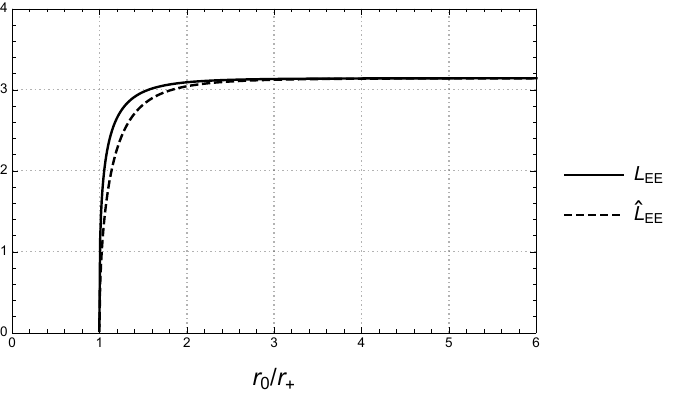}
    \end{minipage}
    \begin{minipage}{.45\textwidth}
    \centering
    \includegraphics[width=.9\linewidth]{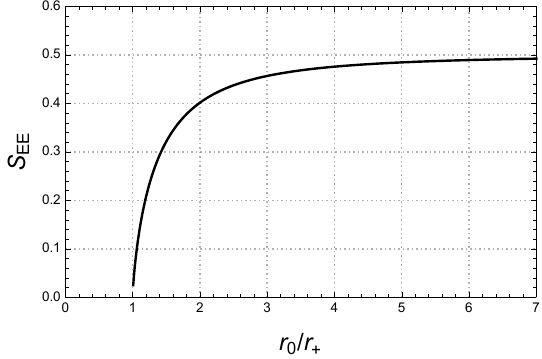}
    \end{minipage}
%
 %   \begin{minipage}{.5\textwidth}
   % \centering
%    5\includegraphics[width=.9\linewidth]{Wilson_EQQvsLQQ.pdf}
    %\end{minipage}
%
%
    %\begin{minipage}{.4\textwidth}
    %\centering
    %\includegraphics[width=.9\linewidth]{Wilson_String_Profile.pdf}
    %\end{minipage}
%
\caption{Left: Plot comparing the exact expression for the separation of the entangled regions  $L_{EE}$ in eq.(\ref{LEEexact}) with the approximate one $\hat{L}_{EE}$ in eq.(\ref{LEEE}). Right: The plot of $S_{EE}(r_0)$.
% the profiles of different strings as they enter the bulk. Strings with small separation of the monopole pair penetrate deeper into the bulk. There is a maximum separation for the pair of monopoles, associated with the Little String Theory scale. The energy of the configuration grows as the separation of the monopole pair grows. This points to a phase transtion to a disconnected configuration.  This behaviour is usual of the backgrounds dual to a screening behaviour for the magnetically charged objects.
}\label{figura3}
\end{figure}

\begin{figure}
\centering
    \begin{minipage}{.5\textwidth}
    \centering
    \includegraphics[width=1.0\linewidth]{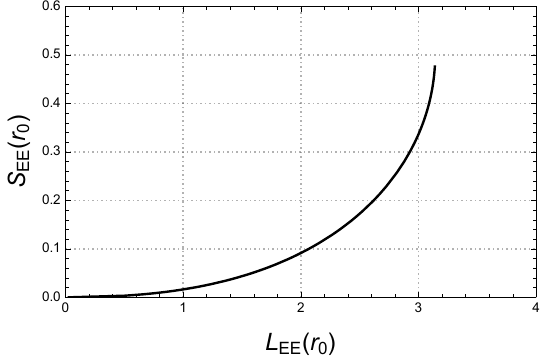}
    \end{minipage}
   \begin{minipage}{.45\textwidth}
    \centering
    \includegraphics[width=.9\linewidth]{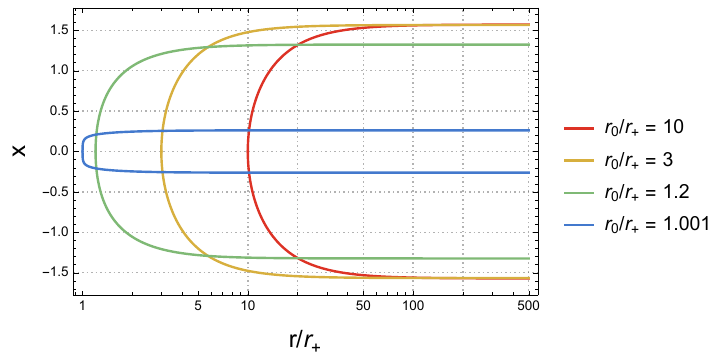}
    \end{minipage}
%
 %   \begin{minipage}{.5\textwidth}
   % \centering
%    5\includegraphics[width=.9\linewidth]{Wilson_EQQvsLQQ.pdf}
    %\end{minipage}
%
%
    %\begin{minipage}{.4\textwidth}
    %\centering
    %\includegraphics[width=.9\linewidth]{Wilson_String_Profile.pdf}
    %\end{minipage}
%
\caption{Left: The plot of $S_{EE}(L_{EE})$ showing the 'downwards' concavity (instability). Right: The profile of the effective strings used in computing the EE. Note the analogous behaviour with the 't Hooft loops strings.
% the profiles of different strings as they enter the bulk. Strings with small separation of the monopole pair penetrate deeper into the bulk. There is a maximum separation for the pair of monopoles, associated with the Little String Theory scale. The energy of the configuration grows as the separation of the monopole pair grows. This points to a phase transtion to a disconnected configuration.  This behaviour is usual of the backgrounds dual to a screening behaviour for the magnetically charged objects.
}\label{figura4}
\end{figure}

Before discussing the presence (or absence) of phase transitions we write the analytic expressions for the values of the separation between the two entangled regions $L_{EE}(r_0)$ and the Entanglement Entropy $S_{EE}(r_0)$. These expressions are explicitly derived in Appendix \ref{appendix3}.  Using the definitions in eqs.(\ref{definitionsEK}),(\ref{definitionelliptic2})  and recalling that $\lambda_-^2=-\frac{r_-^2}{r_+^2}$, we find
\begin{eqnarray}
L_{EE}\left( r_{0}\right)  &=&2\sqrt{\frac{8}{e_{A}^{2}+e_{B}^{2}}}\sqrt{%
\frac{r_{0}^{2}/r_{+}^{2}-1}{r_{0}^{2}/r_{+}^{2}+\lambda _{-}^{2}}}\left[ i%
\mathbf{K}\left( \frac{\left( r_{0}^{2}/r_{+}^{2}-1\right) ^{2}}{\left(
r_{0}^{2}/r_{+}^{2}+\lambda _{-}^{2}\right) ^{2}}\right) \right.  \nonumber\\
&&\left. +\mathbf{F}\left( \left. \arcsin \sqrt{\frac{%
r_{0}^{2}/r_{+}^{2}-r_{-}^{2}/r_{+}^{2}}{1-r_{-}^{2}/r_{+}^{2}}}\right\vert 
\frac{\left( 1-r_{-}^{2}/r_{+}^{2}\right) \left(
-r_{-}^{2}/r_{+}^{2}+2r_{0}^{2}/r_{+}^{2}-1\right) }{\left(
r_{0}^{2}/r_{+}^{2}-r_{-}^{2}/r_{+}^{2}\right) ^{2}}\right) \right] \ , 
\label{LEEexact}
\end{eqnarray}%
and 
\begin{equation}
S_{EE}^{BPS}\left( r_{0}\right) =\frac{\mathcal{N}}{G_{N}}r_{+}^{2}\left[ 
\frac{r_{+}^{2}}{2r_{0}^{2}}\left( -r_{0}^{4}/r_{+}^{4}\mathbf{E}\left(
r_{+}^{4}/r_{0}^{4}\right) -\mathbf{K}\left( r_{+}^{4}/r_{0}^{4}\right)
+r_{0}^{4}/r_{+}^{4}\mathbf{K}\left( r_{+}^{4}/r_{0}^{4}\right) \right) +%
\frac{1}{2}\right] \ .\label{SEEexact}
\end{equation}
To analyse these expressions, it is useful to show some plots. First, we check that the approximate $\hat{L}_{EE}(r_0)$ in eq.(\ref{LEEE}) approximates well the analytic expression in eq.(\ref{LEEexact}), see  the left panel of Figure \ref{figura3}.
We also plot $S_{EE}(r_0)$, see  the right panel of Figure \ref{figura3}. The plot of $S_{EE}$ in terms of $L_{EE}$  in the left panel of  Figure \ref{figura4}, shows an upwards concavity indicating  that the configuration is unstable. This follows the prediction of  \cite{Kol:2014nqa} that indicates that new configurations should appear as we introduce a UV-cutoff in the geometry. The profiles of the effective strings shown in  the right  panel of Figure \ref{figura4}, display a behaviour similar to the one we encountered in the study of 't Hooft loops (and opposite to that of the Wilson loop), again suggesting the need for a phase transition. 

\begin{figure}
\centering
    \begin{minipage}{.5\textwidth}
    \centering
    \includegraphics[width=1.0\linewidth]{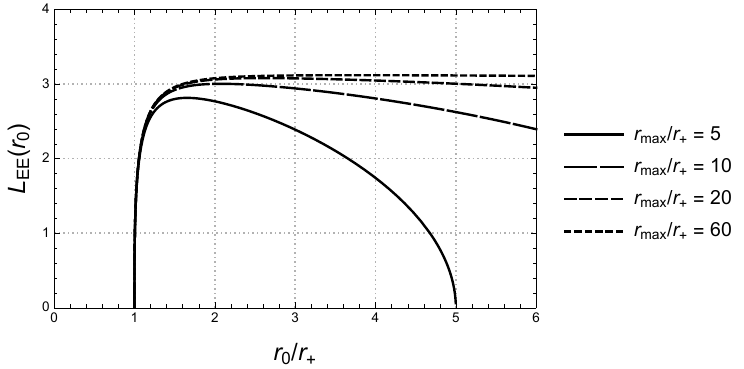}
    \end{minipage}
   \begin{minipage}{.45\textwidth}
    \centering
    \includegraphics[width=.9\linewidth]{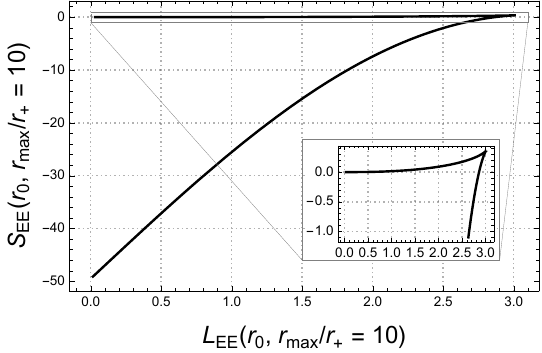}
    \end{minipage}
%
 %   \begin{minipage}{.5\textwidth}
   % \centering
%    5\includegraphics[width=.9\linewidth]{Wilson_EQQvsLQQ.pdf}
    %\end{minipage}
%
%
    %\begin{minipage}{.4\textwidth}
    %\centering
    %\includegraphics[width=.9\linewidth]{Wilson_String_Profile.pdf}
    %\end{minipage}
%
\caption{Left: Plot of $L_{EE}$ in eq.(\ref{LEEexact}) imposing the presence of a UV-cutoff. Right: The plot of $S_{EE}(L_{EE})$. Notice that the concavity has changed, indicating the stability of the configuration. The phase transition becomes apparent thanks to the cutoff, avoiding the non-local behaviour
% the profiles of different strings as they enter the bulk. Strings with small separation of the monopole pair penetrate deeper into the bulk. There is a maximum separation for the pair of monopoles, associated with the Little String Theory scale. The energy of the configuration grows as the separation of the monopole pair grows. This points to a phase transtion to a disconnected configuration.  This behaviour is usual of the backgrounds dual to a screening behaviour for the magnetically charged objects.
}\label{figura5}
\end{figure}

In analogy with the case of the 't Hooft loop, if we introduce a UV-cutoff, the separation between the two entangled regions becomes multiple-valued, as shown in Figure \ref{figura5} (right panel). This is at the root of the phase transition. The plot of $S_{EE} (L_{EE})$  shows the correct concavity and the presence of a transition to the disconnected configuration is clearly displayed. See Figure \ref{figura5}

%\textcolor{red}{a good discussion of EE, phase transitions and confinement, why it does not happen here, etc. This should be followed by a numerical analysis of the integrals in eq.(\ref{EE211}), with and without cutoff . make plots, introduce cutoff, re do plots, see that the same happens as in \cite{Kol:2014nqa}, study of the concavity of $S(L)$ without the cutoff, make the case that this is concave-up, the wrong concavity for stability--see sfetsos-brandhuber, bachas.}
%\subsection{One form symmetry}
%If this works, we can have this.
Following the findings of \cite{Kol:2014nqa}, we state that if a field theoretical UV completion to our system (that is completed by a LST) were found, the phase transitions for the 't Hooft loops and the EE would become apparent. In this sense the UV-cutoff captures the correct dynamics.

\section{Sigma Model}\label{sigmamodelsection}

In this section we review some results \cite{Itzhaki:2005tu,Giveon:2019twx}, concerning the string $\sigma$-model on the  background in eq.(\ref{background1})\footnote{We gratefully acknowledge conversations with Lewis Cole and Gast\'on Giribet  on the topics discussed here.}. To properly study the string action we  move to the S-dual frame, and work in terms of NS5 branes. After suitable coordinate changes detailed in Appendix \ref{appendix2}, the background reads,
    \begin{equation}\label{LinearDilatonBackground}
    \begin{aligned}
        ds^{2}_{10D} &= -dt^2+ dx^2+d\varphi^2+ d\rho^{2} 
        + N_{B} ds^{2}(S^{3}_{A})
        + N_{A} ds^{2}(S^{3}_{B}),\\
        H_{3} &= 2N_{B} \Vol(S^{3}_{A}) + 2N_{A}\Vol(S^{3}_{B}),\\
        \Phi &= -\sqrt{\frac{1}{N_{A}}+\frac{1}{N_{B}}} \, \rho.
    \end{aligned}
    \end{equation}
%
%where performed the change of coordinates $r = e^{\sqrt{\frac{1}{N_{A}}+\frac{1}{N_{B}}}\rho}$.
For a careful derivation we refer the reader to Appendices \ref{appendix2} and \ref{appendix28}. The background is a product space of the form
    \begin{equation}
        \mathbb{R}^{2,1} \times \mathbb{R}_{\rho} \times S^{3}_{N_{A}} \times S^{3}_{N_{B}},
    \end{equation}
where $\mathbb{R}_{\rho}$ denotes the direction (with the linear dilaton), and the subscript on the $S^{3}$ denotes the square of their radius. The metric on the spheres together with the $H_{3}$ flux on each of them allows us to write the $\sigma$-model on them as a WZW model on $SU(2)$. This is due to the fact the $S^{3}$ is a group manifold. Naively, the contribution of the $\rho$ coordinate to the string action is (for clarity we reinstate the $\alpha'$-factor)
    \begin{equation}
        S_{\rho} = \frac{1}{4\pi \alpha'}\int d^{2}\sigma \sqrt{-h}
        \left( h^{ab}\partial_{a}\rho\partial_{b}\rho - \alpha' \sqrt{\frac{1}{N_{A}} + \frac{1}{N_{B}} }R^{(2)}\rho \right),
    \end{equation}
where $R^{(2)}$ is the world-sheet Ricci scalar. However when $\rho\rightarrow -\infty$ the theory becomes non-perturbative, since $g_{s}\sim e^{-\rho}$. In order to avoid the strong coupling region, it is necessary to add the tachyon operator $e^{2b\rho}$ to the action, so that
    \begin{equation}
        S_{\rho} = \frac{1}{4\pi \alpha'}\int d^{2}\sigma \sqrt{-h}
        \left( h^{ab}\partial_{a}\rho\partial_{b}\rho - \alpha' Q R^{(2)}\rho + \Lambda e^{2b\rho}\right).
    \end{equation}
Here, $b$ is related to the background charge $Q = \sqrt{\frac{1}{N_{A}} + \frac{1}{N_{B}} }$ as 
\begin{equation}
Q = b+\frac{1}{b},\nonumber
\end{equation} 
such that it does not have a strong coupling region. Thus the contribution of the $\rho$ direction to the $\sigma$-model corresponds to a Liouville field. The complete $\sigma$-model on this geometry is
    \begin{equation}
        U(1)^{3} \times \text{Liouville} \times SU(2)_{N_{A}} \text{WZW} \times SU(2)_{N_{B}} \text{WZW}
    \end{equation}

where the subscripts denote the WZW level. One can check that this is indeed a good $\sigma$-model by computing the central charge. Here we use
    \begin{equation}
        c(U(1))=1,\quad c(\text{Liouville}) = 1+ 6Q^{2},\quad c(SU(2)_{k}) = \frac{3(k-2)}{k}.
    \end{equation}
In total we have
    \begin{equation}
        c_{\text{total}}=3 + 1 + 6 \left( \frac{1}{N_{A}} + \frac{1}{N_{B}} \right) 
        + \frac{3(N_{A}-2)}{N_{A}}+ \frac{3(N_{B}-2)}{N_{B}} = 10.
    \end{equation}
To this we should add the ten free fermions, that contribute to the central charge $c_{ferm}=5$. The central charge of the SUSY system $c_{SUSY}=15$, is then cancelled by the $b-c$ and $\beta-\gamma$ ghosts.
\\
\\
We now study the interesting case  $Q_A=Q_B=0$, but with $m>0$. The configuration of interest is obtained by S-dualising 
eq.\eqref{background 211}, 
    \begin{equation}\label{NSBH}
    \begin{aligned}
        ds^{2}_{10D} &= -dt^2+ dx^2+ f_s(r)d\varphi^2+ \frac{4 dr^{2}}{r^2 f_s(r)} 
        + N_{B}\, ds^{2}(S^{3}_{A}) + N_{A}\, ds^{2}(S^{3}_{B}), \\
       H_{3} &= 2N_{B} \Vol(S^{3}_{A}) + 2N_{A}\Vol(S^{3}_{B}),\\
        \Phi &= - \log(r).
    \end{aligned}
    \end{equation}

Recall that here $\varphi \sim \varphi + \pi/Q^{2}$, see eq.\eqref{EuclideanPeriod}, with $Q^2=\frac{1}{N_A}+\frac{1}{N_B}$, as above. Before proceeding to the $\sigma$-model analysis, it is convenient to perform some changes of variables. First, we want to rewrite the background \eqref{NSBH} in such a way that it reduces to \eqref{LinearDilatonBackground} when $m=0$. For this use   
    \begin{equation}
        r  = e^{Q \rho}, \quad \tilde{m} = \frac{m}{4Q^{2}},
    \end{equation}
which leads to
    \begin{equation}
    \begin{aligned}
        ds^{2}_{10D} &= -dt^2+ dx^2+ 4Q^{2}\left(1-\tilde{m} e^{-2Q\rho}\right)d\varphi^2 
        + \frac{d\rho^{2}}{1-\tilde{m} e^{-2Q\rho}} 
        + N_{B}\, ds^{2}(S^{3}_{A}) + N_{A}\, ds^{2}(S^{3}_{B}), \\
       H_{3} &= N_{B} \Vol(S^{3}_{A}) + N_{A}\Vol(S^{3}_{B}),\\
        \Phi &= - Q \rho .
    \end{aligned}
    \end{equation}
We change coordinates as,
    \begin{equation}
        \tanh^{2}(\lambda) = 1- \tilde{m} e^{-2Q\rho}, \quad \varphi = \frac{\phi}{2Q^{2}},
    \end{equation}
which puts the geometry in the usual cigar form (note that $\phi$ has period $2\pi$)
    \begin{equation}
    \begin{aligned}
        ds^{2}_{10D} &= -dt^2+ dx^2+ \frac{1}{Q^{2}}\left( \tanh^{2}(\lambda)d\phi^{2} + d\lambda^{2} \right) 
        + N_{B}\, ds^{2}(S^{3}_{A}) + N_{A}\, ds^{2}(S^{3}_{B}), \\
       H_{3} &= 2N_{B} \Vol(S^{3}_{A}) + 2N_{A}\Vol(S^{3}_{B}),\\
        \Phi &= - \log\left( \cosh(\lambda) \right) - \frac{1}{2}\log(\tilde{m}) .
    \end{aligned}
    \end{equation}

As explained in \cite{Witten:1991yr}, for more details see Section 2 of \cite{Giveon:2019twx}, the above backgrounds leads to an exact $\sigma$-model
\begin{equation}
U(1)^2\times \frac{SL(2,\mathbb{R})_{k}}{U(1)}\times SU(2)_{N_B}~\text{WZW}\times SU(2)_{N_A}~\text{WZW},
\end{equation}
where $k^{-1} = Q^{2}=\frac{1}{N_A}+\frac{1}{N_B}$. To check that this is also a good string $\sigma$-model we use
    \begin{equation}
        c\left( SL(2,\mathbb{R})_{k} \right) = \frac{3(k+2)}{k}, \quad c\left( G/H \right) = c(G)-c(H),
    \end{equation}
so that
    \begin{equation}
        c\left( \frac{SL(2,\mathbb{R})_{k}}{U(1)} \right) = 2 + \frac{6}{k}=2+\frac{6}{N_A}+\frac{6}{N_B} ,
    \end{equation}
from where is easy to see that $c_{\text{total}}=10$. We leave for future research the study of the $\sigma$-model in the background with $m, Q_A,Q_B$ arbitrary.

\section{Conclusions and future research}\label{concl}
The I-brane QFT, defined as the $(1+1)$ field theory on the intersection of two stacks of D5 branes, was studied in \cite{Itzhaki:2005tu}. The field theory has the remarkable behaviour that as the coupling is increased, the system gains one more dimension and enhances its SUSY (with a peculiar SUSY algebra in $(2+1)$ dimensions \cite{Lin:2005nh}). The background dual to this strongly coupled QFT was written in  \cite{Itzhaki:2005tu}, see our eq.(\ref{background1}). This presents a singular behaviour for large values of the radial coordinate, where the dilaton diverges and string coupling effects cannot be neglected. This is solved by performing an S-duality and working with the NS branes system. The background of eq.(\ref{background1}) is also singular for small values of the radial coordinate, $r\to 0$, as indicated by eq.(\ref{riccisingular}).  This ill-defined IR behaviour is amended by our background in eq.(\ref{background 211}). Our simple and explicit solution describes the holographic dual to  a $(2+1)$ QFT that gets compactified to $(1+1)$ dimensions, preserving four supercharges and ending the flow with a confining and gapped  behaviour.

We holographically studied different aspects of this peculiar QFT. Maldacena-Wilson loops, 't Hooft loops, Entanglement entropy were discussed in dedicated sections,  with emphasis on  the effects of the UV-completion in terms of LST. R-symmetry and its breaking, a suitably defined gauge coupling and a quantity measuring the number of degrees of freedom as a function of the radial coordinate (the energy) are presented and discussed. 
By a double Wick rotation, a black membrane solution is found. Also, some of the NS string $\sigma$-model aspects are briefly mentioned.

It would be interesting to dedicate future efforts to
\begin{itemize}
\item{The careful study of the black membrane solution. In particular if its entropy can be computed in terms of a $(2+1)$ field theory compactified on a torus.}
\item{The study of the string $\sigma$-model for the full solution in eq.(\ref{background 211}), in the NS5 branes frame}
\item{To achieve a cleaner understanding of the R-symmetry breaking in terms of anomalies in two dimensional QFT. To relate this to the Chern Simons coefficients discussed in eq.(\ref{CS2+1}). Note that while we find an anomalous breaking $U(1)\times U(1) \rightarrow \mathbb{Z}_{N_{A}}\times \mathbb{Z}_{N_{B}}$, we do not find a further spontaneous breaking to $\mathbb{Z}_{2}\times \mathbb{Z}_{2}$, as it normally occurs in holographic models to four dimensional ${\cal N}=1$ dynamics.}
\item{Geometrically, it would be interesting to generalise the metric and fluxes in eq.(\ref{background 211}) adding warp factors in front of the $\omega_i$'s and more general fibrations. Finding a more general classification is of interest. It may be possible to relate this to the material in \cite{Lin:2005nh}.}
\item{A fair amount of papers have been written studying the background in eq.(\ref{background1}). See for example \cite{Hung:2006nn},\cite{Kluson:2005eb}-\cite{Biswas:2023lkj}. It would be interesting to understand the effects of the resolution provided in eq.(\ref{background 211}) on some of these observables.}
\end{itemize}
We hope to report on some of these problems in the near future.

\section*{Acknowledgements: }
The contents and presentation of this work much benefitted from extensive discussion with various colleagues. We are very happy to thank: Andres Anabalon, Adi Armoni, Fabrizio Canfora, Lewis Cole, Gaston Giribet, Nicolas Grandi, Nabil Iqbal, Prem Kumar, Juan Maldacena, Anibal Neira, Leo Pando Zayas, Julio Oliva, Niels Obers,  Dibakar Roychowdhury, Kostas Skenderis, Christoph Uhlemann who shared their knowledge with us.
We are supported by  STFC  grant  ST/T000813/1. The work of M.O. is partially funded by Beca ANID de
Doctorado 21222264. The work of R.S. is supported by STFC grant ST/W507878/1. The authors have applied to a Creative Commons Attribution (CC BY) licence.

\appendix

\section{Details of the Supergravity Backgrounds}\label{appendix1}
In this appendix we set some of the conventions used in this paper and  study the SUSY preserved by the background in eq.(\ref{background 211}).

\subsection{Type IIB Supergravity}

We start this appendix by explicitly writing the Type IIB Supergravity action
and its SUSY variations. The field content of Type IIB is split into two sector. In the NS-NS sector we have: the metric $g_{\mu\nu}$, 2-form potential $B_{2}$ with field strength $H_{3}$, and the Dilaton $\Phi$. In the R-R sector we have a set  of Abelian p-form gauge fields: $C_{0}$, $C_{2}$, $C_{4}$. By defining
    \begin{equation}
        F_{1} = dC_{0}, \quad F_{3} = dC_{2} - C_{0}\wedge H_{3},\quad 
        F_{5} = dC_{4} - C_{2}\wedge H_{3}.
    \end{equation}

The bosonic part of the Type IIB action in String frame is
    \begin{equation}
    \begin{aligned}
        S_{\text{IIB}} &= \frac{1}{2\kappa^{2}}\int d^{10}x\sqrt{-g} \left[e^{-2\Phi}\left( R + 4\partial_{\mu}\Phi\partial^{\mu}\Phi - \frac{1}{2} |H_{3}|^{2}\right) - \frac{1}{2}|F_{1}|^{2} - \frac{1}{2}|F_{3}|^{2} 
        - \frac{1}{4}|F_{5}|^{2} \right] \\
       &\phantom{=} -  \frac{1}{4\kappa^{2}}\int\, C_{4}\wedge H_{3}\wedge F_{3}
    \end{aligned}
    \end{equation}

where $|F_{p}|^{2} = F_{\mu_{1}...\mu_{p}}F^{\mu_{1}...\mu_{p}}/p!$ and analogously for $H_{3}$. On the solutions of this theory we need to impose \footnote{To be precise, there is no covariant action for the effective theory of the Type IIB Superstring, but the presented here is close enough. The issue is that is not possible to implement the self-duality condition for $F_{5}$ at the level of the action.} self-duality of the $F_{5}=\star F_{5}$. The equations of motion of this theory are
    \begin{equation}
    \begin{aligned}
        &\nabla^{2}\Phi - \nabla_{\mu}\Phi\nabla^{\mu}\Phi + \frac{1}{4}R - \frac{1}{8}|H_{3}|^{2}=0,\\
        &d\left( e^{-2\Phi}\star H_{3} \right) = -  F_{5}\wedge F_{3} - F_{1}\wedge F_{7},\\
        &d F_{5}- H_{3}\wedge F_{3} = 0,\\
        &d F_{7} - H_{3}\wedge F_{5} = 0,\\
        &d F_{9} - H_{3}\wedge F_{7} = 0,\\
        & R_{\mu\nu} + 2\nabla_{\mu}\nabla_{\nu}\Phi 
        - \frac{1}{2}|H_{3}|^{2}_{\, \mu\nu} 
        - \frac{e^{2\Phi}}{2}\left( |F_{1}|^{2}_{\, \mu\nu} 
        + |F_{3}|^{2}_{\, \mu\nu} 
        +\frac{1}{2}|F_{5}|^{2}_{\, \mu\nu} -\frac{1}{2}g_{\mu\nu}
        \left( |F_{1}|^{2}+ |F_{3}|^{2}  \right)\right) = 0
    \end{aligned}
    \end{equation}

where $|F_{p}|^{2}_{\mu\nu} = F_{\mu \nu_{1}...\nu_{p-1}}F_{\nu}^{\phantom{\mu}\nu_{1}...\nu_{p-1}}/(p-1)!$, similarly for $H_{3}$, and
    \begin{equation}
        F_{7} = - \star F_{3}, \quad F_{9} = \star F_{1}.
    \end{equation}

The equations of motion are complemented by the Bianchi Identities
    \begin{equation}
        dF_{1} = 0, \quad dF_{3} - H_{3}\wedge F_{1} = 0.
    \end{equation}

Due to the self-duality of $F_{5}$, its  equation of motion and its the Bianchi identity are the same. 

Solutions of the purely bosonic part of Type IIB has all the fermionic partners, the dilatino $\lambda$ and gravitino $\Psi_{\mu}$, set to zero. If we are interested in finding SUSY solutions, we need to be consistent with the fact that we turn off the fermions by asking for the SUSY variations of these fields to vanish. In string frame, the SUSY variations of the fermionic fields  are \cite{Martucci:2005rb},
    \begin{align}
        &\delta\lambda = \frac{1}{2}\left( \Gamma^{\mu}\partial_{\mu}\Phi
        + \frac{1}{2\cdot 3!}H_{\mu\nu\lambda}\Gamma^{\mu\nu\lambda}\sigma^{3} -e^{\Phi}\left(F_{\mu}\Gamma^{\mu}(i\sigma_{2}) + \frac{1}{2\cdot 3!}F_{\mu\nu\lambda}\Gamma^{\mu\nu\lambda}\sigma^{1} \right)\right)\epsilon,\\ 
        &\begin{aligned}
            \delta \Psi_{\mu} &= \partial_{\mu}\epsilon 
        + \frac{1}{4}\omega^{\phantom{\mu}ab}_{\mu}\Gamma_{ab}\epsilon + \frac{1}{4\cdot 2!}H_{\mu\nu\lambda}\Gamma^{\nu\lambda}\sigma^{3}\epsilon \\
        &\phantom{=} +\frac{e^{\Phi}}{8}\left( 
        F_{\nu}\Gamma^{\nu}(i\sigma_{2}) 
        + \frac{1}{3!}F_{\nu\lambda\rho}\Gamma^{\nu\lambda\rho}\sigma^{1} 
        + \frac{1}{2\cdot5!}F_{\nu\lambda\rho\sigma\tau}\Gamma^{\nu\lambda\rho\sigma\tau}(i\sigma_{2})\right)\Gamma_{\mu}\epsilon.
        \end{aligned}
    \end{align}

Here $\omega^{\phantom{\mu}ab}_{\mu}$ is the spin connection of the 10D background, where the $a,b$ indexes are flat space ones, and $\sigma^{1}$, $\sigma^{2}$ and $\sigma^{3}$ are Pauli matrices. Also 
    \begin{equation}
        \Gamma^{\mu_{1}...\mu_{p}} = \Gamma^{[\mu_{1}}...\Gamma^{\mu_{p}]}
    \end{equation}

Here $\epsilon$ is a 64 component spinor, 
    \begin{equation}
        \epsilon = \begin{pmatrix}
           \epsilon_{1} \\ \epsilon_{2} 
        \end{pmatrix}
    \end{equation}

where both 32-component parts are left-handed. 

\subsection{Checking SUSY for the Fibered Background}

Here we aim to compute how many supercharges are preserved by the backgrounds presented in this paper. For the un-fibered background in eq.(\ref{background1}) we refer the reader to \cite{Itzhaki:2005tu}, \cite{Lin:2005nh}, where it is shown that this solution preserves 16 Supercharges in an interesting way: the anti-commutator of two supercharges includes the $R$-Symmetry generators. Now we present the analysis for the fibered background in eq.(\ref{background 211}). We perform all the analysis in the S-dual system, in terms of NS5 branes, where we only have $H_{3}$ flux.

First, note that the dilatino variation is a matrix equation of the form $M\epsilon=0$. In order to have non-trivial solutions to this equation, we require $M$ to be non-invertible, for which we need to impose $\text{det}(M)=0$. It is also possible to obtain a matrix equation from the gravitino variation. Noting that we can write the gravitino variation as a covariant derivative, for which we define the connection
    \begin{equation}
        W_{\mu} = \frac{1}{4}\omega^{\phantom{\mu}ab}_{\mu}\Gamma_{ab} + \frac{1}{4\cdot 2!}H_{\mu\nu\lambda}\Gamma^{\nu\lambda}\sigma^{3} +\frac{e^{\Phi}}{8}\left( F_{\mu}\Gamma^{\mu}(i\sigma_{2}) 
        + \frac{1}{3!}F_{\mu\nu\lambda}\Gamma^{\mu\nu\lambda}\sigma^{1} 
        + \frac{1}{2\cdot5!}F_{\mu\nu\lambda\rho\sigma}\Gamma^{\mu\nu\lambda\rho\sigma}(i\sigma_{2})\right)\Gamma_{\mu},
    \end{equation}
then we can write the gravitino variation as
    
    \begin{equation}
        \delta\psi_{\mu}dx^{\mu} = \left( \partial_{\mu}\epsilon 
        + W_{\mu}\epsilon\right)dx^{\mu} \equiv \mathcal{D}\epsilon.
    \end{equation}

We can get rid of the partial derivative of the spinor by acting with $\mathcal{D}$ a second time
    \begin{equation}
        \mathcal{D}\wedge \mathcal{D} \epsilon = \left( dW + W\wedge W  \right)\epsilon = \frac{1}{2}\Theta_{\mu\nu}dx^{\mu}\wedge dx^{\nu} \epsilon.
    \end{equation}

Each of the components of $\Theta_{\mu\nu}$ defines a matrix equation, giving a total of 45 independent equations. We need to make sure that $\text{det}(\Theta_{\mu\nu})=0$ for each of the components. The equations
    \begin{equation}
        M \epsilon = 0 , \quad \Theta_{\mu\nu}\epsilon =0,
    \end{equation}
constrain the number of independent components of the spinor. After this procedure we use the gravitino variation to solve the dependence of the spinor on the spacetime coordinates.

Specialising to our background, the determinant of the Dilatino variation for the background in eq.(\ref{background 211}) reads
    \begin{equation}
        \text{det}(M) \sim \left( 4(e_{B}Q_{A}-e_{A}Q_{B})^{2}+m^{2} \right)^{8} \left(4(e_{B}Q_{A}+e_{A}Q_{B})^{2}+m^{2}\right)^{8}.
    \end{equation}
In order to have non-trivial solutions we need to impose the following BPS conditions on the parameters of the background
    \begin{equation}
        e_{A}Q_{B} = \pm e_{B}Q_{A}, \quad m = 0.
    \end{equation}
With this conditions it is possible to check that $\text{det}(\Theta_{\mu\nu})=0$ is also satisfied. Solving these matrix equations shows that the spinor has 8 independent components. Then,  solving for the gravitino variation shows that these components are not independents, and in fact, the total number of independent components its reduced to 4. The solution for the spinor is
    \begin{equation}
        \epsilon_{1} = \vec{0} 
    \end{equation}
and
    \begin{equation}
        \epsilon_{2} = \frac{1}{r}\begin{pmatrix}
            c_{1} e^{-\frac{1}{4} i \text{$\varphi $} \left(e_{A}^2+e_{B}^2\right)} \sqrt{e_{A} r^2+2 Q_{A}} \\
            0\\
            0\\
            c_{1} e^{-\frac{1}{4} i \varphi \left(e_{A}^2+e_{B}^2\right)} \sqrt{e_{A} r^2+2 Q_{A}}\\
            0\\
            i c_{2} e^{\frac{1}{4} i \varphi \left(e_{A}^2+e_{B}^2\right)} \sqrt{e_{A} r^2-2 Q_{A}}\\
            i c_{2} e^{\frac{1}{4} i \varphi \left(e_{A}^2+e_{B}^2\right)} \sqrt{e_{A} r^2-2 Q_{A}} \\
            0\\
            0\\
            \frac{c_{1} e^{-\frac{1}{4} i \varphi \left(e_{A}^2+e_{B}^2\right)} \sqrt{\left(e_{A}^2+e_{B}^2\right) \left(e_{A} r^2-2 Q_{A}\right)}}{ (e_{A}-i e_{B})}\\
           -\frac{c_{1} e^{-\frac{1}{4} i \varphi \left(e_{A}^2+e_{B}^2\right)} \sqrt{\left(e_{A}^2+e_{B}^2\right) \left(e_{A} r^2-2 Q_{A}\right)}}{ (e_{A}-i e_{B})}\\
            0\\
            i c_{2} (e_{A}+i e_{B}) e^{\frac{1}{4} i \varphi \left(e_{A}^2+e_{B}^2\right)} \sqrt{\frac{e_{A} r^2+2 Q_{A}}{e_{A}^2+e_{B}^2}}\\
            0\\
            0\\
            c_{2} (e_{B}-i e_{A}) e^{\frac{1}{4} i \varphi \left(e_{A}^2+e_{B}^2\right)} \sqrt{\frac{e_{A} r^2+2 Q_{A}}{e_{A}^2+e_{B}^2}}\\
            0\\
            c_{3} e^{-\frac{1}{4} i \varphi \left(e_{A}^2+e_{B}^2\right)} \sqrt{e_{A} r^2+2 Q_{A}}\\
            -c_{3} e^{-\frac{1}{4} i \varphi \left(e_{A}^2+e_{B}^2\right)} \sqrt{e_{A} r^2+2 Q_{A}}\\
            0\\
            i c_{4} e^{\frac{1}{4} i \varphi \left(e_{A}^2+e_{B}^2\right)} \sqrt{e_{A} r^2-2 Q_{A}}\\
            0\\
            0\\
            -i c_{4} e^{\frac{1}{4} i \varphi \left(e_{A}^2+e_{B}^2\right)} \sqrt{e_{A} r^2-2 Q_{A}}\\
            -\frac{c_{3} e^{-\frac{1}{4} i \varphi \left(e_{A}^2+e_{B}^2\right)} \sqrt{\left(e_{A}^2+e_{B}^2\right) \left(e_{A} r^2-2 Q_{A}\right)}}{ (e_{A}+i e_{B})}\\
            0\\
            0\\
            -\frac{c_{3} e^{-\frac{1}{4} i \varphi \left(e_{A}^2+e_{B}^2\right)} \sqrt{\left(e_{A}^2+e_{B}^2\right) \left(e_{A} r^2-2 Q_{A}\right)}}{ (e_{A}+i e_{B})}\\
            0\\
            -i c_{4} (e_{A}-i e_{B}) e^{\frac{1}{4} i \varphi \left(e_{A}^2+e_{B}^2\right)} \sqrt{\frac{e_{A} r^2+2 Q_{A}}{e_{A}^2+e_{B}^2}}\\
            -i c_{4} (e_{A}-i e_{B}) e^{\frac{1}{4} i \varphi \left(e_{A}^2+e_{B}^2\right)} \sqrt{\frac{e_{A} r^2+2 Q_{A}}{e_{A}^2+e_{B}^2}}\\
            0
        \end{pmatrix}
    \end{equation}

We have found a spinor with four arbitrary constants $(c_1,c_2,c_3,c_4)$.  Being the spinor complex, we count four preserved supercharges.
\section{How are the backgrounds obtained}\label{appendix2}
In this appendix we describe the procedure followed to obtain the backgrounds in eqs.(\ref{background1}) and (\ref{background 211}). These solutions are originally obtained in gauged supergravity, together with a lift procedure. Below, we review these steps.
\subsection{4D \texorpdfstring{$\mathcal{N}=4\, SU(2)\times SU(2)$}{N=4 SU(2)xSU(2)} Gauged Supergravity}
The action of the bosonic part of the Freedman-Schwarz (FS) gauged Supergravity is
    \begin{equation}
    \begin{aligned}
        S_{\text{FS}} &= \int d^{4}x\sqrt{-g_{(4)}}\left(-\frac{R^{(4)}}{4} + \frac{1}{2}(\partial\phi)^{2} 
        + \frac{1}{2}e^{4\phi}(\partial{\textbf{a}})^{2} - V(\phi) 
    \right. \\
    &\phantom{=}\left.
    -\frac{e^{-2\phi}}{4}\text{Tr}\left( F_{(A)mn}F_{(A)}^{mn} 
    + F_{(B)mn}F_{(B)}^{mn} \right)
    - \frac{\textbf{a}}{2}\,\text{Tr}\left( \tilde{F}_{(A)mn}F_{(A)}^{mn} 
    + \tilde{F}_{(B)mn}F_{(B)}^{mn} \right) \right).
    \end{aligned}
    \end{equation}

Here $g_{(4)}$ and $R^{(4)}$ are the determinant of the 4D metric and the 4D Ricci scalar, $\phi$ is the 4D dilaton, \textbf{a} is a pseudo-scalar called axion and  $F_{(A)mn}$ and $F_{(B)mn}$ are the field strengths of two $SU(2)$ gauge fields $A_{m}$ and $B_{m}$,  
    \begin{align}
        F^{i}_{(A)mn} &= \partial_{m}A^{i}_{n} - \partial_{n}A^{i}_{m} + e_{A}\epsilon_{ijk}A^{j}_{m}A^{k}_{n}\\
        F^{i}_{(B)mn} &= \partial_{m}B^{i}_{n} - \partial_{n}B^{i}_{m} + e_{B}\epsilon_{ijk}B^{j}_{m}B^{k}_{n}
    \end{align}

where $e_{A}$ and $e_{B}$ are the gauge couplings of $A_{m}$ and $B_{m}$ respectively. Here the index $i=1,2,3$ transform in the adjoint of each of the copies of $SU(2)$. Also, the duals of the field strengths are
    \begin{equation}
        \tilde{F}_{(A)mn} = \frac{1}{2\sqrt{-g}}\epsilon_{mn\lambda\rho} F_{(A)mn}, \quad
         \tilde{F}_{(B)mn} = \frac{1}{2\sqrt{-g}}\epsilon_{mn\lambda\rho} F_{(B)mn},
    \end{equation}

\subsubsection{A BPS Solution}

This theory admits a series of BPS solution preserving some amount of supersymmetry. Our main focus is the 1/4 BPS soliton presented in \cite{Canfora:2021nca}. As we will review, this solution is particularly interesting because it manages to resolve a singularity by introducing a thermal cycle that preserves some SUSY (in the usual case, the non-extremal factor completely breaks SUSY). The field configuration of the Soliton is\footnote{In \cite{Canfora:2021nca} the solution was presented in the mostly-minus signature. We write the solution in the mostly-plus one.}
    \begin{align}
        ds^{2}_{4D} &= -\rho dt^{2} +  \frac{d\rho^{2}}{g(\rho)} + g(\rho)d\varphi^{2} +\rho dx^{2},\\
        \phi(\rho) &= -\frac{1}{2}\log(\rho),\\
        A^{1}&=0,\quad A^{2}=0, \quad A^{3}= Q_{A}\zeta(\rho)d\varphi,\\
        B^{1}&=0, \quad B^{2}=0, \quad B^{3}= Q_{B}\zeta(\rho)d\varphi,
    \end{align}

where
    \begin{equation}
        g(\rho) = \frac{e^{2}_{A}+e^{2}_{B}}{2}\rho - m -2\frac{Q^{2}_{A}+Q^{2}_{B}}{\rho}
    \end{equation}

which has a zeros
    \begin{equation}
        \rho_{\pm} = \frac{m\pm\sqrt{4(e^{2}_{A}+e^{2}_{B})(Q^{2}_{A}+Q^{2}_{B})+m^{2}}}{(e^{2}_{A}+e^{2}_{B})}.
    \end{equation}

Also 
    \begin{equation}
        \zeta(\rho) = \frac{1}{\rho} - \frac{1}{\rho_{+}},
    \end{equation}

where the last term ensures that both of the gauge fields vanish at $\rho=\rho_{+}$.

In order for the cycle $\varphi$ to close smoothly at $\rho=\rho_{+}$, the period of $\varphi$ needs to be
    \begin{equation}
        \beta_{\varphi} = \frac{4\pi}{g'(\rho_{+})}
        = \frac{8\pi\rho^{2}_{+}}{(e^{2}_{A}+e^{2}_{B})\rho^{2}_{+}+
        4(Q^{2}_{A}+Q^{2}_{B})}.
    \end{equation}
This solution was shown to preserve 4 supercharges \cite{Canfora:2021nca}, when the parameters $(e_{A},e_{B}, Q_{A}, Q_{B},m)$ satisfy
    \begin{equation}
        e_{A}Q_{B} = \pm e_{B}Q_{A}, \quad m=0.
    \end{equation}

In what follows it is convenient to perform the change of coordinates $\rho = r^{2}$. The background configuration now reads
    \begin{align}
        ds^{2}_{4D} &= -r^{2} dt^{2} +  \frac{4dr^{2}}{f(r)} + r^{2}f(r)d\varphi^{2} +r^{2} dx^{2},\\
        \phi(r) &= -\log(r),\\
        A^{1}&=0,\quad A^{2}=0, \quad A^{3}= Q_{A}\zeta(r)d\varphi,\\
        B^{1}&=0 \quad B^{2}=0, \quad B^{3}= Q_{B}\zeta(r)d\varphi,
    \end{align}    
with
    \begin{equation}
        f(r) = \frac{e^{2}_{A}+e^{2}_{B}}{2} - \frac{m}{r^{2}} -2\frac{Q^{2}_{A}+Q^{2}_{B}}{r^{4}}, \quad 
        \zeta(r) = \frac{1}{r^{2}}-\frac{1}{r^{2}_{+}}.
    \end{equation}

\subsection{Lift to 10D Supergravity}

It was shown in \cite{Chamseddine:1997mc} that the FS Supergravity has
a Kaluza-Klein interpretation as a compactification of $\mathcal{N}=1$ Supergravity in 10D on the group manifold $S^{3}\times S^{3}$. The action of the 10D theory is 
    \begin{equation}\label{10Dfail}
       S_{10D} = -\frac{1}{4}\int d^{10}x\sqrt{-g}\left( R
       -2\partial_{\mu}\tilde{\Phi}\partial^{\mu}\tilde{\Phi}
       -\frac{e^{2\tilde{\Phi}}}{3}\tilde{H}_{\mu\nu\lambda}\tilde{H}^{\mu\nu\lambda}\right),
    \end{equation}

where $R$ is the 10D Ricci scalar, $\tilde{\Phi}$ the 10D Dilaton and $\tilde{H}_{\mu\nu\lambda}$ is a 3-form field strength $\tilde{H}_{3}=d\tilde{B}_{2}$. We split the indexes as
    \begin{equation}
        x^{\mu} =\left\{ x^{m}=t,r,\varphi,x \, ; \, 
        z_{A}^{i} = \psi_{A},\theta_{A},\phi_{A}\, ; \,
        z_{B}^{i} = \psi_{B},\theta_{B},\phi_{B}
        \right\} \ .
    \end{equation}

The of the lift to 10D is
    \begin{equation}
        ds^{2}_{10D}=e^{3\phi /2}ds^{2}_{4D}+2e^{-\phi /2}\left( 
        \Theta_{A}^{i}\Theta_{A}^{i}
        +\Theta_{B}^{i}\Theta_{B}^{i}\right)
    \end{equation}

where the 1-forms 
    \begin{align}
        \Theta^{i}_{A} &= A^{i}-\frac{1}{e_{A}}\omega_{A}^{i}, \\
        \Theta^{i}_{B} &= B^{i}-\frac{1}{e_{B}}\omega_{B}^{i},
    \end{align}

with $\omega_{A}^{i}$ and $\omega_{B}^{i}$ are the Maurer-Cartan forms of the two different $SU(2)$,  is so that $\Theta^{i}_{A}$ and $\Theta^{i}_{B}$ realise a fibration of the two 3-spheres. 

The 10D Dilaton $\tilde{\Phi}$ is written in terms of the four dimensional one $\phi$
    \begin{equation}
        \tilde{\Phi}=-\frac{\phi}{2},
    \end{equation}

while the 3-form field strength is written in term of the non-Abelian gauge fields and the $SU(2)$ Maurer-Cartan forms as
    \begin{equation}
        \tilde{H}_{3}= -\sum^{3}_{i=1}F_{A}^{i}\wedge\Theta_{A}^{i}
        -\sum^{3}_{i=1}F_{B}^{i}\wedge \Theta_{B}^{i}
        + e_{A}\Theta_{A}^{1}\wedge\Theta_{A}^{2}\wedge\Theta_{A}^{3}
        +e _{B}\Theta_{B}^{1}\wedge\Theta_{B}^{2}\wedge\Theta_{B}^{3}.
    \end{equation}

\subsubsection{A Note on Conventions}

%Since we want to give a String Theory interpretation of the lift of the 4D background, we need the 10D action to be an effective theory of String Theory. In our case, 
We are interested in lifting the theory to Type II Supergravity, when the field content is purely of the NS-NS sector. The action  in eq.\eqref{10Dfail} can be mapped to Type II, after the field redefinitions 
    \begin{equation}
        \tilde{\Phi} \rightarrow \Phi =- 2\tilde{\Phi}, \quad 
        \tilde{H}_{3} \rightarrow H_{3}= 2 \tilde{H}_{3}.
    \end{equation}
The action corresponds to the Type II in Einstein frame
    \begin{equation}
        S_{\text{Type II, E}}=  -\frac{1}{4}\int d^{10}x\sqrt{-g}\left(R
        -\frac{1}{2} \partial_{\mu}\Phi\partial^{\mu}\Phi 
        - \frac{e^{-\hat{\phi}}}{12} \hat{H}_{\mu\nu\rho}\hat{H}^{\mu\nu\rho}\right).
    \end{equation}

We move to String frame by $g^{(S)}_{\mu\nu} = e^{\frac{1}{2}\Phi}g^{(E)}_{\mu\nu}$, then the action  reads
    \begin{equation}\label{action string}
        S_{\text{Type II, S}}=-\frac{1}{4}\int d^{10}x\sqrt{-g} e^{-2\Phi}\left( R 
        + 4\partial_{\mu}\Phi\partial^{\mu}\Phi
        -\frac{1}{2\cdot 3!}H_{\mu \nu \rho} H^{{\mu \nu \rho}} \right).
    \end{equation}

In this frame, the lift of the 4D FS Supergravity reads
    \begin{align}
        ds^{2}_{10D} &= e^{2\phi}ds^{2}_{4D}+2\left( 
        \Theta_{A}^{i}\Theta_{A}^{i}
        +\Theta_{B}^{i}\Theta_{B}^{i}\right),\\
        H_{3} &= 2\left(-\sum^{3}_{i=1}F_{A}^{i}\wedge\Theta_{A}^{i}
        -\sum^{3}_{i=1}F_{B}^{i}\wedge \Theta_{B}^{i}
        + e_{A}\Theta_{A}^{1}\wedge\Theta_{A}^{2}\wedge\Theta_{A}^{3}
        +e _{B}\Theta_{B}^{1}\wedge\Theta_{B}^{2}\wedge\Theta_{B}^{3}.\right)
        ,\\
        \Phi &= \phi(x^{m}).
    \end{align}

It is convenient to write the lift in the S-dual frame, where instead of $H_{3}$ flux, we have a $F_{3}$ flux, the Dilaton is $\Phi'=-\Phi$ and the metric now is $g_{\mu\nu}' = e^{-\Phi}g_{\mu\nu}$, explicitly this is
    \begin{align}
        ds^{2}_{10D} &= e^{\phi}ds^{2}_{4D}+2e^{-\phi}\left( 
        \Theta_{A}^{i}\Theta_{A}^{i}
        +\Theta_{B}^{i}\Theta_{B}^{i}\right) ,\label{b32}\\
        F_{3} &= 2\left(-\sum^{3}_{i=1}F_{A}^{i}\wedge\Theta_{A}^{i}
        -\sum^{3}_{i=1}F_{B}^{i}\wedge \Theta_{B}^{i}
        + e_{A}\Theta_{A}^{1}\wedge\Theta_{A}^{2}\wedge\Theta_{A}^{3}
        +e _{B}\Theta_{B}^{1}\wedge\Theta_{B}^{2}\wedge\Theta_{B}^{3}.\right)
        ,\\
        \Phi &= -\phi(x^{m}).
    \end{align}

\subsubsection{Lift of the BPS Solution}
Following the explicit construction of the lift, we now read the lift of the 4D solution that preserves 4 Supercharges. In the S-dual frame of eqs. (\ref{b32}), we have
\begin{eqnarray}
ds^{2}_{st} &=&r\left\{ -dt^{2}+dx^{2}+f_{s}\left( r\right) d\varphi ^{2}+\frac{4}{r^{2}f_{s}\left( r\right) }%
dr^{2}+\frac{2}{e_{A}^{2}}\left[ \hat{%
\omega}_{1}^{2}+\hat{\omega}_{2}^{2}+\left( \hat{\omega}_{3}-e_{A}Q_{A}\zeta
(r)d\varphi \right) ^{2}\right] \right.   \notag \\
&&\left. +\frac{2}{e_{B}^{2}}\left[ \tilde{\omega}_{1}^{2}+\tilde{\omega}%
_{2}^{2}+\left( \tilde{\omega}_{3}-e_{B}Q_{B}\zeta \left( r\right)
d\varphi \right) ^{2}\right] \right\} \ ,  \\
F_{3} &=& dC_2= 2 \zeta'(r)dr\wedge d\varphi \wedge \left( 
\frac{Q_{A}}{e_{A}}\hat{\omega}_{3}+\frac{Q_{B}}{e_{B}}\tilde{\omega}%
_{3}\right) +\frac{2}{e_{A}^{2}}\hat{\omega}_{1}\wedge \hat{\omega}%
_{2}\wedge \left( e_{A}Q_{A}\zeta (r)d\varphi -\hat{\omega}_{3}\right)  \notag\\
&&+\frac{2}{e_{B}^{2}}\tilde{\omega}_{1}\wedge \tilde{\omega}_{2}\wedge
\left( e_{B}Q_{B}\zeta (r)d\varphi -\tilde{\omega}_{3}\right) \ , \notag \\
C_2&=&
{\psi_A}\left( \frac{2Q_{A}}{e_{A}}\zeta ^{\prime }\left(
r\right) dr\wedge d\varphi -\frac{2}{e_{A}^{2}}\sin {\theta_A}d{\theta_A}%
\wedge d{\phi_A}\right) +\frac{2}{e_{A}}\cos {\theta_A}Q_{A}\zeta
\left( r\right) d\varphi \wedge d{\phi_A}\nonumber \\
&&+{\psi_B}\left( \frac{2Q_{B}}{e_{B}}\zeta ^{\prime }\left( r\right)
dr\wedge d\varphi -\frac{2}{e_{B}^{2}}\sin {\theta_B}d{\theta_B}%
\wedge d{\phi_B}\right) +\frac{2}{e_{B}}\cos {\theta_B}Q_{B}\zeta
\left( r\right) d\varphi \wedge d{\phi_B}\ .\nonumber\\
C_6&=& -\frac{16e_{A}r^{2}}{e^{3}_{B}}dt\wedge dx\wedge d\varphi \wedge \Vol(S^{3}_{B}) 
+\frac{16e_{B}r^{2}}{e^{3}_{A}}dt\wedge dx\wedge d\varphi \wedge \Vol(S^{3}_{A})   ,\nonumber\\
    &\phantom{=}&  -\frac{64Q_{B}}{e^{3}_{A}e^{2}_{B}}\cos(\theta_{B})dt\wedge dx \wedge \Vol(S^{3}_{A})\wedge d\phi_{B}
    +\frac{64Q_{A}}{e^{3}_{B}e^{2}_{A}}\cos(\theta_{A})dt\wedge dx \wedge \Vol(S^{3}_{B})\wedge d\phi_{A}, \nonumber \\
\Phi  &=&\log r\ .  \notag
%H_{3} &=&0.
\end{eqnarray}%
This is the background in eq.(\ref{background 211}).
In the case for which $Q_{A}=Q_{B}=m=0$,  the background fields read (note that we S-dualise moving to the NS5 brane frame),
    \begin{equation}
    \begin{aligned}
        ds^{2}_{10D} &= -dt^{2}+dx^{2}+ \frac{e^{2}_{A}+e^{2}_{B}}{2}d\varphi^{2} + \frac{8}{e^{2}_{A}+e^{2}_{B}}\frac{dr^{2}}{r^{2}} 
        + \frac{8}{e^{2}_{A}}ds^{2}(S^{3}_{A})
        + \frac{8}{e^{2}_{B}}ds^{2}(S^{3}_{B}),\\
        H_{3} &= -\frac{16}{e^{2}_{A}} \Vol(S^{3}_{A}) -\frac{16}{e^{2}_{B}} \Vol(S^{3}_{B}),\\
        \Phi &= -\log(r).
    \end{aligned}
    \end{equation}
This is the background in eq.(\ref{background1}).
We can rescale $\varphi$ to absorb the prefactor. Also, it is convenient to set  as in eq.(\ref{charges})
    \begin{equation}
        N_{A} = \frac{8}{e^{2}_{B}},\quad  N_{B} = \frac{8}{e^{2}_{A}},
    \end{equation}
and perform the change of coordinates
    \begin{equation}
        r = e^{\sqrt{\frac{1}{N_{A}}+\frac{1}{N_{B}}}\,\rho},
    \end{equation}
after changing $H_{3}\rightarrow -H_{3}$, the background  reads
    \begin{equation}
    \begin{aligned}
        ds^{2}_{10D} &= - dt^2 + dx^{2} + d\varphi^2 + d\rho^{2} 
        + N_{B} ds^{2}(S^{3}_{A})
        + N_{A}ds^{2}(S^{3}_{B}),\\
        H_{3} &= 2N_{A} \Vol{S^{3}_{A}} + 2N_{B}\Vol{S^{3}_{B}},\\
        \Phi &= -\sqrt{\frac{1}{N_{A}}+\frac{1}{N_{B}}} \, \rho.
    \end{aligned}
    \end{equation}

This is the background written in Section \ref{sigmamodelsection} to study the string $\sigma$-model on this field configuration.

\section{On the Unfibered Geometry}\label{appendix28}

Here, we review a different derivation of the background \eqref{LinearDilatonBackground}. The S-dual of this background, \eqref{background1}, was first introduced in \cite{Itzhaki:2005tu}, \cite{Khuri:1993ii}. Here we review the derivation of the pure NS-NS frame for simplicity. Let us consider two stacks of $NS5$-branes, the first extended in $(t,x,y_{1},y_{2},y_{3},y_{4})$ and while the second one spans $(t,x,w_{1},w_{2},w_{3},w_{4})$. These stacks intersect in the $(t,x)$ directions, thus in the weak coupling regime, the effective theory on the intersection is 1+1 dimensional and preserves 8 Supercharges. 

We now move to the  strong coupling regime. For this, we write the space $\mathbb{R}^{4}_{y} = (y_{1},y_{2},y_{3},y_{4})$ in spherical coordinates $(r_{A}, S^{3}_{A})$, and similarly for $\mathbb{R}^{4}_{w}$ we use $(r_{B}, S^{3}_{B})$. In terms of the harmonic functions
    \begin{equation}
        H_{A}(r_{A}) = 1 + \frac{N_{B}}{r^{2}_{A}}, \quad 
        H_{B}(r_{B}) = 1 + \frac{N_{A}}{r^{2}_{B}},
    \end{equation}
the backreacted fields are given by
    \begin{equation}
    \begin{aligned}
        ds^{2}_{10D} &= dx^{2}_{1,1} 
        + H_{A}(r_{A})\left( dr^{2}_{A} + r^{2}_{A}ds^{2}(S^{3}_{A})\right)
        + H_{B}(r_{B})\left( dr^{2}_{B} + r^{2}_{B}ds^{2}(S^{3}_{B})\right),\\
        H_{3} &= 2N_{B} \Vol(S^{3}_{A}) + 2N_{A} \Vol(S^{3}_{B}),\\
        \Phi &= \frac{1}{2}\log(H_{A}(r_{A})H_{B}(r_{B})).
    \end{aligned}
    \end{equation}

By taking the near-horizon geometry we are led to 
    \begin{equation}
    \begin{aligned}
        ds^{2}_{10D} &= dx^{2}_{1,1} 
        + N_{B}\frac{dr^{2}_{A}}{r^{2}_{A}} + N_{A}\frac{dr^{2}_{B}}{r^{2}_{B}}
        + N_{B}ds^{2}(S^{3}_{A}) + N_{A}ds^{2}(S^{3}_{B}),\\
        H_{3} &= 2N_{B} \Vol(S^{3}_{A}) + 2N_{A} \Vol(S^{3}_{B}),\\
        \Phi &= -\log(r_{A}) -\log(r_{B}) + \frac{1}{2}\log(N_{A}N_{B}).
    \end{aligned}
    \end{equation}

Under the change of coordinates $r_{A} = e^{\rho_{A}/\sqrt{N_{B}}}$ and $r_{B} = e^{\rho_{B}/\sqrt{N_{A}}}$ we obtain 
    \begin{equation}\label{Unfibered1}
    \begin{aligned}
        ds^{2}_{10D} &= dx^{2}_{1,1} 
        + d\rho^{2}_{A} + d\rho^{2}_{B}
        + N_{B}ds^{2}(S^{3}_{A}) + N_{A}ds^{2}(S^{3}_{B}),\\
        H_{3} &= 2N_{B} \Vol(S^{3}_{A}) + 2N_{A} \Vol(S^{3}_{B}),\\
        \Phi &= -\sqrt{\frac{1}{N_{B}}}\rho_{A} -\sqrt{\frac{1}{N_{A}}}\rho_{B} + \frac{1}{2}\log(N_{A}N_{B}).
    \end{aligned}
    \end{equation}

Finally, taking the linear combination
    \begin{equation}
    \begin{aligned}
        \sqrt{\frac{1}{N_{A}}+\frac{1}{N_{B}}} \rho_{A} &= \sqrt{\frac{1}{N_{B}}}\rho - \sqrt{\frac{1}{N_{A}}}\varphi,\\
        \sqrt{\frac{1}{N_{A}}+\frac{1}{N_{B}}} \rho_{B} &= \sqrt{\frac{1}{N_{A}}}\rho + \sqrt{\frac{1}{N_{B}}}\varphi,
    \end{aligned}
    \end{equation}
we reach the background fields in \eqref{LinearDilatonBackground}
    \begin{equation}\label{Unfibered2}
    \begin{aligned}
        ds^{2}_{10D} &= -dt^2+ dx^2+d\varphi^2+ d\rho^{2} 
        + N_{B}\, ds^{2}(S^{3}_{A})
        + N_{A}\, ds^{2}(S^{3}_{B}),\\
        H_{3} &= 2N_{B}\, \Vol(S^{3}_{A}) + 2N_{A}\, \Vol(S^{3}_{B}),\\
        \Phi &= -\sqrt{\frac{1}{N_{A}}+\frac{1}{N_{B}}} \, \rho,
    \end{aligned}
    \end{equation}
where we cancelled the constant term of the Dilaton by a suitable shift of its zero mode. Note that the Dilaton in \eqref{Unfibered1} has functional dependence in two of the coordinates of the background, while the one in \eqref{Unfibered2} only depends only on one coordinate. This allows us to interpret the extra flat direction of \eqref{Unfibered2} as being part of the Field Theory ones. In this way, we see that in the strong coupling regime, the theory on the intersection acquires an extra dimension, becoming (2+1) dimensional. There is also a SUSY enhancement from 8 to 16 supercharges. We refer the reader to \cite{Itzhaki:2005tu} to cover this matter. 

\section{R-Symmetry Breaking}\label{RSymmetryBreaking}

In this appendix we give a detailed derivation of the symmetry breaking pattern of three $U(1)$ directions present in our background. In order to do this, we gauge this symmetries by introducing a gauge field $A$ and a scalar $\epsilon$. The presence of a mass term in the effective action of the gauge field, i.e. an explicit symmetry breaking of the gauge symmetry, signals the breaking  of this $U(1)$ symmetry on the dual field theory. In the QFT, the breaking of the global symmetry  can be either spontaneous or anomalous.

\subsection{\texorpdfstring{$U(1)$}{U(1)} R-Symmetry of \texorpdfstring{$\psi_{A}$}{psi A} and \texorpdfstring{$\psi_{B}$}{psi B}}

Let us start by recalling the R-R $C_{2}$ potential
    \begin{eqnarray}\label{C21}
       & & C_{2} = {\psi_A}\left( \frac{2Q_{A}}{e_{A}}\zeta ^{\prime }\left(
r\right) dr\wedge d\varphi -\frac{2}{e_{A}^{2}}\sin {\theta_A}d{\theta_A}%
\wedge d{\phi_A}\right) +\frac{2}{e_{A}}\cos {\theta_A}Q_{A}\zeta
\left( r\right) d\varphi \wedge d{\phi_A}\nonumber \\
&&+{\psi_B}\left( \frac{2Q_{B}}{e_{B}}\zeta ^{\prime }\left( r\right)
dr\wedge d\varphi -\frac{2}{e_{B}^{2}}\sin {\theta_B}d{\theta_B}%
\wedge d{\phi_B}\right) +\frac{2}{e_{B}}\cos {\theta_B}Q_{B}\zeta
\left( r\right) d\varphi \wedge d{\phi_B}\ .
    \end{eqnarray}

Since this potential is not invariant under $\psi_{A,B} \rightarrow \psi_{A,B}+4\pi$, we expect this symmetry to be broken in the dual field theory. We gauge these isometries by doing the following replacements in the R-R potential and the metric
    \begin{equation}
        d\psi_{A,B} \rightarrow d\psi_{A,B} + A_{A,B}, \quad 
        \psi_{A,B} \rightarrow  \psi_{A,B} + \epsilon_{A,B}.
    \end{equation}

Where $A_{A,B}$ is a $U(1)$ gauge field and $\epsilon_{A,B}$ is a scalar charged under the gauged $U(1)$ symmetry, which makes the combination $D_{A,B}\epsilon = \partial_{A,B} \epsilon_{A,B} - A_{A,B}$ is gauge invariant (here $\partial_{A,B} = \partial/\partial\psi_{A,B}$). These fields only depend on the coordinates of the field theory directions, i.e. $(t,x)$. After these replacements the metric and the R-R 3-form read
    \begin{align}
        &\begin{aligned}
            ds^{2} = &ds^{2}_{(0)} +\frac{4r}{e^{2}_{A}}\left( \hat{\omega}_{3}-e_{A}Q_{A}\xi(r)d\varphi \right) A_{A\, \mu}dx^{\mu} + \frac{2r}{e^{2}_{A}}A_{A\, \mu} A_{A\, \nu} dx^{\mu}dx^{\nu} \\
            & +\frac{4r}{e^{2}_{B}}\left( \tilde{\omega}_{3}-e_{B}Q_{B}\xi(r)d\varphi \right) A_{B\, \mu}dx^{\mu} + \frac{2r}{e^{2}_{B}}A_{B\, \mu} A_{B\, \nu} dx^{\mu}dx^{\nu}
        \end{aligned} \\
        &\begin{aligned}
            F_{3} = &F^{(0)}_{3} - 2 d\epsilon_{A}\wedge \left( -\frac{Q_{A}}{e_{A}}\xi'(r)dr\wedge d\varphi  + \frac{1}{e^{2}_{A}} \Vol(S^{2}_{A}) \right) \\
        &- 2 d\epsilon_{B}\wedge \left( -\frac{Q_{B}}{e_{B}}\xi'(r)dr\wedge d\varphi  + \frac{1}{e^{2}_{B}} \Vol(S^{2}_{B}) \right) 
        \end{aligned}
    \end{align}

where $ds^{2}_{(0)}$ and $F^{(0)}_{3}$ denotes the metric and the 3-form of the configuration before gauging the $U(1)$ symmetries and $\Vol(S^{2}_{A,B}) = \sin(\theta_{A,B})d\theta_{A,B}\wedge d\phi_{A,B}$. Now we want to obtain an effective lagrangian for $A_{A,B}$ and $\epsilon_{A,B}$. In order to do this, we consider how the Ricci scalar and the kinetic term of $F_{3}$ change under the gauging of the symmetry. Explicitly, the Ricci scalar transforms as
    \begin{equation}
        R = R^{(0)} - \frac{1}{4} \frac{2r}{e^{2}_{A}} F^{2}_{A} - \frac{1}{4} \frac{2r}{e^{2}_{B}} F^{2}_{B},
    \end{equation}

where $F^{2}_{A,B} = F_{A,B \,\mu\nu}F^{\mu\nu}_{A,B}$ and $F_{A,B \, \mu\nu}$ is the field strengh of $A_{A,B\, \mu}$, while the kinetic term of the R-R potential reads
    \begin{equation}
    \begin{aligned}
        \frac{1}{12} F_{\mu\nu\lambda}F^{\mu\nu\lambda}  
        =  &\frac{1}{12} F^{(0)}_{\mu\nu\lambda}F_{(0)}^{\mu\nu\lambda}
        + \frac{1}{2r^{2}}\left( \frac{Q^{2}_{A}r^{2}}{e^{2}_{A}}\xi'(r)^{2}+1\right)\left( A_{A}-d\epsilon_{A} \right)^{2} \\
        &+ \frac{1}{2r^{2}}\left( \frac{Q^{2}_{B}r^{2}}{e^{2}_{B}}\xi'(r)^{2}+1\right)\left( A_{B}-d\epsilon_{B} \right)^{2} + \frac{Q_{A}Q_{B}}{e_{A}e_{B}}\left( A_{A}-d\epsilon_{A} \right)\cdot\left( A_{B}-d\epsilon_{B} \right).
    \end{aligned}
    \end{equation}

Finally, replacing this expression in the Type IIB action (in string frame), leads to the following effective lagrangian
    \begin{equation}
    \begin{aligned}
        \mathcal{L} &=  
        - \frac{1}{4} \frac{2}{e^{2}_{A}r} F^{2}_{A} - \frac{1}{4} \frac{2}{e^{2}_{B}r}F^{2}_{B}
        - \frac{1}{2r^{2}}\left( \frac{Q^{2}_{A}r^{2}}{e^{2}_{A}}\xi'(r)^{2}+1\right)\left( A_{A}-d\epsilon_{A} \right)^{2} \\
        &\phantom{=} - \frac{1}{2r^{2}}\left( \frac{Q^{2}_{B}r^{2}}{e^{2}_{B}}\xi'(r)^{2}+1\right)\left( A_{B}-d\epsilon_{B} \right)^{2} 
        - \frac{Q_{A}Q_{B}}{e_{A}e_{B}}\left( A_{A}-d\epsilon_{A} \right)\cdot\left( A_{B}-d\epsilon_{B} \right)
    \end{aligned}
    \end{equation}

Due to the coupling between $A_{A,B}$ and $\epsilon_{A,B}$ the gauge field obtains a mass. This is the same as the Stueckelberg mechanism. Defining $W_{A,B} = A_{A,B} - d\epsilon_{A,B}$, we obtain an action for the massive gauge field. 
    \begin{equation}
    \begin{aligned}
        \mathcal{L} &=  
        - \frac{1}{4} \frac{2}{e^{2}_{A}r} F^{2}_{A} - \frac{1}{4} \frac{2}{e^{2}_{B}r}F^{2}_{B}
        - \frac{1}{2r^{2}}\left( \frac{Q^{2}_{A}r^{2}}{e^{2}_{A}}\xi'(r)^{2}+1\right)
        W_{A\, \mu}W^{\mu}_{A} \\
        &\phantom{=} - \frac{1}{2r^{2}}\left( \frac{Q^{2}_{B}r^{2}}{e^{2}_{B}}\xi'(r)^{2}+1\right)W_{B\, \mu}W^{\mu}_{B}
        - \frac{Q_{A}Q_{B}}{e_{A}e_{B}}W_{A\, \mu}W^{\mu}_{B}
    \end{aligned}
    \end{equation}

\subsection{\texorpdfstring{$U(1)$}{U(1))} R-Symmetry of \texorpdfstring{$\varphi$}{varphi}}

Now we repeat the same procedure as above for the $\varphi$ direction. The only difference is at the starting point. The potential \eqref{C21} does not depend on $\varphi$. We need to perform a gauge transformation to give it $\varphi$ dependance, after which
    \begin{equation}
    \begin{aligned}
        C_{2} &= \frac{2Q_{A}}{e_{A}}\varphi\left[ \xi'(r)\left( d\psi_{A} + \cos\left(\theta_{A}\right)d\phi_{A}\right)\wedge dr + \xi(r) \Vol(S^{2}_{A})\right] - \frac{2}{e^{2}_{A}}\psi_{A} \Vol(S^{2}_{A})\\ 
        &\phantom{=} + \frac{2Q_{B}}{e_{B}}\varphi\left[ \xi'(r)\left( d\psi_{B} + \cos\left(\theta_{B}\right)d\phi_{b}\right)\wedge dr + \xi(r) \Vol(S^{2}_{b})\right] - \frac{2}{e^{2}_{A}}\psi_{B} \Vol(S^{2}_{B}).
    \end{aligned}
    \end{equation}

As before, we gauge the symmetry along $\varphi$ by shifting the metric and the R-R potential as follows
    \begin{equation}
        d\varphi \rightarrow d\varphi  + A_{\varphi}, \quad 
        \varphi \rightarrow  \varphi + \epsilon_{\varphi}.
    \end{equation}

Repeating the procedure of the previous section lead to the following shifts for Ricci scalar 
    \begin{equation}
        R = R^{(0)} - \frac{1}{4}\left( r f_{s}(r) + 2(Q^{2}_{A}+Q^{2}_{B})r\xi(r)^{2}  \right)F^{2}_{\varphi}
    \end{equation}

where $F^{2}_{\varphi} = F_{\varphi \, \mu\nu}F^{\mu\nu}_{\varphi}$, with $F_{\varphi \, \mu\nu}$ the field strenght of $A_{\varphi}$, and the kinetic term of the R-R potential
    \begin{equation}
    \begin{aligned}
        \frac{1}{12} F_{\mu\nu\lambda}F^{\mu\nu\lambda}  
        &=  \frac{1}{12} F^{(0)}_{\mu\nu\lambda}F_{(0)}^{\mu\nu\lambda} \\
        &\phantom{=} +\frac{1}{4}\left( \frac{2}{r^{2}}\xi(r)^{2}\left( e^{2}_{A}Q^{2}_{A} + e^{2}_{B}Q^{2}_{B} 
        + (Q^{2}_{A} + Q^{2}_{B})^{2}r^{2}\xi'(r)^{2}\right) + (Q^{2}_{A} + Q^{2}_{B})f_{s}(r)\xi'(r)^{2}  \right) \left( A_{\varphi} - d\epsilon_{\varphi}\right)^{2}
    \end{aligned}
    \end{equation}

which leads to the effective lagrangian
    \begin{equation}
    \begin{aligned}
        \mathcal{L} &= - \frac{1}{4}\left( f_{s}(r) + 2(Q^{2}_{A}+Q^{2}_{B})\xi(r)^{2} \right)F^{2}_{\varphi} \\
        &\phantom{=} -\frac{1}{4}\left( \frac{2}{r^{2}}\xi(r)^{2}\left( e^{2}_{A}Q^{2}_{A} + e^{2}_{B}Q^{2}_{B} 
        + (Q^{2}_{A} + Q^{2}_{B})^{2}r^{2}\xi'(r)^{2}\right) + (Q^{2}_{A} + Q^{2}_{B})f_{s}(r)\xi'(r)^{2}  \right) \left( A_{\varphi} - d\epsilon_{\varphi}\right)^{2}
    \end{aligned}
    \end{equation}

As before, we see from the action that after a gauge transformation the gauge field obtains a mass via Stueckelberg mechanism. Explicitly by defining $W_{\varphi} = A_{\varphi} - d\epsilon_{\varphi}$ we obtain
    \begin{equation}
    \begin{aligned}
        \mathcal{L} &= - \frac{1}{4}\left( f_{s}(r) + 2(Q^{2}_{A}+Q^{2}_{B})\xi(r)^{2} \right)F^{2}_{\varphi} \\
        &\phantom{=} -\frac{1}{4}\left( \frac{2}{r^{2}}\xi(r)^{2}\left( e^{2}_{A}Q^{2}_{A} + e^{2}_{B}Q^{2}_{B} 
        + (Q^{2}_{A} + Q^{2}_{B})^{2}r^{2}\xi'(r)^{2}\right) + (Q^{2}_{A} + Q^{2}_{B})f_{s}(r)\xi'(r)^{2}  \right) W_{\varphi \, \mu}W^{\mu}_{\varphi}
    \end{aligned}
    \end{equation}

\section{Maldacena-Wilson, 't Hooft loops and EE. Detailed calculations}\label{appendix3}

In this appendix, we study the integrals needed to compute the Wilson loops, 't Hooft loops and Entanglement Entropy.  We express the analytic results in terms of $r_0$.

Let us define the following quantities that allow us to write the integrals
in a simpler way%
\begin{eqnarray}
\lambda _{0} &=&\frac{r_{0}}{r_{+}}\ ,\quad \xi =\frac{r}{r_{+}}\ ,\quad
\eta =\frac{e_{A}}{e_{B}}\ ,  \label{new parameters} \\
\lambda _{-}^{2} &=&-\frac{r_{-}^{2}}{r_{+}^{2}}\equiv 1-\frac{m}{%
r_{+}^{2}\left( e_{A}^{2}+e_{B}^{2}\right) }\ ,
\end{eqnarray}%
where $\lambda _{-}\in \left[ 0,1\right] $, $\lambda _{0}>1$ and $\eta >0$.

All the problems we will address here can be reduced to a one-dimensional
problem for the function $r=r\left( x\right) $ which minimises the functional
in eq.(\ref{NGwilson}) once we impose that the parameter $x(\sigma)=\sigma $. Then, the equation for the function  $r(\sigma)=r(x)$ reduces to%
\begin{equation}
\frac{dr}{dx}=\pm V_{eff}\left( r\right) \ ,
\label{the equation for the profile}
\end{equation}%
for a suitable effective potential which is case-dependent. In most 
cases the function $r\left( x\right) $ can be interpreted as a string (or a
section of a higher dimensional surface) with end points at $r\rightarrow
\infty $. The profile of the string subject to the initial condition $%
x\left( r_{0}\right) =0$, can be obtained by performing the integral%
\begin{equation}
x\left( r\right) =\pm \int_{r_{0}}^{r}\frac{dr}{V_{eff}\left( r\right) }\ .
\label{profile integral}
\end{equation}%
From here we compute the end points separation as%
\begin{equation}
L\left( r_{0}\right) \equiv \lim_{r\rightarrow \infty }2x\left( r\right) \ .
\label{definition length from profile}
\end{equation}%
The definition  in eq.(\ref{definition length from profile}) coincides with the
quark-anti-quark separation, monopole-anti-monopole separation and the
interval length for the Maldacena-Wilson loop, t' Hooft loop and
entanglement entropy, respectively.

The results of the integrals that we compute analytically are given
in terms of elliptic integrals. The elliptic integral of first
kind $\mathbf{F}\left( \phi |m\right) $ and the complete elliptic integral
of first kind $\mathbf{K}\left( m\right) $ are defined as%
\begin{eqnarray}
\mathbf{F}\left( \phi |m\right)  &=&\int_{0}^{\phi }d\theta \frac{1}{\sqrt{%
1-m\sin ^{2}\theta }}\ , \\
\mathbf{K}\left( m\right)  &=&\mathbf{F}\left( \left. \frac{\pi }{2}%
\right\vert m\right) \ ,
\end{eqnarray}%
respectively for $-\frac{\pi }{2}<\phi <\frac{\pi }{2}$. The elliptic
integral of second kind $\mathbf{E}\left( \phi |m\right) $ and the complete
elliptic integral $\mathbf{E}\left( m\right) $ are defined respectively as%
\begin{eqnarray}
\mathbf{E}\left( \phi |m\right)  &=&\int_{0}^{\phi }\sqrt{1-m\sin ^{2}\theta 
}d\theta \ , \\
\mathbf{E}\left( m\right)  &=&\mathbf{E}\left( \left. \frac{\pi }{2}%
\right\vert m\right) \ ,
\end{eqnarray}%
where $-\frac{\pi }{2}<\phi <\frac{\pi }{2}$.

\subsection{Wilson loop}

The effective potential in terms of the variable $\xi $ defined in (\ref{new
parameters}) reads%
\begin{equation*}
V_{eff}\left( \xi \right) =\sqrt{\frac{e_{A}^{2}+e_{B}^{2}}{8}}\frac{r_{+}}{%
\lambda _{0}\xi }\sqrt{\left( \xi ^{2}-\lambda _{0}^{2}\right) \left( \xi
^{2}+\lambda _{-}^{2}\right) \left( \xi ^{2}-1\right) }\ .
\end{equation*}%
The string profile, considering the change of variables in eq.(\ref{new
parameters})%
\begin{eqnarray}
x\left( \xi \right) &=&\pm \sqrt{\frac{8}{e_{A}^{2}+e_{B}^{2}}}\int_{\lambda
_{0}}^{\xi }\frac{\lambda _{0}\xi d\xi }{\sqrt{\left( \xi ^{2}-\lambda
_{0}^{2}\right) \left( \xi ^{2}+\lambda _{-}^{2}\right) \left( \xi
^{2}-1\right) }}\ ,  \notag \\
&=&\pm \sqrt{\frac{8}{e_{A}^{2}+e_{B}^{2}}}\frac{\lambda _{0}}{\sqrt{\lambda
_{0}^{2}-1}}  \label{profile wilson loop} \\
&&\times \left[ -\mathbf{F}\left( \left. \arcsin \sqrt{\frac{\lambda
_{0}^{2}-1}{\xi ^{2}-1}}\right\vert \frac{1+\lambda _{-}^{2}}{1-\lambda
_{0}^{2}}\right) +\mathbf{K}\left( \frac{1+\lambda _{-}^{2}}{1-\lambda
_{0}^{2}}\right) \right] \ .  \notag
\end{eqnarray}%
The definition of the quark-antiquark separation given in eq.(\ref{Lqq}) can
be expressed in terms of the limit (\ref{definition length from profile}) of
the string profile. Replacing in eq.(\ref{profile wilson loop}) we find%
\begin{equation*}
L_{QQ}\left( \lambda _{0}\right) =2\sqrt{\frac{8}{e_{A}^{2}+e_{B}^{2}}}\frac{%
\lambda _{0}}{\sqrt{\lambda _{0}^{2}-1}}\mathbf{K}\left( \frac{1+\lambda
_{-}^{2}}{1-\lambda _{0}^{2}}\right) \ .
\end{equation*}%
This is our result in eq.(\ref{lqq2}).

In order to get an analytic expression for the energy in eq.(\ref{Eqq}) we compute the integrals%
\begin{equation}
E_{QQ}=F\left( r_{0}\right) L_{QQ}\left( r_{0}\right) +I_{2}+I_{3}
\label{energy WL appendix}
\end{equation}%
where%
\begin{eqnarray}
I_{2} &=&\sqrt{\frac{32}{e_{A}^{2}+e_{B}^{2}}}\int_{r_{0}}^{+\infty }dz\ z%
\sqrt{\frac{z^{2}-r_{0}^{2}}{\left( z^{2}-r_{-}^{2}\right) \left(
z^{2}-r_{+}^{2}\right) }}\ , \\
I_{3} &=&-\sqrt{\frac{32}{e_{A}^{2}+e_{B}^{2}}}\int_{r_{+}}^{\infty }dz\frac{%
z^{2}}{\sqrt{\left( z^{2}-r_{-}^{2}\right) \left( z^{2}-r_{+}^{2}\right) }}\
.
\end{eqnarray}%
Considering the change of variable $\xi =z/r_{+}$ and the definitions (\ref%
{new parameters}) the integrals become%
\begin{eqnarray}
I_{2} &=&\frac{4\sqrt{2}r_{+}}{\sqrt{e_{A}^{2}+e_{B}^{2}}}\int_{\lambda
_{0}}^{+\infty }d\xi \ \xi \sqrt{\frac{\xi ^{2}-\lambda _{0}^{2}}{\left( \xi
^{2}+\lambda _{-}^{2}\right) \left( \xi ^{2}-1\right) }}\ , \\
I_{3} &=&-\frac{4\sqrt{2}r_{+}}{\sqrt{e_{A}^{2}+e_{B}^{2}}}\int_{1}^{\infty
}d\xi \frac{\xi ^{2}}{\sqrt{\left( \xi ^{2}+\lambda _{-}^{2}\right) \left(
\xi ^{2}-1\right) }}\ .
\end{eqnarray}%
We perform the indefinite integral of $I_{2}$ giving%
\begin{eqnarray}
\int^{\xi }d\xi \ \xi \sqrt{\frac{\xi ^{2}-\lambda _{0}^{2}}{\left( \xi
^{2}+\lambda _{-}^{2}\right) \left( \xi ^{2}-1\right) }} &=&\frac{\sqrt{%
\left( \xi ^{2}-\lambda _{0}^{2}\right) \left( \lambda _{-}^{2}+\xi
^{2}\right) }}{\sqrt{\xi ^{2}-1}} \\
&&+\sqrt{\left( \lambda _{0}^{2}-1\right) }\mathbf{EE}\left( \left. \arcsin 
\sqrt{\frac{\lambda _{0}^{2}-1}{\xi ^{2}-1}}\right\vert \frac{1+\lambda
_{-}^{2}}{1-\lambda _{0}^{2}}\right) \ .  \notag
\end{eqnarray}%
Taking the limits%
\begin{eqnarray}
\lim_{\xi \rightarrow \infty }\int^{\xi }d\xi \ \xi \sqrt{\frac{\xi
^{2}-\lambda _{0}^{2}}{\left( \xi ^{2}+\lambda _{-}^{2}\right) \left( \xi
^{2}-1\right) }} &=&\lim_{\xi \rightarrow \infty }\xi +\mathcal{O}\left( 
\frac{1}{\xi }\right) \ , \\
\lim_{\xi \rightarrow \lambda _{0}}\int^{\xi }d\xi \ \xi \sqrt{\frac{\xi
^{2}-\lambda _{0}^{2}}{\left( \xi ^{2}+\lambda _{-}^{2}\right) \left( \xi
^{2}-1\right) }} &=&\sqrt{\left( \lambda _{0}^{2}-1\right) }\mathbf{EE}%
\left( \frac{1+\lambda _{-}^{2}}{1-\lambda _{0}^{2}}\right) \ .
\end{eqnarray}%
Hence,%
\begin{equation}
I_{2}=\frac{4\sqrt{2}r_{+}}{\sqrt{e_{A}^{2}+e_{B}^{2}}}\left[ \lim_{\xi
\rightarrow \infty }\xi -\sqrt{\left( \lambda _{0}^{2}-1\right) }\mathbf{EE}%
\left( \frac{1+\lambda _{-}^{2}}{1-\lambda _{0}^{2}}\right) \right].
\end{equation}%
The indefinite integral of $I_{3}$ gives%
\begin{equation}
\int^{\xi }d\xi \frac{\xi ^{2}}{\sqrt{\left( \xi ^{2}+\lambda
_{-}^{2}\right) \left( \xi ^{2}-1\right) }}=i\lambda _{m}\left[ \mathbf{E}%
\left( \arcsin \xi \left\vert -\frac{1}{\lambda _{-}^{2}}\right. \right) -%
\mathbf{F}\left( \arcsin \xi \left\vert -\frac{1}{\lambda _{-}^{2}}\right.
\right) \right] \ .
\end{equation}%
Computing the limits we get%
\begin{eqnarray*}
\lim_{\xi \rightarrow \infty }\int^{\xi }d\xi \frac{\xi ^{2}}{\sqrt{\left(
\xi ^{2}+\lambda _{-}^{2}\right) \left( \xi ^{2}-1\right) }} &=&\lim_{\xi
\rightarrow \infty }\xi +\mathcal{I}_{\lambda _{-}}+O\left( \xi ^{-1}\right)
\ , \\
\lim_{\xi \rightarrow 1}\int^{\xi }d\xi \frac{\xi ^{2}}{\sqrt{\left( \xi
^{2}+\lambda _{-}^{2}\right) \left( \xi ^{2}-1\right) }} &=&i\lambda _{-}%
\left[ \mathbf{E}\left( -\lambda _{-}^{-2}\right) -\mathbf{K}\left( -\lambda
_{-}^{-2}\right) \right] \ ,
\end{eqnarray*}%
where%
\begin{eqnarray*}
\mathcal{I}_{\lambda _{-}} &=&\left[ i\lambda _{-}\mathbf{E}\left( -\lambda
_{-}^{-2}\right) -\mathbf{E}\left( -\lambda _{-}^{2}\right) -\lambda _{-}%
\mathbf{K}\left( 1+\lambda _{-}^{-2}\right) \right. \\
&&\left. -2i\lambda _{-}\mathbf{K}\left( -\lambda _{-}^{-2}\right) +\mathbf{K%
}\left( -\lambda _{-}^{2}\right) +\lambda _{-}^{2}\mathbf{K}\left( -\lambda
_{-}^{2}\right) \right].
\end{eqnarray*}%
Therefore, the integral becomes%
\begin{eqnarray*}
I_{3} &=&-\frac{4\sqrt{2}r_{+}}{\sqrt{e_{A}^{2}+e_{B}^{2}}}[\lim_{\xi
\rightarrow \infty }\xi -\mathbf{E}\left( -\lambda _{-}^{2}\right) -\lambda
_{-}\mathbf{K}\left( 1+\lambda _{-}^{-2}\right) \\
&&\qquad \left. -i\lambda _{-}\mathbf{K}\left( -\lambda _{-}^{-2}\right) +%
\mathbf{K}\left( -\lambda _{-}^{2}\right) +\lambda _{-}^{2}\mathbf{K}\left(
-\lambda _{-}^{2}\right) \right] \ .
\end{eqnarray*}%
Replacing into the energy in eq. (\ref{energy WL appendix}), we find the result in eq.(\ref{eqq2})%
\begin{eqnarray*}
E_{QQ}\left( \lambda _{0}\right) &=&2r_{+}\sqrt{\frac{8}{e_{A}^{2}+e_{B}^{2}}%
}\left[ \frac{\lambda _{0}^{2}}{\sqrt{\lambda _{0}^{2}-1}}\mathbf{K}\left( 
\frac{1+\lambda _{-}^{2}}{1-\lambda _{0}^{2}}\right) \right. \\
&&+\mathbf{E}\left( -\lambda _{-}^{2}\right) +\lambda _{-}\mathbf{K}\left(
1+\lambda _{-}^{-2}\right) +i\lambda _{-}\mathbf{K}\left( -\lambda
_{-}^{-2}\right) \\
&&\left. -\sqrt{\left( \lambda _{0}^{2}-1\right) }\mathbf{E}\left( \frac{%
1+\lambda _{-}^{2}}{1-\lambda _{0}^{2}}\right) -\left( 1+\lambda
_{-}^{2}\right) \mathbf{K}\left( -\lambda _{-}^{2}\right) \right]
\end{eqnarray*}

\subsection{t' Hooft loop}
The effective potential $V_{eff}$ is given in eq.%TCIMACRO{\TeXButton{refveffmm}{(\ref{veffmm})} }%
%BeginExpansion
(\ref{veffmm}).
%EndExpansion
Replacing explicitly the functions and using the definition in eq. (\ref{new
parameters}) leads to
\begin{eqnarray}
V_{eff} &=&\frac{r_{+}e_{B}}{2}\sqrt{\eta ^{2}+1}\frac{1}{\xi }\sqrt{\left(
\xi ^{2}-\lambda _{0}^{2}\right) \left( \xi ^{2}-1\right) \left( \xi
^{2}+\lambda _{-}^{2}\right) } \\
&&\times \sqrt{\frac{\left( 4Q_{B}^{2}r_{+}^{-4}e_{B}^{-2}\left( \xi
^{2}+\lambda _{0}^{2}-2\right) +\left( \eta ^{2}+1\right) \left( \xi
^{2}+\lambda _{0}^{2}-1+\lambda _{-}^{2}\right) \right) }{2\left( \lambda
_{0}^{2}-1\right) \left[ 4Q_{B}^{2}r_{+}^{-4}e_{B}^{-2}\left( \lambda
_{0}^{2}-1\right) +\left( \eta ^{2}+1\right) \left( \lambda _{0}^{2}+\lambda
_{-}^{2}\right) \right] }} . \notag
\end{eqnarray}%
We compute analytically the integrals in the BPS bound in
which $\lambda _{-}=1$ and $Q_{B}=\pm \frac{e_{B}}{e_{A}}Q_{A}$ implying $%
Q_{A}=\frac{e_{A}}{2}r_{+}^{2}$. In this limit the effective potential
simplifies to

\begin{eqnarray*}
V_{eff}^{BPS} &=&r_{+}\frac{1}{2}\sqrt{\frac{e_{A}^{2}+e_{B}^{2}}{2\left(
\lambda _{0}^{2}-1\right) \left( \left( e_{A}^{2}+2e_{B}^{2}\right) \lambda
_{0}^{2}+e_{A}^{2}\right) }} \\
&&\times \sqrt{\frac{1}{\xi ^{2}}\left( \xi ^{2}-\lambda _{0}^{2}\right)
\left( \xi ^{4}-1\right) \left( e_{A}^{2}\left( \xi ^{2}+\lambda
_{0}^{2}\right) +2e_{B}^{2}\left( \xi ^{2}+\lambda _{0}^{2}-1\right) \right) .
}
\end{eqnarray*}%
The indefinite integral (\ref{profile integral}) gives%
\begin{eqnarray*}
\int^{\xi }\frac{d\xi r_{+}}{V_{eff}^{BPS}} &=&\frac{2}{e_{B}}\sqrt{\frac{%
2\left( \lambda _{0}^{2}-1\right) \left( \left( \eta ^{2}+2\right) \lambda
_{0}^{2}+\eta ^{2}\right) }{\eta ^{2}+1}} \\
&&\times \int \frac{\xi d\xi }{\sqrt{\left( \xi ^{2}-\lambda _{0}^{2}\right)
\left( \xi ^{4}-1\right) \left( \eta ^{2}\left( \xi ^{2}+\lambda
_{0}^{2}\right) +2\left( \xi ^{2}+\lambda _{0}^{2}-1\right) \right) }}\ , \\
&=&\frac{1}{e_{B}}\sqrt{\frac{2\left( \lambda _{0}^{2}-1\right) \left(
\left( \eta ^{2}+2\right) \lambda _{0}^{2}+\eta ^{2}\right) }{\left( \eta
^{2}+1\right) \left( \lambda _{0}^{2}\left( \eta ^{2}+2\right) -1\right) }}
\\
&&\times \mathbf{F}\left( \left. \arcsin \sqrt{\frac{2\left( -1+\left(
2+\eta ^{2}\right) \lambda _{0}^{2}\right) \left( \xi ^{2}-1\right) }{\left(
2\lambda _{0}^{2}+\eta ^{2}\left( 1+\lambda _{0}^{2}\right) \right) \left(
\xi ^{2}-\lambda _{0}^{2}\right) }}\right\vert \frac{\left( 1+\lambda
_{0}^{2}\right) \left( 2\lambda _{0}^{2}+\eta ^{2}\left( 1+\lambda
_{0}^{2}\right) \right) }{-4+4\left( 2+\eta ^{2}\right) \lambda _{0}^{2}}%
\right).
\end{eqnarray*}%
Taking the limit $\xi \rightarrow r_{0}$ we find%
\begin{eqnarray}
&&\lim_{\xi \rightarrow \lambda _{0}}r_{+}\int^{\xi }\frac{d\xi }{V_{eff}} \\
&=&\frac{1}{e_{B}}\sqrt{\frac{2\left( \lambda _{0}^{2}-1\right) \left(
\left( \eta ^{2}+2\right) \lambda _{0}^{2}+\eta ^{2}\right) }{\left( \eta
^{2}+1\right) \left( \lambda _{0}^{2}\left( \eta ^{2}+2\right) -1\right) }}%
\left( -i\right) \mathbf{K}\left( 1-\frac{\left( 1+\lambda _{0}^{2}\right)
\left( 2\lambda _{0}^{2}+\eta ^{2}\left( 1+\lambda _{0}^{2}\right) \right) }{%
-4+4\left( 2+\eta ^{2}\right) \lambda _{0}^{2}}\right) \ .  \notag
\end{eqnarray}%
Therefore, the profile of the string is%
\begin{eqnarray}
\pm x\left( \xi \right) &=&\frac{1}{e_{B}}\sqrt{\frac{2\left( \lambda
_{0}^{2}-1\right) \left( \left( \eta ^{2}+2\right) \lambda _{0}^{2}+\eta
^{2}\right) }{\left( \eta ^{2}+1\right) \left( \lambda _{0}^{2}\left( \eta
^{2}+2\right) -1\right) }} \\
&&\times \left[ \mathbf{F}\left( \left. \arcsin \sqrt{\frac{2\left(
-1+\left( 2+\eta ^{2}\right) \lambda _{0}^{2}\right) \left( \xi
^{2}-1\right) }{\left( 2\lambda _{0}^{2}+\eta ^{2}\left( 1+\lambda
_{0}^{2}\right) \right) \left( \xi ^{2}-\lambda _{0}^{2}\right) }}%
\right\vert \frac{\left( 1+\lambda _{0}^{2}\right) \left( 2\lambda
_{0}^{2}+\eta ^{2}\left( 1+\lambda _{0}^{2}\right) \right) }{-4+4\left(
2+\eta ^{2}\right) \lambda _{0}^{2}}\right) \right.  \notag \\
&&\qquad \left. +i\mathbf{K}\left( 1-\frac{\left( 1+\lambda _{0}^{2}\right)
\left( 2\lambda _{0}^{2}+\eta ^{2}\left( 1+\lambda _{0}^{2}\right) \right) }{%
-4+4\left( 2+\eta ^{2}\right) \lambda _{0}^{2}}\right) \right] \ .  \notag
\end{eqnarray}%
The monopole-anti-monopole separation can be deduced easily from the above
expression by taking the limit (\ref{definition length from profile}). We obtain,

\begin{eqnarray}
L_{MM}^{BPS}\left( \lambda _{0}\right)  &=&\frac{2}{e_{B}}\sqrt{\frac{%
2\left( \lambda _{0}^{2}-1\right) \left( \left( \eta ^{2}+2\right) \lambda
_{0}^{2}+\eta ^{2}\right) }{\left( \eta ^{2}+1\right) \left( \lambda
_{0}^{2}\left( \eta ^{2}+2\right) -1\right) }} \\
&&\times \left[ \mathbf{F}\left( \left. \arcsin \sqrt{\frac{2\left(
-1+\left( 2+\eta ^{2}\right) \lambda _{0}^{2}\right) }{\left( 2\lambda
_{0}^{2}+\eta ^{2}\left( 1+\lambda _{0}^{2}\right) \right) }}\right\vert 
\frac{\left( 1+\lambda _{0}^{2}\right) \left( 2\lambda _{0}^{2}+\eta
^{2}\left( 1+\lambda _{0}^{2}\right) \right) }{-4+4\left( 2+\eta ^{2}\right)
\lambda _{0}^{2}}\right) \right.   \notag \\
&&\qquad \left. +i\mathbf{K}\left( 1-\frac{\left( 1+\lambda _{0}^{2}\right)
\left( 2\lambda _{0}^{2}+\eta ^{2}\left( 1+\lambda _{0}^{2}\right) \right) }{%
-4+4\left( 2+\eta ^{2}\right) \lambda _{0}^{2}}\right) \right] \ .  \notag
\end{eqnarray}%
We compare it with the approximate function 
%TCIMACRO{\TeXButton{refapproxLQQ}{(\ref{approxLQQ})} }%
%BeginExpansion
(\ref{approxLQQ})
%EndExpansion
for the separation, replacing the functions explicitly we find%
\begin{equation}
\hat{L}_{MM}\left( r_{0}\right) =\frac{\pi \sqrt{2}}{\sqrt{%
e_{A}^{2}+e_{B}^{2}}}\frac{\left( e_{A}^{2}+2e_{B}^{2}\right)
r_{0}^{2}/r_{+}^{2}+e_{A}^{2}}{\left( e_{A}^{2}+2e_{B}^{2}\right)
r_{0}^{2}/r_{+}^{2}-e_{B}^{2}}\sqrt{\frac{r_{0}^{2}/r_{+}^{2}-1}{%
r_{0}^{2}/r_{+}^{2}+1}}\ .
\end{equation}

For the energy of the t' Hooft loop we have a similar expression to the one obtained when computing the 
energy of the Wilson loop,%
\begin{equation}
E_{MM}\left( r_{0}\right) =F\left( r_{0}\right) L_{MM}\left( r_{0}\right)
+I_{2}+I_{3}\ ,
\end{equation}%
\bigskip where 
\begin{eqnarray}
I_{2} &=&\frac{4r_{+}^{2}}{\sqrt{\eta ^{2}+1}}\int_{\lambda _{0}}^{+\infty
}d\xi \sqrt{\frac{\xi ^{2}\left( \xi ^{2}-\lambda _{0}^{2}\right) \left(
\eta ^{2}\left( \xi ^{2}+\lambda _{0}^{2}\right) +2\left( \xi ^{2}+\lambda
_{0}^{2}-1\right) \right) }{\xi ^{4}-1}}\ , \\
I_{3} &=&-2\int_{r_{+}}^{+\infty }dzG\left( z\right) =-\frac{4r_{+}^{2}}{%
\sqrt{\left( 1+\eta ^{2}\right) }}\int_{1}^{\infty }d\xi \ \xi \sqrt{\frac{%
2\xi ^{2}+\eta ^{2}\left( \xi ^{2}+1\right) }{\left( 1+\xi ^{2}\right) }}\ .
\end{eqnarray}%
\qquad \qquad These integrals are quite involved and present technical
difficulties to be performed in terms of known functions. Therefore we
compute them numerically up to a large value of the upper limit $\tilde{\xi}%
_{\max }$. Since $\tilde{\xi}_{\max }$ is finite the integrals are
convergents and we can write them together in terms of a  single integral which depends
on $\tilde{\xi}_{\max }$%
\begin{eqnarray}
&&E_{MM}\left( r_{0}\right) \left. =\right. F\left( r_{0}\right)
L_{MM}\left( r_{0}\right) -\left[ \frac{4r_{+}^{2}}{\sqrt{1+\eta ^{2}}}%
\int_{1}^{\lambda _{0}}\ \xi \sqrt{\frac{2\xi ^{2}+\eta ^{2}\left( \xi
^{2}+1\right) }{\left( 1+\xi ^{2}\right) }}\right. \\
&&-\lim_{\tilde{\xi}_{\max }\rightarrow \infty }\int_{\lambda _{0}}^{\tilde{%
\xi}_{\max }}d\xi \left( \sqrt{\frac{\xi ^{2}\left( \xi ^{2}-\lambda
_{0}^{2}\right) \left( \eta ^{2}\left( \xi ^{2}+\lambda _{0}^{2}\right)
+2\left( \xi ^{2}+\lambda _{0}^{2}-1\right) \right) }{\xi ^{4}-1}}\right. 
\notag \\
&&\qquad \qquad \qquad \qquad \qquad \left. \left. -\xi \sqrt{\frac{2\xi
^{2}+\eta ^{2}\left( \xi ^{2}+1\right) }{\left( 1+\xi ^{2}\right) }}\right) %
\right] \ .  \notag
\end{eqnarray}%
We verify that the integral converges to a limiting value for $\tilde{\xi}%
_{\max }$ large enough and much bigger than $\lambda _{0}=r_{0}/r_{+}$.

The limiting value of the function $L_{MM}\left( r_{0}\right) $ when $%
r_{0}\rightarrow \infty $ is non-zero and is given in terms of a
characteristic length of the Little String Theory (LST). This asymptotic
behavior matches with the asymptotic behavior of the background 
%TCIMACRO{\TeXButton{background1}{(\ref{background1})}}%
%BeginExpansion
(\ref{background1})%
%EndExpansion
, to see this fact explicitly we compute the separation length of the 't
Hooft loop of the background 
%TCIMACRO{\TeXButton{background1}{(\ref{background1})} }%
%BeginExpansion
(\ref{background1})
%EndExpansion
by taking the limit of the end-points separations 
%TCIMACRO{%
%\TeXButton{definition length from profile}{(\ref{definition length from profile})}}%
%BeginExpansion
(\ref{definition length from profile})%
%EndExpansion
. The relevant functions for the t' Hooft loop of the background 
%TCIMACRO{\TeXButton{background1}{(\ref{background1})} }%
%BeginExpansion
(\ref{background1})
%EndExpansion
are the metric function%
\begin{equation}
f_{s}\left( r\right) =\frac{1}{2}\left( e_{A}^{2}+e_{B}^{2}\right) \ ,
\end{equation}%
and the effective potential%
\begin{equation}
V_{eff}\left( r,r_{0}\right) =\frac{\sqrt{2\left( e_{A}^{2}+e_{B}^{2}\right) 
}}{4r_{0}^{2}}\sqrt{r^{2}\left( r^{4}-r_{0}^{4}\right) }\ .
\end{equation}%
Then, the profile of the string in the bulk is 
\begin{equation}
\pm x\left( r\right) =\sqrt{\frac{2}{e_{A}^{2}+e_{B}^{2}}}\arctan \left( 
\frac{\sqrt{r^{4}-r_{0}^{4}}}{r_{0}^{2}}\right) \ .
\end{equation}%
The end-point separation is given by the limit 
%TCIMACRO{%
%\TeXButton{definition length from profile}{(\ref{definition length from profile})} }%
%BeginExpansion
(\ref{definition length from profile})
%EndExpansion
and leads to the following constant%
\begin{equation*}
L_{MM}\left( r_{0}\right) =\pi \sqrt{\frac{2}{e_{A}^{2}+e_{B}^{2}}}\ .
\end{equation*}%
Thus, all the strings in the background 
%TCIMACRO{\TeXButton{background1}{(\ref{background1})} }%
%BeginExpansion
(\ref{background1})
%EndExpansion
that explores the bulk has the same end points separation. This value
coincides with the limiting value of the separation length of the background 
%TCIMACRO{\TeXButton{fibred background}{(\ref{background 211})} }%
%BeginExpansion
(\ref{background 211})
%EndExpansion
and with a LST characteristic length. Therefore the UV behavior of the dual
theory is driving by the LST. To capture the field theory behavior we
introduce a cut-off to rule out the non-local effects of the LST. In that
case the system present a phase transition between the unstable
configurations to the short strings configuration.

The energy of the t' Hooft loop of the background 
%TCIMACRO{\TeXButton{background1}{(\ref{background1})} }%
%BeginExpansion
(\ref{background1})
%EndExpansion
is given by 
%TCIMACRO{\TeXButton{QQ energy general}{(\ref{QQ energy})} }%
%BeginExpansion
(\ref{QQ energy})
%EndExpansion
with 
\begin{equation}
F\left( r\right) =\sqrt{\frac{e_{A}^{2}+e_{B}^{2}}{2}}\ r^{2},\qquad G\left(
r\right) =2\sqrt{\frac{e_{A}^{2}+2e_{B}^{2}}{e_{A}^{2}+e_{B}^{2}}}r\ .
\end{equation}%
We find that the energy is zero. This implies that energy of the t' Hooft
loop of the disconnected solution is the same to the connected one.

\subsection{Entanglement entropy}

The profile of the 8-dimensional surface is governed by the function $%
r=r\left( x\right) $ with equations coming from the minimisation of eq.
(\ref{EE211}) which gives an equation like (\ref{the equation for the
profile}) for the effective potential%
\begin{equation}
V_{eff}=\frac{1}{2r_{0}^{2}\sqrt{f_{s}\left( r_{0}\right) }}\sqrt{%
r^{2}f_{s}\left( r\right) \left( r^{4}f_{s}\left( r\right)
-r_{0}^{4}f_{s}\left( r_{0}\right) \right) }\ .  \label{Veff EE}
\end{equation}%
The integration of (\ref{Veff EE}) for this potential subject to the initial
condition $x\left( r_{0}\right) =0$ gives%
\begin{eqnarray*}
\pm x\left( \xi \right) &=&\int_{r_{0}}^{r}\frac{2r_{0}^{2}\sqrt{f_{s}\left(
r_{0}\right) }dr}{\sqrt{r^{2}f_{s}\left( r\right) \left( r^{4}f_{s}\left(
r\right) -r_{0}^{4}f_{s}\left( r_{0}\right) \right) }} \\
&=&\sqrt{\frac{8}{e_{A}^{2}+e_{B}^{2}}}\sqrt{\left( \lambda
_{0}^{2}-1\right) \left( \lambda _{0}^{2}+\lambda _{-}^{2}\right) } \\
&&\times \int_{\lambda _{0}}^{\xi }\frac{\xi d\xi }{\sqrt{\left( \xi
^{2}-\lambda _{0}^{2}\right) \left( \xi ^{2}-1\right) \left( \xi
^{2}+\lambda _{-}^{2}\right) \left( \xi ^{2}+\lambda _{0}^{2}+\lambda
_{-}^{2}-1\right) }}.
\end{eqnarray*}%
Performing the indefinite integral%
\begin{eqnarray*}
\mathcal{I}_{L}\left( \xi \right) &\equiv &\int^{\xi }\frac{\xi d\xi }{\sqrt{%
\left( \xi ^{2}-\lambda _{0}^{2}\right) \left( \xi ^{2}+\lambda
_{-}^{2}\right) \left( \xi ^{2}-1\right) \left( \xi ^{2}+\lambda
_{0}^{2}+\lambda _{-}^{2}-1\right) }} \\
&=&\frac{1}{\lambda _{0}^{2}+\lambda _{-}^{2}}\mathbf{F}\left( \left.
\arcsin \sqrt{\frac{\left( \lambda _{0}^{2}+\lambda _{-}^{2}\right) \left(
\xi ^{2}-1\right) }{\left( \lambda _{-}^{2}+1\right) \left( \xi ^{2}-\lambda
_{0}^{2}\right) }}\right\vert \frac{\left( 1+\lambda _{-}^{2}\right) \left(
\lambda _{-}^{2}+2\lambda _{0}^{2}-1\right) }{\left( \lambda
_{0}^{2}+\lambda _{-}^{2}\right) ^{2}}\right) \ .
\end{eqnarray*}%
The limit $\xi \rightarrow \lambda _{0}$ gives%
\begin{equation}
\lim_{\xi \rightarrow \lambda _{0}}\mathcal{I}_{L}\left( \xi \right) =-\frac{%
i}{\lambda _{0}^{2}+\lambda _{-}^{2}}\mathbf{K}\left( \frac{\left( \lambda
_{0}^{2}-1\right) ^{2}}{\left( \lambda _{0}^{2}+\lambda _{-}^{2}\right) ^{2}}%
\right) \ .
\end{equation}%
Thus, the profile of the surface in the bulk is%
\begin{eqnarray}
\pm x\left( \xi \right) &=&\sqrt{\frac{8}{e_{A}^{2}+e_{B}^{2}}}\sqrt{\frac{%
\lambda _{0}^{2}-1}{\lambda _{0}^{2}+\lambda _{-}^{2}}}\left[ i\mathbf{K}%
\left( \frac{\left( \lambda _{0}^{2}-1\right) ^{2}}{\left( \lambda
_{0}^{2}+\lambda _{-}^{2}\right) ^{2}}\right) \right. \\
&&\left. +\mathbf{F}\left( \left. \arcsin \sqrt{\frac{\left( \lambda
_{0}^{2}+\lambda _{-}^{2}\right) \left( \xi ^{2}-1\right) }{\left( \lambda
_{-}^{2}+1\right) \left( \xi ^{2}-\lambda _{0}^{2}\right) }}\right\vert 
\frac{\left( 1+\lambda _{-}^{2}\right) \left( \lambda _{-}^{2}+2\lambda
_{0}^{2}-1\right) }{\left( \lambda _{0}^{2}+\lambda _{-}^{2}\right) ^{2}}%
\right) \right].  \notag
\end{eqnarray}%
The length of the interval is given by the limit  in eq.(\ref{definition length
from profile}). In term of the variables in eq.(\ref{new parameters}) gives%
\begin{eqnarray}
L_{EE}\left( \lambda _{0}\right) &=&2\sqrt{\frac{8}{e_{A}^{2}+e_{B}^{2}}}%
\sqrt{\frac{\lambda _{0}^{2}-1}{\lambda _{0}^{2}+\lambda _{-}^{2}}}\left[ i%
\mathbf{K}\left( \frac{\left( \lambda _{0}^{2}-1\right) ^{2}}{\left( \lambda
_{0}^{2}+\lambda _{-}^{2}\right) ^{2}}\right) \right. \\
&&\left. +\mathbf{F}\left( \left. \arcsin \sqrt{\frac{\left( \lambda
_{0}^{2}+\lambda _{-}^{2}\right) }{\left( \lambda _{-}^{2}+1\right) }}%
\right\vert \frac{\left( 1+\lambda _{-}^{2}\right) \left( \lambda
_{-}^{2}+2\lambda _{0}^{2}-1\right) }{\left( \lambda _{0}^{2}+\lambda
_{-}^{2}\right) ^{2}}\right) \right] \ .  \notag
\end{eqnarray}
This is our expression in eq.(\ref{LEEexact}).

We cannot see a phase transition in this background. However, if we put a
cutoff at $\xi _{\text{cutoff}}=\frac{r_{\text{cutoff}}}{r_{+}}$ the
coordinate $\lambda _{0}<\xi <\xi _{\text{cutoff}}$, the double-valued character of $L_{EE}$ shows, as in Figure \ref{figura5}.

The renormalized EE in eq.(\ref{SEE}) written in the variables of eq.(\ref{new
parameters}) reads%
\begin{equation*}
S_{EE}\left( \lambda _{0}\right) =\frac{\mathcal{N}}{G_{N}}r_{+}^{2}\left[
\int_{\lambda _{0}}^{\infty }d\xi \sqrt{\frac{\xi ^{2}\left( \xi
^{2}+\lambda _{-}^{2}\right) \left( \xi ^{2}-1\right) }{\left( \xi
^{2}-\lambda _{0}^{2}\right) \left( \xi ^{2}+\lambda _{0}^{2}+\lambda
_{-}^{2}-1\right) }}-\int_{1}^{\infty }\xi d\xi \right] \ .
\end{equation*}%
This integral can be done analytically. In the BPS limit it becomes
particularly simple,%
\begin{equation}
S_{EE}^{BPS}\left( \lambda _{0}\right) =\frac{\mathcal{N}}{G_{N}}r_{+}^{2}%
\left[ \int_{\lambda _{0}}^{\infty }d\xi \sqrt{\frac{\xi ^{2}\left( \xi
^{4}-1\right) }{\left( \xi ^{2}-\lambda _{0}^{2}\right) \left( \xi
^{2}+\lambda _{0}^{2}\right) }}-\int_{1}^{\infty }\xi d\xi \right] \ .
\label{EE for BPS A}
\end{equation}%
The indefine integral reads%
\begin{eqnarray}
\mathcal{I}_{S}^{BPS}\left( \xi \right)  &=&\int^{\xi }d\xi \sqrt{\frac{\xi
^{2}\left( \xi ^{4}-1\right) }{\left( \xi ^{2}-\lambda _{0}^{2}\right)
\left( \xi ^{2}+\lambda _{0}^{2}\right) }}\nonumber\\
&=&\frac{1}{2}\mathbf{E}\left( \left. \arcsin \frac{\xi ^{2}}{\lambda
_{0}^{2}}\right\vert \lambda _{0}^{4}\right) \ .
\end{eqnarray}%
Expanding for large $\xi $ and $\xi \rightarrow \lambda _{0}$ we find%
\begin{eqnarray}
\lim_{\xi \rightarrow \infty }\mathcal{I}_{S}\left( \xi \right) 
&=&\lim_{\xi \rightarrow \infty }\frac{1}{2}\xi ^{2}+\frac{1}{2\lambda
_{0}^{2}}\left[ -\lambda _{0}^{4}\mathbf{E}\left( \lambda _{0}^{-4}\right)
+\lambda _{0}^{2}\mathbf{E}\left( \lambda _{0}^{4}\right) -\mathbf{K}\left(
\lambda _{0}^{-4}\right) +\lambda _{0}^{4}\mathbf{K}\left( \lambda
_{0}^{-4}\right) \right]   \notag \\
&&+\mathcal{O}\left( \xi ^{-2}\right) \ , \\
\lim_{\xi \rightarrow \lambda _{0}}\mathcal{I}_{S}\left( \xi \right)  &=&%
\frac{1}{2}\mathbf{E}\left( \lambda _{0}^{4}\right) \ .
\end{eqnarray}%
Replacing in these expressions in the entanglement entropy of the BPS
configuration (\ref{EE for BPS A}) we obtain the expression in eq.(\ref{SEEexact}),%
\begin{equation}
S_{EE}^{BPS}\left( \lambda _{0}\right) =\frac{\mathcal{N}}{G_{N}}r_{+}^{2}%
\left[ \frac{1}{2\lambda _{0}^{2}}\left( -\lambda _{0}^{4}\mathbf{E}\left(
\lambda _{0}^{-4}\right) -\mathbf{K}\left( \lambda _{0}^{-4}\right) +\lambda
_{0}^{4}\mathbf{K}\left( \lambda _{0}^{-4}\right) \right) +\frac{1}{2}\right]
\ .
\end{equation}

The limiting of the interval length 
%TCIMACRO{\TeXButton{LEEexact}{(\ref{LEEexact})} }%
%BeginExpansion
(\ref{LEEexact})
%EndExpansion
value when $r_{0}\rightarrow \infty $  is non-zero and coincides with the
characteristic length of the Little String Theory. The background 
%TCIMACRO{\TeXButton{background1}{(\ref{background1})} }%
%BeginExpansion
(\ref{background1})
%EndExpansion
and the fibred one coincides in the UV leading to a regime in which the LST
dominates the behavior of dual theory. To verify this point we compute
profiles of the strings in the bulk%
\begin{equation}
\pm x\left( r\right) =\sqrt{\frac{2}{e_{A}^{2}+e_{B}^{2}}}\arctan \left( 
\frac{\sqrt{r^{4}-r_{0}^{4}}}{r_{0}^{2}}\right) \ ,
\end{equation}%
which end points separation at $r\rightarrow \infty $ gives interval length%
\begin{equation}
L_{EE}\left( r_{0}\right) =\pi \sqrt{\frac{2}{e_{A}^{2}+e_{B}^{2}}}\ .
\end{equation}%
One again, in order to capture the field theoretical behaviour of the dual
theory we add a cut-off to the observables which allow us to recover
expected behaviour of a confining field theory.

\section{Charges of the black membrane background}\label{appendixCharges}
We compute the charges of the configuration considering the Noether-Wald method
\cite{Wald:1993nt}. The bulk action principle in string frame of IIB in the metric-dilaton-$%
F_{3}$ sector is%
\begin{equation}
S_{\text{IIB,bulk}}=\frac{1}{2\kappa ^{2}}\int d^{10}x\sqrt{-g}\left( R-%
\frac{1}{2}\left( \partial \Phi \right) ^{2}-\frac{1}{12}e^{\Phi }F_{\mu \nu
\rho }F^{\mu \nu \rho }\right) =\int d^{10}x\sqrt{-g}\mathcal{L}\ ,
\label{bulk action}
\end{equation}%
A general variation of the action gives%
\begin{equation}
\delta S_{\text{IIB,bulk}}=\int d^{10}x\sqrt{-g}\left[ \delta g^{\mu \nu }%
\mathcal{E}_{\mu \nu }^{\left( g\right) }+\delta \Phi \mathcal{E}^{\left(
\Phi \right) }+\delta C_{\nu \rho }\mathcal{E}_{\left( F_{3}\right) }^{\nu
\rho }+\nabla _{\mu }\Theta ^{\mu }\left( \mathbf{f},\delta \mathbf{f}%
\right) \right] \ .
\end{equation}%
where $\mathbf{f}$ denotes the fields collectively and 
\begin{eqnarray}
\Theta ^{\mu }\left( \mathbf{f},\delta \mathbf{f}\right)  &=&\frac{1}{%
2\kappa ^{2}}\left( g^{\delta \eta }\delta \Gamma _{\ \eta \delta }^{\mu
}-g^{\delta \mu }\delta \Gamma _{\ \lambda \delta }^{\lambda }-\delta \Phi
\partial ^{\mu }\Phi -e^{\Phi }\delta C_{\nu \rho }F^{\mu \nu \rho }\right)
\ , \\
\mathcal{E}_{\mu \nu }^{\left( g\right) } &=&\frac{1}{2\kappa ^{2}}\left[
R_{\mu \nu }-\frac{1}{2}g_{\mu \nu }R\right.  \\
&&-\frac{1}{2}\left( \partial _{\mu }\Phi \partial _{\nu }\Phi -\frac{1}{2}%
g_{\mu \nu }\partial _{\rho }\Phi \partial ^{\rho }\Phi \right)   \notag \\
&&\left. -\frac{1}{2}e^{\Phi }\left( \frac{1}{2}F_{\mu \delta \rho }F_{\nu
}^{\ \delta \rho }-\frac{1}{12}g_{\mu \nu }F_{\delta \rho \sigma }F^{\delta
\rho \sigma }\right) \right] \ ,  \notag \\
&=&\frac{1}{2\kappa ^{2}}\left( G_{\mu \nu }-\frac{1}{2}T_{\mu \nu }^{\left(
\Phi \right) }-\frac{1}{2}T_{\mu \nu }^{\left( F_{3}\right) }\right) \ , 
\notag \\
\mathcal{E}^{\left( \Phi \right) } &=&\frac{1}{2\kappa ^{2}}\left( \nabla
^{\rho }\nabla _{\rho }\Phi -\frac{1}{12}e^{\Phi }F_{\mu \nu \rho }F^{\mu
\nu \rho }\right) \ , \\
\mathcal{E}_{\left( F_{3}\right) }^{\nu \rho } &=&\frac{1}{2\kappa ^{2}}%
\nabla _{\mu }\left( e^{\Phi }F^{\mu \nu \rho }\right) \ .
\end{eqnarray}%
The Noether current is defined by%
\begin{equation}
J^{\mu }=\Theta ^{\mu }\left( \mathbf{f},\mathcal{L}_{\xi }\mathbf{f}\right)
-\xi ^{\mu }\mathcal{L}\mathbb{\ },  \label{def noether current}
\end{equation}%
where $\mathcal{L}$ is the Lagrangian scalar under diffeomorphisms in (\ref%
{bulk action}) and $\mathcal{L}_{\xi }$ is the Lie derivative along the
vector $\xi $. The Noether current is conserved on-shell, thus it can be
written locally as $J^{\mu }=\nabla _{\nu }q^{\mu \nu }$. The Noether
current (\ref{def noether current}) for our system gives%
\begin{eqnarray}
J^{\mu } &=&-\frac{1}{\kappa ^{2}}\nabla _{\nu }\left( \nabla ^{\lbrack \mu
}\xi ^{\nu ]}+\frac{1}{2}e^{\Phi }2C_{\lambda \rho }\xi ^{\lambda }F^{\mu
\nu \rho }\right)   \label{our noether current} \\
&&+2\xi ^{\lambda }\mathcal{E}_{\left( g\right) \lambda }^{\mu }-2\xi
^{\lambda }C_{\lambda \rho }\mathcal{E}_{\left( F_{3}\right) }^{\mu \rho }\ ,
\notag
\end{eqnarray}%
on-shell it defines the Noether pre-potential:%
\begin{equation}
q^{\mu \nu }\left( \xi \right) =-\frac{1}{\kappa ^{2}}\left( \nabla
^{\lbrack \mu }\xi ^{\nu ]}+\frac{1}{2}e^{\Phi }2C_{\lambda \rho }\xi
^{\lambda }F^{\mu \nu \rho }\right) \ .
\end{equation}%
The Hodge dual of the Noether pre-potential gives the 8-form%
\begin{equation}
\boldsymbol{Q}\left[ \xi \right] =\frac{1}{2}\frac{1}{8!}\sqrt{-g}\epsilon
_{\mu \nu \rho _{1}\dots \rho _{8}}q^{\mu \nu }dx^{\rho _{1}}\wedge \dots
\wedge dx^{\rho _{8}}\ ,
\end{equation}%
that in differential forms is%
\begin{equation}
\boldsymbol{Q}\left[ \xi \right] =-\frac{1}{\kappa ^{2}}\left( \star d\xi
+e^{\Phi }\xi \lnot C_{2}\wedge \star F_{3}\right) \ .
\end{equation}
$\lnot $ stands for the contraction operator. 

The boundary term that allow us to have a well posed action principle and
finite mass is%
\begin{equation}
S_{\text{full}}=S_{\text{IIB,bulk}}+\int_{\partial M}d^{9}x\sqrt{-h}\frac{1}{\kappa ^{2}}\left( 
\mathcal{K}-e^{-\frac{1}{4}\Phi }\right) \ ,
\end{equation}%
where the first term in the integral is the Gibbons-Hawking-York term and
the last term corresponds to a counter term, that depends only on intrinsic
quantities, that allow us to renormalize the mass term. The extrinsic
curvature is defined in terms of the normal unit outwards  vector $n^{\mu }$
to the boundary of the spacetime by%
\begin{equation}
\mathcal{K}_{\mu \nu }=h_{\ \mu }^{\rho }h_{\ \nu }^{\sigma }\nabla _{\rho
}n_{\sigma }\ ,
\end{equation}%
and the induced metric is $h_{\mu \nu }=g_{\mu \nu }-n_{\mu }n_{\nu }$ for
our case.

Following \cite{Wald:1993nt} the energy, angular momentum and entropy are defined by%
\begin{eqnarray}
\mathcal{E}\left[ \boldsymbol{t}\right]  &=&\int_{\infty }\left( \boldsymbol{Q}\left[ 
\boldsymbol{t}\right] -\xi \lnot \boldsymbol{B}\right) \ ,
\label{def energy} \\
\mathcal{J}\left[ \boldsymbol{\psi }\right]  &=&-\int_{\infty }\boldsymbol{Q}\left[ 
\boldsymbol{\psi }\right] \ ,  \label{def J} \\
S\left[ \boldsymbol{\xi }\right]  &=&\frac{1}{T}\int_{\mathcal{H}}%
\boldsymbol{Q}\left[ \boldsymbol{\xi }\right] \ .  \label{def S}
\end{eqnarray}%
The boundary terms are in the 9-form 

\begin{equation}
\boldsymbol{B}=-\frac{1}{\kappa ^{2}}\left( \mathcal{K}-e^{-\frac{1}{4}\Phi
}\right) \star n\ .
\end{equation}%
$\boldsymbol{t}$ is the time-like Killing vector at infinity properly
normalized, $\boldsymbol{\psi }$ is the rotation generator and $\boldsymbol{%
\xi }$ is the horizon generator%
\begin{equation}
\boldsymbol{\xi }=\boldsymbol{t}+\Omega \boldsymbol{\psi \ .}
\end{equation}%
$\boldsymbol{\xi }$ is null at the Horizon which defines the angular
velocity $\Omega $ and satisfies the geodesic equation at the horizon%
\begin{equation}
\xi ^{\mu }\nabla _{\mu }\xi ^{\nu }=\kappa _{s}\xi ^{\nu }\ ,
\label{geodesic}
\end{equation}%
defining the surface gravity $\kappa _{s}$ that is related to the
temperature as $T=\frac{\kappa _{s}}{2\pi }$. In this case we are in general
relativity, therefore the entropy give one-quarter of the horizon area. 

Let us consider the black membrane configuration\ in Einstein frame \eqref{blackhole1}, with
\begin{equation}
    \zeta(r)=\frac{1}{r^2} ,\qquad \Phi \to \Phi-2\log\left(\frac{e_A^2+e_B^2}{2}\right) ,\qquad F_3 \to \frac{e_A^2+e_B^2}{2}F_3 \,\, .
\end{equation}
The in-going Eddington-Finkelstein coordinates are well-defined
at the horizonl, their defined by%
\begin{eqnarray}
dt &=&dv-\frac{2dr}{rf_{bh}\left( r\right) }\ , \\
d\psi _{A} &=&d\psi _{A}^{\prime }-\frac{2e_{A}Q_{A}\zeta \left( r\right) }{%
rf_{bh}\left( r\right) }dr\ , \\
d\psi _{B} &=&d\psi _{B}^{\prime }-\frac{2e_{B}Q_{B}\zeta \left( r\right) }{%
rf_{bh}\left( r\right) }dr\ .
\end{eqnarray}%
Then, the metric becomes%
\begin{eqnarray}
ds_{E}^{2} &=&\sqrt{r}\left\{ dy^{2}+dx^{2}-f_{bh}\left( r\right) dv^{2}+%
\frac{4}{r}drdv\right.  \\
&&+\frac{2}{e_{A}^{2}}\left[ d\theta _{A}^{2}+\sin ^{2}\theta _{A}d\phi
_{A}^{2}+\left( d\psi _{A}^{\prime }+\cos \theta _{A}d\phi
_{A}-e_{A}Q_{A}\zeta \left( r\right) dv\right) ^{2}\right]   \notag \\
&&\left. +\frac{2}{e_{B}^{2}}\left[ d\theta _{B}^{2}+\sin ^{2}\theta
_{B}d\phi _{B}^{2}+\left( d\psi _{B}^{\prime }+\cos \theta _{B}d\phi
_{B}-e_{B}Q_{B}\zeta \left( r\right) dv\right) ^{2}\right] \right\} \ , 
\notag
\end{eqnarray}%
In this coordinates we consider the vector%
\begin{equation}
\boldsymbol{\xi }=\boldsymbol{t}+\Omega _{A}\boldsymbol{\psi }_{A}+\Omega
_{B}\boldsymbol{\psi }_{B}
\end{equation}%
where%
\begin{equation}
\boldsymbol{t}=-\frac{1}{2}\frac{\partial }{\partial v}\ ,\qquad \boldsymbol{%
\psi }_{A}=\frac{\partial }{\partial \psi _{A}}\ ,\qquad \boldsymbol{\psi }%
_{B}=\frac{\partial }{\partial \psi _{B}}\ .
\end{equation}%
The vector $\boldsymbol{\xi }$ is null at the horizon located at $r_{+}$
when 
\begin{equation}
\Omega _{A}=\frac{e_{A}Q_{A}}{r_{+}^{2}}\ ,\qquad Q_{B}=\frac{e_{B}Q_{B}}{%
r_{+}^{2}}\ .
\end{equation}%
Due to the fact that the spacetime that we are considering is not asymptotically Minkowski times $S^3\times S^3$, instead is conformal to Minkowski times $S^3\times S^3$, it is not clear how we should normalize the vector $\boldsymbol{t}$ time-like at infinity. This ambiguity propagates to the energy and the temperature. Therefore we expect to obtain the temperature in 4D up to a factor. 
The energy (\ref{def energy}), angular momentum (\ref{def J}), temperature
defined through (\ref{geodesic}) and the entropy give%
\begin{eqnarray}
E &=&\mathcal{E}[\boldsymbol{t}]=\frac{2m}{e_{A}^{3}e_{B}^{3}r_{0}^{2}\kappa ^{2}}\left( 16\pi
^{2}\right) ^{2}L_{x}L_{y}\ , \\
J_{A} &=&\mathcal{J}[\boldsymbol{\psi_{A}}]=-\frac{8Q_{A}}{e_{A}^{4}e_{A}^{3}\kappa ^{2}}\left( 16\pi
^{2}\right) ^{2}L_{x}L_{y}\ ,\\
J_{B} &=&\mathcal{J}[\boldsymbol{\psi_{B}}]=-\frac{8Q_{B}}{e_{A}^{4}e_{A}^{3}\kappa ^{2}}\left( 16\pi
^{2}\right) ^{2}L_{x}L_{y}\ , \\
T &=&\frac{e_{A}^{2}+e_{B}^{2}}{16\pi }-\frac{4\left(
Q_{A}^{2}+Q_{B}^{2}\right) }{16\pi r_{+}^{4}}\ , \quad
S =\frac{2r_{0}^{2}}{e_{A}^{3}e_{B}^{3}G_{10}}\left( 16\pi \right)
^{2}L_{x}L_{y}\ .
\end{eqnarray}%
\newline
They satisfy the first law of thermodynamics%
\begin{equation}
dE=TdS+\Omega _{A}dJ_{A}+\Omega _{B}dJ_{B}\ .
\end{equation}

\end{document}